\documentclass[aps,prc,reprint,showpacs,eqsecnum,floatfix]{revtex4-1}

\usepackage[utf8]{inputenc}
\usepackage{amsmath,amssymb,latexsym,amsfonts}
\usepackage{cancel}
\usepackage{color}
\usepackage{combelow}
\usepackage{amssymb}
\usepackage{bm}
\usepackage{graphicx}
\usepackage{epstopdf}
\usepackage{braket}

\def\rtin#1{\text{\tiny #1}}
\def\feq{\ensuremath{f^{\rtin{(eq)}}}}
\def\bq{\begin{eqnarray}}
\def\eq{\end{eqnarray}}
\def\nn{\nonumber}

\def\omegabar{\overline{\omega}}
\def\wp{\overline{p}}
\def\vu{\bm{u}}
\def\vp{\bm{p}}
\def\vv{\bm{v}}
\def\arcsinh{{\rm arcsinh}\,}
\def\etas{(\overline{\eta/s})}

\def\wi{{\widetilde{\imath}}}
\def\wj{{\widetilde{\jmath}}}
\def\hatt{{\hat{t}}}
\def\hati{{\hat{\imath}}}
\def\hatx{{\hat{x}}}
\def\haty{{\hat{y}}}
\def\hatz{{\hat{z}}}

\def\halpha{{\hat{\alpha}}}
\def\hbeta{{\hat{\beta}}}
\def\hgamma{{\hat{\gamma}}}
\def\hsigma{{\hat{\sigma}}}
\def\htau{{\hat{\tau}}}
\def\heta{{\hat{\eta}}}
\def\hlambda{{\hat{\lambda}}}

\def\nn{{\nonumber}}

\begin{document}

\title{High-order quadrature-based lattice Boltzmann models \\
for the flow of ultrarelativistic rarefied gases}


\author{Victor E. \surname{Ambru\cb{s}}}
\thanks{Corresponding author} 
\email[E-mail: ]{victor.ambrus@e-uvt.ro}
\affiliation{Department of Physics, West University of Timi\cb{s}oara,\\
Vasile P\^arvan Avenue 4, 300223 Timi\cb{s}oara, Romania}

\author{Robert Blaga}
\email[E-mail: ]{robert.blaga@e-uvt.ro}

\affiliation{Department of Physics, West University of Timi\cb{s}oara,\\
Vasile P\^arvan Avenue 4, 300223 Timi\cb{s}oara, Romania}

\date{\today}

\begin{abstract}
We present a systematic procedure for the construction of relativistic lattice Boltzmann models (R-SLB) 
appropriate for the simulation of flows of massless particles. Quadrature rules are used for the discretization of the momentum
space in spherical coordinates. The models are optimized for one-dimensional flows. The applications 
considered in this paper are the Sod shock tube and the one-dimensional boost invariant expansion 
(Bjorken flow). Our models are tested against exact solutions in the inviscid and ballistic limits. 
At intermediate relaxation times (finite viscosity), we compare with the results obtained 
using the Boltzmann approach of multiparton scattering model for the Sod shock tube problem, as well as 
with a semi-analytic solution for the non-ideal Bjorken flow. 
In all cases our models give remarkably good results. We define a convergence test in order to find the 
quadrature order necessary to obtain convergence at a predefined accuracy. We show that, while in the hydrodynamic 
regime the number of velocities is comparable to that required for the more popular collision-streaming 
type models, as we go towards the ballistic regime, the size of the velocity set must be substantially 
increased in order to accurately reproduce the analytic profiles.
\end{abstract}

\maketitle

\section{Introduction}\label{sec:intro}

Kinetic theory has long been used for the description of flows far from equilibrium, where 
the hydrodynamic description based on the Navier-Stokes-Fourier equations is no longer applicable.
A few examples of nonrelativistic nonequilibrium flows comprise plasmas \cite{balescu05}, 
flows of rarefied gases \cite{sone06}, as well as flows at microscale \cite{gadelhaq06hbook,gadelhaq06app}.

In extreme conditions, relativistic effects become important, so that the Navier-Stokes-Fourier equations,
based on Galilean relativity, have to be replaced by the equations of relativistic hydrodynamics
\cite{rezzolla13}. Applications of relativistic hydrodynamics comprise flows in special and general relativity, 
such as accretion problems \cite{banyuls97}, stellar collapse \cite{fryer04} or cosmology \cite{ellis12},
relativistic jets emitted by active galactic nuclei \cite{begelman84,marti15}, gamma-ray bursts \cite{kouveliotou12,marti15} or the evolution 
of pulsar wind nebulae, \cite{hester95,marti15}  
as well as the quark-gluon plasma encountered in high-energy particle colliders \cite{jacak12,romatschke10}.

One of the most prolific arenas where relativistic hydrodynamics has been applied is the study of 
the quark-gluon plasma (QGP).  
It has become well established that in heavy-ion collisions performed at modern colliders (RHIC, LHC), a new form of matter,
called quark-gluon plasma, forms as a transitory state \cite{heinz2002,gyulassy2005}. 
In this phase of matter the quarks are deconfined from their hadronic prisons,
and form a plasma made of quarks and gluons. For a recent review on the subject see 
\cite{heinz2013,romatschke17}.

The study of the QGP system is very difficult due to the small temporal and spatial scales associated with the plasma. Experimentally, we only 
have access to the initial state, containing two nuclei,
and the final state, containing the resultants from the collision, as measured by 
the detectors. On the theoretical side, it is difficult to study the system because at these energies the QCD matter is strongly coupled and 
perturbative methods cannot be applied. This is rather unfortunate as both at lower energies (where the quarks are confined into baryons and mesons) 
and at high energies (where the coupling is weak) the system is approachable with traditional methods.
In the present experimental setups however, the QGP is at energies close to the phase transition point in phase space, where 
the coupling is strong \cite{monnai2014}. At such energies, present lattice QCD simulations are computationally costly and have 
large errors \cite{meyer2007,meyer2008,meyer2009,meyer2011}. 

In these circumstances, the preferred approach is to describe the system in a coarse grained way using fluid dynamics. This is backed 
by experimental \cite{romatschke2007} and theoretical \cite{policastro2001,kovtun2005} evidence that the QGP behaves as a nearly perfect fluid. The basic assumption 
in order to use hydrodynamics is that the QGP is at, or close to, thermodynamic equilibrium. This is a reasonable
assumption, as the strong coupling between the constituents draws the system rapidly towards thermal equilibrium.
In this approach it is assumed that the microscopic degrees of freedom have effect only in a statistical way, the system
being faithfully describable through a (small) number of macroscopic fields. 

The complete process is extremely complex, 
containing the collision of two highly Lorentz contracted nuclei, the subsequent formation 
and thermalization of the QGP, the collective evolution of the QGP and, finally, (kinetic) freeze-out, when the energy density drops sufficiently 
so that quarks become (re)confined into baryons or mesons \cite{jaiswal2016}. A complete simulation should contain the whole chain of processes, with one 
stage representing the initial conditions for the following one.  In recent years it has become clear that dissipative effects should 
also be taken into account in order to obtain a full picture. In the dissipative case, 
thermodynamic equilibrium is not fully achieved, and 
the existing gradients of the thermodynamic fields give rise to transport phenomena. This approach has led to a number of remarkable 
successes over the last two decades \cite{adams2005,adcox2005,song2014,adare2011,chatrchyan2014,aad2012}. The basic problem remains however, that the theory of relativistic dissipative hydrodynamics is not fully 
established yet. In particular, the values of the 
transport coefficients are heavily dependent on the way the theory is deduced \cite{el2010,denicol2012,jaiswal2013a,jaiswal2013b,jaiswal2013c,jaiswal2013d,jaiswal2013e,jaiswal2014a,jaiswal2014b,florkowski2015}. 
The best approach is based on relativistic kinetic theory, where all macroscopic fields are derived from the microscopic distribution function,
which evolves according to the Boltzmann equation. 

Such an approach has the advantage that 
the set of hydrodynamic conservation equations, which is highly nonlinear in the viscous 
regime, emerges from the relativistic Boltzmann equation, where the advection is performed 
in a simple manner. In particular, for flows not far from equilibrium, the hydrodynamic limit of the 
Boltzmann equation can be obtained through the Chapman-Enskog expansion \cite{cercignani02}.

Apart from naturally encompassing the nonequilibrium physics induced by 
rarefaction effects, the (relativistic) Boltzmann equation has the advantage that 
the advection term is performed along the momenta of the fluid constituents, rather than 
along the macroscopic flow velocity, making its implementation much simpler than 
required in order to solve the highly nonlinear hydrodynamic equations.
The caveat of the kinetic theory approach is that in the mesoscopic description, 
the usual four-dimensional (4D) space-time is replaced by the 7D phase space 
on which the one-particle distribution function is defined.

The lattice Boltzmann (LB) method offers a prescription for the discretization of the 
momentum space, which is based on the recovery of the macroscopic moments of the distribution 
function, such that the conservation equations are exactly satisfied \cite{succi01}.
It can be shown, via the Chapman-Enskog procedure \cite{cercignani02,cercignani88},
that the LB method correctly recovers the viscous hydrodynamic equations, provided that 
the moments of sufficiently high order of the equilibrium distribution function $\feq$
(the Maxwell-Boltzmann or Maxwell-J\"uttner distributions in the nonrelativistic and 
relativistic regimes, respectively) are exactly recovered. The access to higher-order 
moments of $\feq$ is granted invariably by enriching the velocity set.

The LB approach has been successfully applied for modeling problems as diverse as turbulent flows \cite{chen03}, multiphase \cite{sofonea04} and multicomponent \cite{swift96} flows,
conductivity in graphene \cite{mendoza13} and the statistical properties of solar flares \cite{mendoza14}. These problems 
are all treatable in the context of Newtonian physics. Recently, the first relativistic \cite{mendoza10prl,hupp11,mohseni13,mendoza13prd,gabbana2017pre,gabbana2017} and 
general relativistic \cite{romatschke11}
lattice Boltzmann models have been developed, that can be applied for describing the ultrarelativistic quark-gluon plasma and astrophysical flows.

Traditionally, LB models are developed using the collision-streaming paradigm \cite{succi01},
according to which the particle velocities, lattice spacing, and time step are chosen such that 
at each iteration, each fluid constituent travels via exact streaming to another fluid node. 
This approach is highly attractive due to the simplicity and efficiency of its implementation, 
but in general, it does not allow extensions to high orders. Indeed, increasing the number of velocities 
requires that particles hop over an increasing number of neighbors \cite{philippi06},
which in turn makes the implementation of boundary conditions cumbersome 
\cite{zhang05,meng11jcp,meng14jcp}.
Another approach to extending the collision-streaming models to higher orders 
is to employ multiple distribution functions \cite{lallemand03}.

An alternative approach for the construction of high-order LB models is to give up 
the collision-streaming paradigm and instead employ off-lattice velocity sets.
Here we mention the 2D shell models introduced by Watari and Tsutahara \cite{watari03}
and their 3D extensions \cite{watari04,watari06}, 
as well as the models based on the Gauss-Hermite quadrature proposed by Shan {\it et\,al}.~\cite{shan06}.
An extension of the shell paradigm from two to three dimensions was presented in Ref.~\cite{ambrus12}
for nonrelativistic flows and in Ref.~\cite{romatschke11} for the relativistic flows of massless
particles.
In this approach, the velocity set is obtained by finding the roots of orthogonal polynomials, 
which are in general irrational numbers. Thus, the beauty of the collision-streaming approach is lost, 
but instead one can use the whole spectrum of finite difference, finite element or finite volume 
methods available for general-purpose hydrodynamic codes \cite{rezzolla13,leveque02} in order to 
perform the advection step.

Our aim in this paper is to extend the results in Refs.~\cite{romatschke11,ambrus12}
by providing a systematic recipe for the construction of high-order 
lattice Boltzmann models which employ quadrature techniques for the recovery of 
moments with respect to the spherical coordinate system in the momentum space. 
Specifically, the momentum space is represented using the coordinates 
$(p, \theta, \varphi)$ and the integrals over these coordinates are recovered using the 
Gauss-Laguerre and Gauss-Legendre quadrature methods on the $p$ and $\xi = \cos\theta$ 
coordinates \cite{hildebrand87,shizgal15}, while the 
azimuthal degree of freedom is integrated analytically, due to the symmetries of the 
flows considered in this paper, namely the 1D Riemann problem 
(the Sod shock tube) and the 
one-dimensional boost-invariant expansion flow (the non-ideal Bjorken flow).

First, we note that Ref.~\cite{romatschke11} proposes a radial quadrature 
with respect to the weight function $e^{-\wp} \wp^3$, which is only 
suitable for the recovery of the evolution of the stress-energy tensor $T^{\mu\nu}$. 
While such a quadrature is sufficient for the case when the chemical potential (or fugacity) 
is constant during the flow, in flows such as the Riemann problem, this is not the case and 
the fugacity has to be tracked separately. For this, the quadrature should allow the recovery of 
the evolution of the particle flux four-vector $N^\mu$, which we ensure by constructing the 
radial quadrature with respect to the weight function $e^{-\wp} \wp^2$.
Moreover, the expansion of the equilibrium Maxwell-J\"uttner distribution function 
is performed in Ref.~\cite{romatschke11} with respect to the vector polynomials
$P^{(n)}_{i_1, i_2, \dots i_n}$. In this paper, we propose a novel expansion with respect to 
the orthogonal Legendre polynomials, which is easily extendible to arbitrarily high orders. 

The models that we introduce in this paper are in general applicable to any relativistic flows 
of massless particles. In order to test our models, we focus in this 
paper on two flow problems: the Sod shock tube subcase of the Riemann problem and the 
Bjorken flow. Accordingly, after the 
general presentation of the construction of the models, we will focus on a discretization of the 
momentum space suitable for these particular problems. Our aim is to demonstrate that our models 
are indeed capable of providing an accurate solution of the Boltzmann equation (we use the 
Anderson-Witting approximation for the collision term) throughout all degrees of 
rarefaction, from the inviscid limit up to the free molecular flow (ballistic) regime.

The recovery of the inviscid limit is a true challenge for kinetic theory-based implementations.
This is so because the Boltzmann equation describes fluids with finite dissipation.
In the Anderson-Witting approximation, the inviscid limit corresponds to taking 
the vanishing relaxation time limit ($\tau_{\rtin{A-W}} \rightarrow 0$) of the Boltzmann equation. 
However, even for $\tau_{\rtin{A-W}}$ (the relaxation time in units of a reference time 
$\widetilde{t}_{\rm ref}$, which is defined below) as small as $10^{-4}$, there is still finite dissipation 
since the hydrodynamic constitutive equations predict that, whenever the transport coefficients
(i.e., the shear viscosity $\eta$ and the heat conductivity $\lambda_{\rtin{heat}}$) are non-zero, 
the shear stress and 
the heat flux are proportional to the gradients of the velocity and fugacity, respectively.
In the context of the Sod shock tube problem, 
these quantities exhibit strong discontinuities close to the inviscid limit,
giving rise to non-vanishing shear stress and heat flux which are sharply peaked around 
the shock front (the fugacity is discontinuous near the contact discontinuity as well, while the 
shear stress is non-zero throughout the rarefaction wave due to the non-vanishing velocity gradient).
While the width of these peaks is dependent on the relaxation time, the total 
heat flux can be obtained as a numerical integral over the vicinity of the discontinuous regions.
Comparing this integral with the analytic prediction confirms that our models correctly 
implement dissipation in the hydrodynamic limit. A more thorough study of the capability 
of these models to recover dissipative effects is performed in Ref.~\cite{ambrus18prc} in 
the context of the attenuation of longitudinal waves propagating through a 
relativistic gas of massless particles.

In numerical implementations, the limit $\tau_{\rtin{A-W}} \rightarrow 0$ is problematic for two reasons: first, 
the collision term becomes stiff, and second, the numerical viscosity due to the 
finite time step $\delta t$ and lattice spacing $\delta s$ can become dominant 
when compared to the physical viscosity, which is proportional to $\tau_{\rtin{A-W}}$. 
The first problem can be solved either by ensuring that $\delta t < \tau_{\rtin{A-W}}$,
or by employing implicit-explicit algorithms \cite{wang07}.
For simplicity, in this paper we choose the first approach.
In order to tackle the problem of numerical viscosity, we follow 
Refs.~\cite{jiang96,wang07,feng09,yan08,shi01,gan11,hejranfar17pre,busuioc17} 
and employ the fifth-order weighted essentially non-oscillatory (WENO5) scheme 
for the advection, while for the time marching, we use the third-order Runge-Kutta 
(RK3) algorithm \cite{shu88,gottlieb98,henrick05,trangenstein07,busuioc17}.

On the other end of the rarefaction spectrum, we validate our models by comparing 
our simulation results with the analytic solution of the Vlasov (collisionless Boltzmann) equation.
In this regime, the flow constituents stream freely, such that the populations corresponding to 
each discrete velocity move as individual groups. When the number of velocities is small, 
these groups can be distinctly seen, giving rise to staircaselike macroscopic profiles.
In order to correctly recover the collisionless limit of the Boltzmann equation, 
we find that the quadrature order on the $\theta$ coordinate has to be increased 
to values as high as $110$. We note that the ballistic limit of the Boltzmann equation 
has been successfully recovered using a similar number of quadrature points, in the 
context of the discrete ordinate method employed in the unified gas kinetic schemes 
presented in Refs.~\cite{xu10,chen12,wang12,guo13,guo15}.
To achieve such high quadrature orders, we employed 
high-precision algebra to generate the quadrature points and their corresponding weights.
For completeness and for the benefit of our readers, we include with this paper 
two files containing the roots and weights corresponding to the Gauss-Legendre quadrature 
for quadrature orders between $1$ and $1000$ (see the Supplemental Material \footnote{
See Supplemental Material at (URL will be inserted by the publisher)
for the roots of the Legendre polynomials of orders up to 1000 
(rootsP.txt) and the corresponding weights for the Gauss-Legendre 
quadrature (weightsP.txt)}).

As the relaxation time becomes non-negligible, the dissipative effects become important 
and the viscous regime settles in. For the validation of our models in this regime,
we used the results obtained using the Boltzmann approach of multiparton scattering (BAMPS), 
as well as those obtained using the viscous sharp and smooth transport algorithm (vSHASTA), 
which were reported in Refs.~\cite{bouras10,bouras09prl,bouras09nucl}.
We note that good agreement with the above-mentioned results was also obtained 
with the LB method in Refs.~\cite{mendoza10prl,hupp11,mohseni13}. 


In order to better position our approach in the context of quark-gluon plasma flows,
we discuss below two limitations of the present framework.
In a more realistic treatment of the quark-gluon plasma, quantum statistics should 
be employed for the quarks, antiquarks and gluons. 
An analysis of a mixture comprised of quarks and anti-quarks obeying 
Fermi-Dirac statistics was considered in Ref.~\cite{jaiswal2013d} and was subsequently 
extended by adding gluons described using the Bose-Einstein distribution 
in Ref.~\cite{jaiswal15}.
The Chapman-Enskog analysis of this coupled system performed in Ref.~\cite{jaiswal15}
revealed that the quantity $\lambda_\rtin{heat} T / \eta$ (where $T$ is the temperature)
is strongly dependent on the 
ratio $\alpha = \mu / T$ between the chemical potential $\mu$ and temperature $T$,
being proportional to $\alpha^{-2}$ at small and large values of $\alpha$. 
Up to a constant factor, this behavior reproduces the one found in Ref.~\cite{son06} 
through the anti-de Sitter/conformal field theory (adS/CFT)
correspondence for strongly coupled systems. 

The analysis presented in this paper is limited to a single species of
massless particles obeying Maxwell-J\"uttner statistics. Within this 
framework, the ratio $\lambda_\rtin{heat} T / \eta$ is always constant 
(equal to $5/3$ according to the Chapman-Enskog expansion). An immediate 
generalization would be to consider a mixture of three species, however 
one quickly realizes that when all species follow Maxwell-J\"uttner 
statistics, the system behaves like an ideal gas at large values of $\alpha$ 
(the antiquark distribution is exponentially suppressed by 
a factor $e^{-\alpha}$ and the gluon contribution becomes negligible). 
Thus, the correct treatment of the 
mixture of quarks, antiquarks, and gluons must account for quantum 
statistics. In future work, we plan 
to extend the present framework to the case of quantum statistics following the 
work of Ref.~\cite{coelho2017}, where the Fermi-Dirac statistics 
is implemented for massless particles using the lattice Boltzmann method.

In recent studies, it was highlighted that the bulk viscosity can play an important role 
in ultrarelativistic heavy-ion collisions \cite{ryu15}. Since our approach developed in 
this paper is limited to the flow of massless particles, the effects of the bulk viscosity 
cannot be investigated. In future work, the current scheme could be extended to also account 
for particles of non-vanishing mass, as performed, e.g., in Refs.~\cite{romatschke12} (where a 
non-ideal equation of state is modelled using a modified Boltzmann equation to accommodate a 
density and temperature-dependent particle mass) and \cite{gabbana2017b}. 

This paper is organized as follows.
In Sec.~\ref{sec:boltz}, the relativistic Boltzmann equation, the Landau frame 
and the Anderson-Witting approximation for the collision term are briefly reviewed. A discussion 
of the transport coefficients obtained using the Chapman-Enskog and Grad methods is also provided.
Section~\ref{sec:LB} constitutes our main contribution. Here, we discuss the quadrature method employed 
and its simplification in the contexts of the one-dimensional Riemann problem and of the 
one-dimensional boost-invariant expansion. 
Also, we discuss the procedure
for the expansion of the Maxwell-J\"uttner equilibrium distribution with respect to the generalized 
Laguerre and Legendre polynomials and give explicitly the expansion coefficients up to 
the first order with respect to the Laguerre polyomials and up to the sixth order with 
respect to the Legendre polynomials in Appendix~\ref{app:feq}. In Sec.~\ref{sec:Sod}, our models 
are validated in the context of the Sod shock tube problem by comparing our 
simulation results with analytic formulas in the inviscid and ballistic regimes. 
In the viscous regime, the BAMPS and vSHASTA results from 
Refs.~\cite{bouras10,bouras09prl,bouras09nucl} are used to validate our results. 
At the end of Sec.~\ref{sec:Sod}, we present a convergence test which can be used to determine 
the minimum quadrature order and the minimum expansion order for $\feq$ necessary to 
achieve a given degree of accuracy. 
In Sec.~\ref{sec:bjorken}, the nonideal Bjorken flow is treated with the aid of 
the Milne coordinates \cite{culetu10,okamoto16}. The relativistic Boltzmann equation 
is solved with 
respect to the Milne coordinates by employing vielbein (tetrad) fields 
which allow the momentum space to be decoupled from the details of the choice of 
space-time coordinates.
Our simulation results are validated against analytic solutions in the inviscid 
(ideal Bjorken flow) and ballistic regimes, as well as against the semi-analytic 
solution presented in Ref.~\cite{florkowski13} at finite relaxation times. 
Our conclusions are presented in Sec.~\ref{sec:conc}.
This paper comes in extension of Ref.~\cite{blaga16aip}, where preliminary results obtained 
using these models in the context of shock propagation were presented.

All quantities presented in this paper are nondimensionalized with respect to 
fundamental reference quantities, such as the speed of light 
$\widetilde{v}_{\rm ref} = \widetilde{c}$, the reference time $\widetilde{t}_{\rm ref}$,
the reference particle number density $\widetilde{n}_{\rm ref}$, and the reference 
temperature $\widetilde{T}_{\rm ref}$. The values of the reference quantities 
depend on the given problem and will be discussed separately in Secs.~\ref{sec:Sod}
and \ref{sec:bjorken} for the Sod shock tube problem and one-dimensional 
boost-invariant expansion.
Throughout this paper, the $(-,+,+,+)$ metric signature convention is used.

\section{Relativistic Boltzmann equation}\label{sec:boltz}

In this paper, we address solving the relativistic Boltzmann equation 
in the specific contexts of the Sod shock tube and Bjorken flow, 
from the nearly-inviscid regime to the free streaming regime. 

The Sod shock tube scenario represents a particular instance of the Riemann
problem. The initial state for the Riemann problem consists of two 
semi-infinite domains at rest separated by a thin membrane placed at $z = 0$, 
which is suddenly removed at $t = 0$. The system is considered to be completely 
homogeneous along the $x$ and $y$ axes, such that the flow becomes onedimensional. 

The one-dimensional boost-invariant expansion was first proposed by 
Bjorken \cite{bjorken83} as a model for the longitudinal expansion of a quark-gluon plasma 
system, following the head-on collision of two energetic nuclei. In this model, the velocity 
profile is imposed by symmetry requirements. The problem can be simplified 
considerably by considering the Milne coordinates, formed by the proper time 
$\tau = \sqrt{t^2 - z^2}$ and rapidity $\eta_M = \tanh^{-1}(z / t)$, with respect 
to which the flow becomes stationary.
In order to solve the relativistic Boltzmann equation with respect to the Milne 
coordinates, an orthonormal (non-holonomic) tetrad vector field is employed 
with respect to which the resulting metric becomes the Minkowski metric. 
Our models are validated by comparison with the 
analytic solution in the inviscid and ballistic regimes, as well as with the 
semianalytic results reported in Ref.~\cite{florkowski13}.

This section is structured as follows.
Section~\ref{sec:boltz:boltz} briefly reviews the relativistic Boltzmann 
equation in the Anderson-Witting single relaxation time approximation, written 
with respect to arbitrary coordinates. The tetrad formalism is reviewed 
in Sec.~\ref{sec:boltz:tetrad}. 
The evaluation of the tetrad components of the macroscopic (hydrodynamic) 
variables (i.e., the particle four-flow $N^\halpha$ and stress-energy 
tensor $T^{\halpha\hbeta}$) as moments of the Boltzmann distribution 
function is presented in Sec.~\ref{sec:boltz:hydro}, alongside the construction 
of the Landau frame. The transport coefficients arising when the Anderson-Witting 
model is employed are reviewed in Sec.~\ref{sec:boltz:transp}.
Finally, the relativistic Boltzmann equation for the Sod shock tube 
problem and for the Bjorken flow in Milne coordinates are presented in 
Secs.~\ref{sec:boltz:Sod} and \ref{sec:boltz:milne}, respectively.

\subsection{Boltzmann equation}\label{sec:boltz:boltz}

On a general space-time and with respect to arbitrary coordinates, the Boltzmann equation 
in the Anderson-Witting approximation for the collision term reads \cite{cercignani02}:
\begin{equation}
 p^\mu \frac{\partial f}{\partial x^\mu} - \Gamma^i{}_{\mu\nu} p^\mu p^\nu \frac{\partial f}{\partial p^i} = 
 \frac{p \cdot u}{\tau_{\rtin{A-W}}} [f - \feq],
 \label{eq:boltz_gr}
\end{equation}
where $f$ is the Boltzmann distribution function, $p^\mu$ is the particle 
four-momentum and $\tau_{\rtin{A-W}}$ is the relaxation time, while the Christoffel 
symbols $\Gamma^\kappa{}_{\mu\nu} = g^{\kappa\lambda} \Gamma_{\lambda\mu\nu}$ 
can be obtained using the following formula \cite{misner73}:
\begin{equation}
 \Gamma_{\lambda\mu\nu} = \frac{1}{2}(g_{\lambda\mu,\nu} + g_{\lambda\nu,\mu} - g_{\mu\nu,\lambda}).
 \label{eq:christoffel}
\end{equation}
In this section, we consider that the equilibrium distribution $\feq$ is the 
Maxwell-J\"uttner distribution function:
\begin{equation}
 \feq = \frac{n}{8\pi T^3} \exp\left(\frac{p^\mu u_\mu}{T}\right),
 \label{eq:MJ}
\end{equation}
which is valid for particles of vanishing mass. 
In the above, $n$ is the particle number density, $T$ is the local temperature and
$u^\mu$ is the macroscopic four-velocity of the fluid obtained in the 
Landau frame (more details are given in Sec.~\ref{sec:boltz:hydro}).

The components $p^\mu$ of the momentum four-vector are constrained by the 
mass-shell condition, which for a general space-time can be written 
with respect to the metric $g_{\mu\nu}$ as follows:
\begin{equation}
 p^2 = g_{\mu\nu} p^\mu p^\nu = 0,\label{eq:mshell_gr}
\end{equation}
leading to the following expression for $p^0$:
\begin{equation}
 p^0 = \frac{p_0 - g_{0i}p^i}{g_{00}}, \qquad 
 p_0 = -\sqrt{(g_{0i} g_{0j} - g_{00} g_{ij}) p^i p^j}.
 \label{eq:mshell_p0}
\end{equation}
It can be seen that, even for massless particles, $p^0$ is in general a complicated function 
depending not only on the spatial components $p^i$ of the momentum, but also on the space-time 
coordinates through the metric tensor $g_{\mu\nu}$. 
In order to simplify the structure of the momentum space, we follow Ref.~\cite{cardall13} and 
introduce a tetrad field as the interface between the momentum and coordinate spaces, as will be discussed in the 
following subsection.

\subsection{Tetrad fields}\label{sec:boltz:tetrad}

The line element for a general space-time with metric tensor $g_{\mu\nu}$ can be written as:
\begin{equation}
 ds^2 = g_{\mu\nu} dx^\mu dx^\nu.\label{eq:ds2}
\end{equation}
A set of four space-time vectors 
$e_\halpha = e_\halpha^\mu \partial_\mu$ ($\halpha \in \{0, 1, 2, 3\}$)
forms an orthonormal tetrad frame if
\begin{equation}
 g_{\mu\nu} e^\mu_\halpha e^\nu_\hbeta = \eta_{\halpha\hbeta},
\end{equation}
where $\eta_{\halpha\hbeta} = {\rm diag}(-1, 1, 1,1)$ is the Minkowski metric.
The tetrad coframe associated with the vectors $e_\halpha$ is denoted 
$\omega^\halpha = \omega^\halpha_\mu dx^\mu$ and satisfies
\begin{equation}
 \braket{\omega^\halpha, e_\hbeta} = \omega^\halpha_\mu e_\hbeta^\mu = 
 \delta^\halpha{}_\hbeta, \qquad 
 e_\halpha^\mu \omega^\halpha_\nu  = \delta^\mu{}_\nu, 
\end{equation}
while
\begin{equation}
 \eta_{\halpha\hbeta} \omega^\halpha_\mu \omega^\hbeta_\nu = g_{\mu\nu}.\label{eq:eta_omega}
\end{equation}
For the remainder of this section, hatted indices will refer to components 
expressed with respect to the above tetrad, i.e.,
\begin{equation}
 p^\halpha = \omega^\halpha_\mu p^\mu, \qquad
 p^\mu = e^\mu_\halpha p^\halpha.
\end{equation}
The advantage of employing the tetrad field can be seen when considering the 
mass-shell condition \eqref{eq:mshell_gr}, 
which, by virtue of Eq.~\eqref{eq:eta_omega}, reads
\begin{align}
 p^2 = \eta_{\halpha\hbeta} p^\halpha p^\hbeta &= -(p^{\hat{0}})^2 + (p^{\hat{1}})^2  + (p^{\hat{2}})^2  + (p^{\hat{3}})^2  \nn	\\
 &= -(p^{\hat{0}})^2 + \vp^2 = 0.\label{eq:mshell_tetrad}
\end{align}
As shown in Refs.~\cite{cardall13,ambrus16prd}, the relativistic Boltzmann 
equation \eqref{eq:boltz_gr} can be written with respect to the above 
tetrad frame components $p^\halpha$ of the momentum vector as follows:
\begin{equation}
 p^\halpha e_\halpha^\mu \frac{\partial f}{\partial x^\mu} - 
 \Gamma^\hati{}_{\halpha\hbeta} p^\halpha p^\hbeta \frac{\partial f}{\partial p^\hati} = 
 \frac{p \cdot u}{\tau_{\rtin{A-W}}} [f - \feq],
\end{equation}
where the connection coefficients are now given by
\begin{equation}
 \Gamma^{\hat{\lambda}}{}_{\halpha\hbeta} = \eta^{\hat{\lambda}\hgamma} \Gamma_{\hgamma\halpha\hbeta}, \qquad 
 \Gamma_{\hgamma\halpha\hbeta} = \frac{1}{2}(c_{\hgamma\halpha\hbeta} + c_{\hgamma\hbeta\halpha} - c_{\halpha\hbeta\hgamma}).
 \label{eq:Gamma}
\end{equation}
In the above, $c_{\halpha\hbeta\hgamma}$ are linked to the Cartan coefficients 
$c_{\halpha\hbeta}{}^{\hat{\lambda}}$ through
\begin{equation}
 c_{\halpha\hbeta\hat{\lambda}} = c_{\halpha\hbeta}{}^{\hgamma} \eta_{\hgamma\hat{\lambda}}, \qquad 
 c_{\halpha\hbeta}{}^{\hgamma} = \braket{\omega^\hgamma, [e_\halpha, e_\hbeta]},\label{eq:cartan}
\end{equation}
where the commutator of the vector fields $e_\halpha$ and $e_\hbeta$ is given by
\begin{equation}
 [e_\halpha, e_\hbeta] = (e^\nu_\halpha \partial_\nu e_\hbeta^\mu - 
 e^\nu_\hbeta \partial_\nu e_\halpha^\mu) \partial_\mu,
\end{equation}
such that the Cartan coefficients can be written as
\begin{equation}
 c_{\halpha\hbeta}{}^{\hgamma} = \omega^\hgamma_\mu(e^\nu_\halpha \partial_\nu e_\hbeta^\mu - 
 e^\nu_\hbeta \partial_\nu e_\halpha^\mu).\label{eq:cartan2}
\end{equation}

Finally, Eq.~\eqref{eq:mshell_tetrad} allows $p^{\hat{0}}$ to be written as:
\begin{equation}
 p^{\hat{0}} = p, \qquad p = \sqrt{(p^{\hat{1}})^2 + (p^{\hat{2}})^2 + (p^{\hat{3}})^2}.\label{eq:p0}
\end{equation}
The above expression for $p^{\hat{0}}$ prompts the use of the spherical coordinates to represent the 
tetrad components $p^\halpha$ of the momentum space, as follows:
\begin{align}
 p^{\hat{1}} =& p \cos\varphi \sqrt{1 - \xi^2},  \nonumber\\
 p^{\hat{2}} =& p \sin\varphi \sqrt{1 - \xi^2}, \nn \\ 
 p^{\hat{3}} =& p \xi,\label{eq:sph_gr} 
\end{align}
where $\xi = \cos\theta$.
In order to account for this coordinate transformation in the momentum space, it is convenient to 
employ the Boltzmann equation in conservative form \cite{cardall13,ambrus16prd}:
\begin{multline}
 \frac{1}{\sqrt{-g}} \partial_\mu\left(\sqrt{-g} e^{\mu}_\halpha p^\halpha f\right) -
 \frac{p^{\hat{0}}}{\sqrt{\lambda}} \frac{\partial}{\partial p^\wi} 
 \left(P^{\wi}{}_{\hati} \Gamma^\hati{}_{\halpha\hbeta} \frac{p^\halpha p^\hbeta }{p^{\hat{0}}} f \sqrt{\lambda}\right) \\
 = \frac{p \cdot u}{\tau_{\rtin{A-W}}} [f - \feq],
 \label{eq:boltz_cons}
\end{multline}
where $p^\wi \in \{p, \xi, \varphi\}$ represent the new coordinates in the momentum space,
$\lambda^{-1/2} = |{\rm det} P^{\widetilde{\jmath}}{}_{\hat{\imath}}|$ and the matrix 
$P^\wj{}_\hati = \partial p^\wj / \partial p^\hati$ has the following components \cite{ambrus15wut}:
\begin{equation}
 P^\wj{}_\hati =
 \begin{pmatrix}
  \cos\varphi \sqrt{1 - \xi^2} & \sin\varphi \sqrt{1 - \xi^2} & \xi\\
  {\displaystyle -\frac{\xi}{p} \cos\varphi \sqrt{1 - \xi^2} }& 
  {\displaystyle -\frac{\xi}{p} \sin\varphi \sqrt{1 - \xi^2} } & {\displaystyle \frac{1 - \xi^2}{p}}\\
  {\displaystyle -\frac{\sin\varphi}{p\sqrt{1 - \xi^2}}} & {\displaystyle \frac{\cos\varphi}{p\sqrt{1 - \xi^2}}} & 0
 \end{pmatrix}.
\end{equation}

\subsection{Hydrodynamic moments of the distribution function}\label{sec:boltz:hydro}

In this paper, we are only interested in tracking the 
particle four-flow $N^\mu$ and the stress-energy tensor $T^{\mu\nu}$, 
which are defined as follows \cite{cercignani02}:
\begin{gather}
 N^\mu = \sqrt{-g} \int \frac{d^3p}{(-p_0)}\, f\, p^\mu,  \\
 T^{\mu\nu} = \sqrt{-g} \int \frac{d^3p}{(-p_0)} \, f\, p^\mu p^\nu.
 \label{eq:macro_def_coord}
\end{gather}
The above expressions exhibit a nontrivial dependence on the space-time metric 
due to the factor $p_0$, which is given in Eq.~\eqref{eq:mshell_p0}. 
However, the tetrad components 
$N^\halpha$ and $T^{\halpha\hgamma}$ can be easily written in terms of the 
tetrad components of the momentum vector as follows:
\begin{equation}
 N^\halpha = \int \frac{d^3p}{p^{\hat{0}}}\, f\, p^\halpha, \qquad 
 T^{\halpha\hgamma} = \int \frac{d^3p}{p^{\hat{0}}} \, f\, p^\halpha p^\hgamma.
 \label{eq:macro_def}
\end{equation}
On a general space-time, these moments are conserved with respect to the Boltzmann 
equation. Indeed, multiplying Eq.~\eqref{eq:boltz_cons} with $1$ and $p^\halpha$ 
and integrating over the momentum space yields
\begin{align}
 \nabla_\halpha N^\halpha =& 
 e_\halpha^\mu \partial_\mu N^\halpha  + \Gamma^{\halpha}{}_{\hgamma\halpha} N^\hgamma =
 0,\nonumber\\ 
 \nabla_\halpha T^{\halpha\hgamma} =& 
 e_\halpha^\mu \partial_\mu T^{\halpha\hgamma} + 
 \Gamma^\halpha{}_{\hsigma\halpha} T^{\hsigma\hgamma} + \Gamma^\hgamma{}_{\hsigma\halpha} 
 T^{\halpha\hsigma} = 0.
 \label{eq:cons}
\end{align}
The right-hand sides of Eqs.~\eqref{eq:cons} vanish only when $1$ and $p^\halpha$ 
are collision invariants \cite{cercignani02}:
\begin{align}
 u_\hgamma \int \frac{d^3p}{p^{\hat{0}}} [f - \feq] p^\hgamma = u_\hgamma(N^\hgamma - N^\hgamma_{\rtin{eq}}) = 0,\nonumber\\
 u_\hgamma \int \frac{d^3p}{p^{\hat{0}}} [f - \feq] p^\halpha p^\hgamma = 
 u_\hgamma(T^{\halpha\hgamma} - T^{\halpha\hgamma}_{\rtin{eq}}) = 0. \label{eq:collinv}
\end{align}
The above relations are automatically satisfied when $u^\halpha$ is the 
Landau velocity, defined through the following eigenvalue equation:
\begin{equation}
 T^\halpha{}_\hgamma u^\hgamma = -E u^\halpha,
 \label{eq:EL_def}
\end{equation}
subject to the conditions $E > 0$, $u^{\hat{0}} > 0$ and $u^2 = -1$. 
Furthermore, Eq.~\eqref{eq:collinv} imposes that $\feq$ \eqref{eq:MJ} is defined in terms of the Landau energy $E$, 
Landau particle number density $n = -u_\hgamma N^\hgamma$, and Landau velocity $u^\hgamma$. 
Equation \eqref{eq:MJ} can be rewritten with respect to the tetrad frame components $p^\halpha$ and $u^\halpha$:
\begin{equation}
  \feq = \frac{n}{8\pi T^3} \exp\left(\frac{p^\halpha u_\halpha}{T}\right).
  \label{eq:MJ_tetrad}
\end{equation}

Equation \eqref{eq:EL_def} defines the Landau velocity with respect 
to which the Landau (energy) frame is established \cite{landau87}.
The stress-energy tensor $T^{\halpha\hgamma}$ can be decomposed with respect to 
the Landau velocity $u^\halpha$ as
\begin{equation}
 T^{\halpha\hgamma} = E u^\halpha u^\hgamma + (P + \omegabar) \Delta^{\halpha\hgamma} +
 \Pi^{\halpha\hgamma},\label{eq:land_tmunu}
\end{equation}
where $\Delta^{\halpha\hgamma} = \eta^{\halpha\hgamma} + u^\halpha u^\hgamma$ is the
projector on the hypersurface orthogonal to $u^\halpha$.
The macroscopic energy density $E$, hydrostatic pressure $P$,
dynamic pressure $\omegabar$, and pressure deviator $\Pi^{\halpha\hbeta}$ 
can be obtained as follows:
\begin{equation} 
 E = u_\halpha u_\hgamma T^{\halpha\hgamma}, \quad
 P + \omegabar = \frac{1}{3} \Delta_{\halpha\hgamma} T^{\halpha\hgamma}, \quad
 \Pi^{\halpha\hgamma} = T^{\langle\halpha\hgamma\rangle},\label{eq:en_eck}
\end{equation}
where the notation $A^{\langle\halpha\hgamma\rangle}$ stands for:
\begin{equation}
 A^{\langle\halpha\hgamma\rangle} = \left[\frac{1}{2} \left(\Delta^\halpha{}_\hlambda 
 \Delta^\hgamma{}_\hsigma +
 \Delta^\hgamma{}_\hlambda \Delta^\halpha{}_\hsigma \right) -
 \frac{1}{3} \Delta^{\halpha\hgamma} \Delta_{\hlambda\hsigma} \right] A^{\hlambda\hsigma}.
 \label{eq:angular_def}
\end{equation}
For the case of massless particles considered in this paper, $\omegabar$ vanishes 
since $T^{\halpha}{}_{\halpha} = 0$ and $E = 3P$. 
In the case when the flow constituents obey the Maxwell-J\"uttner statistics, the 
temperature $T$ is related to the pressure via the equation of state of the 
ideal gas:
\begin{equation}
 P = nT.\label{eq:T_def}
\end{equation}
The particle number density $n$ can be related to $N^\halpha$ 
through \cite{bouras10,cercignani02}
\begin{equation}
 N^\halpha = n u^\halpha + J^\halpha, \qquad 
 J^\halpha = -\frac{n}{E + P} q^\halpha,\label{eq:qL}
\end{equation}
where $J^\halpha$ and the heat flux $q^\halpha$ are orthogonal to $u^\halpha$. 

In terms of the spherical coordinates \eqref{eq:sph_gr},
the integrals in Eqs.~\eqref{eq:macro_def} become
\begin{equation}
 \begin{pmatrix}
  N^\halpha \\
  T^{\halpha\hgamma} 
 \end{pmatrix}
 = \int_0^\infty dp\, p \int_{-1}^1 d\xi \int_0^{2\pi} d\varphi\, f\, 
 \begin{pmatrix}
  p^\halpha\\
  p^{\halpha} p^{\hgamma}
 \end{pmatrix}.
 \label{eq:macro_def_sph}
\end{equation}

\subsection{Transport coefficients}\label{sec:boltz:transp}

In flows close to equilibrium, the heat flux $q^\halpha$ and pressure deviator 
$\Pi^{\halpha\hbeta}$ can be written as follows \cite{rezzolla13,cercignani02}:
\begin{subequations}\label{eq:hydro}
\begin{align}
 q^\halpha =& -\lambda_{\rtin{heat}} \Delta^{\halpha\hbeta} e_\hbeta^\nu 
 \left(\partial_\nu T - \frac{T}{E+P} \partial_\nu P\right) \label{eq:hydro_q_orig}\\
 =&\ \frac{\lambda_{\rtin{heat}} T}{4} \Delta^{\halpha\hbeta} e_\hbeta^\nu 
 \partial_\nu \ln \overline{\lambda}, \label{eq:hydro_q}\\
 \Pi_{\halpha\hbeta} =& -2\eta \nabla_{<\halpha} u_{\hbeta>},\label{eq:hydro_pi}
\end{align}
\end{subequations}
where the relative fugacity $\overline{\lambda}$ is defined as:
\begin{equation}
 \overline{\lambda} = \frac{P}{T^4}. \label{eq:fugacity}
\end{equation}
Noting that for ultrarelativistic 
flows, the dynamic pressure $\omegabar$ and thus also the coefficient of bulk viscosity vanishes,
the heat conductivity $\lambda_{\rtin{heat}}$ and shear viscosity $\eta$ 
in Eqs.~\eqref{eq:hydro} comprise the 
only non-vanishing transport coefficients for the ultrarelativistic fluid.
When the Anderson-Witting approximation for the collision term is employed,
expressions for the transport coefficients can be obtained in the hydrodynamic 
regime (small $\tau_{\rtin{A-W}}$)
through either the Chapman-Enskog procedure \cite{cercignani02,anderson74b,ambrus17aip} 
or via Grad's 14
moments approach \cite{cercignani02}. While in the nonrelativistic limit, 
the results from the two approaches
coincide, in the ultrarelativistic limit
they give different expression for the shear viscosity and heat conductivity \cite{cercignani02}:
\begin{subequations}\label{eq:tcoeff}
\begin{align}
&\text{Grad's method:} & \eta =& \frac{2}{3}P\tau_{\rtin{A-W}}, & 
\lambda_\rtin{heat} =& \frac{4}{5}\frac{P\tau_{\rtin{A-W}}}{T},\label{eq:tcoeff_grad}\\
&\text{Chapman-Enskog:}& \eta =& \frac{4}{5}P\tau_{\rtin{A-W}}, & 
\lambda_\rtin{heat} =& \frac{4}{3}\frac{P\tau_{\rtin{A-W}}}{T}.\label{eq:tcoeff_ce}
\end{align}
\end{subequations}

In Ref.~\cite{romatschke11}, the Grad expression for $\eta$ is preferred. There is, however, a recent indication in the literature that the Chapman-Enskog expansion
leads to better agreement with the solution of the Boltzmann equation in the
relaxation time approximation for the collision term, such as the Anderson-Witting
model used in this paper \cite{bhalerao14,ryblewski15,florkowski15,gabbana2017,ambrus18prc}.
Thus, for the remainder of this paper, we use the Chapman-Enskog expressions for $\eta$
and $\lambda_{\rtin{heat}}$, unless otherwise specified. 

\subsection{Equation for the Sod shock tube problem}\label{sec:boltz:Sod}

In the case of the Sod shock tube problem, the background space-time 
is the Minkowski space and the tetrad field is trivially given by
\begin{equation}
 e_{\hat{0}} = \partial_t, \qquad e_{\hat{1}} = \partial_x, 
 \qquad e_{\hat{2}} = \partial_y, 
 \qquad e_{\hat{3}} = \partial_z,
\end{equation}
such that all connection coefficients $\Gamma_{\hgamma\halpha\hbeta}$ vanish. Assuming 
that the flow is homogeneous along the $x$ and $y$ directions, Eq.~\eqref{eq:boltz_cons} 
reduces to
\begin{equation}
 \partial_t f + \xi \partial_z f = - \frac{\gamma(1 - \beta \xi)}{\tau_{\rtin{A-W}}} 
 [f - \feq],
 \label{eq:boltz_Sod}
\end{equation}
where the four-velocity was taken as $u^\halpha = (\gamma, 0, 0, \beta \gamma)$,
such that $p \cdot u = -p \gamma (1 - \beta \xi)$. The equilibrium distribution 
simplifies to
\begin{equation}
 \feq = \frac{n}{8\pi T^3} \exp\left[-\frac{p\gamma}{T}(1 - \beta \xi)\right].
\end{equation}

Since the flow in the Sod shock tube problem is unidirectional, 
$f$ can be assumed to be independent of $\varphi$ and the integrals with respect to 
$\varphi$ can be performed automatically in Eq.~\eqref{eq:macro_def_sph}, yielding 
\begin{equation}
 N^\halpha = 
 \begin{pmatrix}
  N^{\hatt} \\ 0\\ 0 \\ N^{\hatz}
 \end{pmatrix},\qquad  
 T^{\halpha\hbeta} =
 \begin{pmatrix}
  T^{\hatt \hatt} & 0 & 0 & T^{\hatt\hatz}\\
  0 & T^{\hatx\hatx} & 0 & 0 \\
  0 & 0 & T^{\haty\haty} & 0 \\
  T^{\hatz\hatt} & 0 & 0 & T^{\hatz\hatz}
 \end{pmatrix},\label{eq:tmunu_comp}
\end{equation}
where $T^{\hatt\hatz} = T^{\hatz\hatt}$ and 
\begin{equation}
 T^{\hatx\hatx} = T^{\haty\haty} = \frac{1}{2}(T^{\hatt\hatt} - T^{\hatz\hatz}).\label{eq:TxxTyy}
\end{equation}

Let us now construct the Landau frame for the stress-energy tensor in Eq.~\eqref{eq:tmunu_comp}.
The solution of Eq.~\eqref{eq:EL_def} is
\begin{equation} \label{eq:E_landau}
 E = \frac{1}{2}\left[T^{\hatt\hatt} - T^{\hatz\hatz} +
 \sqrt{(T^{\hatt\hatt} + T^{\hatz\hatz})^2 - 4(T^{\hatt\hatz})^2}\right],
\end{equation}
while the Landau velocity is given by
\begin{gather} \label{eq:u_landau}
 u^\halpha = \gamma(1, 0, 0, \beta)^T, \qquad \nn
 \beta = \frac{T^{\hatt\hatz}}{E + T^{\hatz\hatz}},  \\
 \gamma = (1 - \beta^2)^{-1/2}.
\end{gather}
In order to satisfy the orthogonality relation $q_\halpha u^\halpha = 0$,
the heat flux $q^\halpha$, defined in Eq.~\eqref{eq:qL}, must have the form:
\begin{equation}
 q^\halpha = q (\beta, 0, 0, 1)^T.\label{eq:q_def}
\end{equation}
Using the definition \eqref{eq:qL} of $q^\halpha$, $q$ can be obtained as follows:
\begin{equation} 
 q = -4T \left(N^\hatz - n \beta \gamma\right).\label{eq:q_val}
\end{equation}
Finally, the restrictions $u_\halpha \Pi^{\halpha\hbeta} = 0$
and $\Pi^\halpha{}_\halpha = 0$ reduce the number of degrees of freedom
of the pressure deviator $\Pi^{\halpha\hbeta}$ to a single number $\Pi$ \cite{bouras10}:
\begin{equation}
 \Pi^{\halpha\hbeta} = \Pi
 \begin{pmatrix}
  \beta^2 \gamma^2  & 0 & 0 & \beta \gamma^2 \\
  0 & -\frac{1}{2} & 0 & 0\\
  0 & 0 & -\frac{1}{2} & 0\\
  \beta \gamma^2 & 0 & 0 & \gamma^2
 \end{pmatrix}.\label{eq:pi_def}
\end{equation}
Substituting the above expression into \eqref{eq:land_tmunu} allows $\Pi$ to be written as:
\begin{equation} 
 \Pi = \frac{2}{3} E -T^{\hatt\hatt} + T^{\hatz\hatz}.\label{eq:pi_val}
\end{equation}

\subsection{Equation for the one-dimensional boost-invariant expansion}\label{sec:boltz:milne}

The formalism described in this section can be specialized to the case of the 
Milne coordinate system, for which the line element \eqref{eq:ds2} reads:
\begin{equation}
 ds^2 = -d\tau^2 + dx^2 + dy^2 + \tau^2 d\eta_M^2,\label{eq:milne_ds2}
\end{equation}
where $\tau = \sqrt{t^2 - z^2}$ is the proper time and 
$\eta_M = \tanh^{-1}(z/t) = \frac{1}{2} \ln \frac{t + z}{t - z}$ is the rapidity \cite{bjorken83}.
The square root of the determinant of the metric $g_{\mu\nu}$ is
\begin{equation}
 \sqrt{-g} = \tau.\label{eq:milne_sqrtg}
\end{equation}
A natural choice for the orthonormal tetrad corresponding to the metric \eqref{eq:milne_ds2} is:
\begin{align}
 \omega^\htau =& d\tau, & 
 \omega^\hatx =& dx, & 
 \omega^\hatx =& dy, &
 \omega^{\heta_M} =& \tau \, d\eta_M,\nonumber\\
 e_\htau =& \partial_\tau, & 
 e_\hatx =& \partial_x, & 
 e_\haty =& \partial_y, &
 e_{\heta_M} =& \tau^{-1} \partial_{\eta_M}.
 \label{eq:milne_tetrad}
\end{align}
The non-vanishing Cartan coefficients \eqref{eq:cartan} are:
\begin{equation}
 c_{\htau\heta_M}{}^{\heta_M} = -c_{\heta_M\htau}{}^{\heta_M} = -\frac{1}{\tau},
\end{equation}
giving rise to the following connection coefficients \eqref{eq:Gamma}:
\begin{equation}
 \Gamma_{\htau\heta_M\heta_M} = - \Gamma_{\heta_M\htau\heta_M} = -\frac{1}{\tau}.
 \label{eq:bjorken_Gamma}
\end{equation}
The Boltzmann equation in conservative form \eqref{eq:boltz_cons} reduces to
\begin{multline}
 \frac{1}{\tau} \partial_\tau (\tau f) -
 \frac{\xi^2}{\tau p^2} \partial_p (fp^3) - \frac{1}{\tau} \partial_\xi[\xi(1-\xi^2) f] \\
 = -\frac{1}{\tau_{\rtin{A-W}}} [f - \feq].\label{eq:boltz_Milne}
\end{multline}
where homogeneity along the spatial directions $x$, $y$, and $\eta_M$
was assumed, such that the Landau velocity is 
\begin{equation}
 u^\halpha = (1,0,0,0)^T\label{eq:bjorken_u}.
\end{equation}
In this case, the equilibrium distribution \eqref{eq:MJ} simplifies to
\begin{equation}
 \feq = \frac{n}{8\pi T^3} e^{-p/T},\label{eq:Bjorken_MB}
\end{equation}
while $N^\halpha = (n, 0, 0,0)^T$ and no heat flux 
is present ($q^\halpha = 0$).
Furthermore, the pressure deviator $\Pi^{\halpha\hgamma}$ reduces to
\begin{equation}
 \Pi^{\halpha\hgamma} = \Pi\, {\rm diag}\left(0, -\frac{1}{2}, -\frac{1}{2}, 1\right).
 \label{eq:bjorken_pi}
\end{equation}
Thus, the stress-energy tensor \eqref{eq:land_tmunu} is diagonal, 
having the following form:
\begin{equation}
 T^{\halpha\hbeta} = {\rm diag}(E, \mathcal{P}_T, \mathcal{P}_T, \mathcal{P}_L),
 \label{eq:bjorken_SET}
\end{equation}
where the transverse pressure $\mathcal{P}_T$ (along the $x$ and $y$ directions) 
and longitudinal pressure $\mathcal{P}_L$ (along the $\eta_M$ direction) are given by 
\cite{florkowski13}
\begin{equation}
 \mathcal{P}_T = P - \frac{\Pi}{2}, \qquad \mathcal{P}_L = P + \Pi.
 \label{eq:bjorken_P}
\end{equation}

\section{Lattice Boltzmann model}\label{sec:LB}

We now wish to employ the lattice Boltzmann (LB) method to solve 
Eqs.~\eqref{eq:boltz_Sod} and \eqref{eq:boltz_Milne} numerically.
Specifically, we are interested in constructing a model which can 
successfully tackle the numerical challenges of simulating flows of all 
degrees of rarefaction, starting from the inviscid limit (small $\tau_{\rtin{A-W}}$)
up to the ballistic regime ($\tau_{\rtin{A-W}} \rightarrow \infty$). We present our solution as a two-step process:
first, a quadrature method is chosen which allows the correct recovery of the
first and second-order moments of $f$; second, $\feq$ in the Anderson-Witting collision term
is replaced by a suitably truncated series expansion such that the relevant moments of
the collision term are correctly recovered.

Since the flows considered in this paper are independent of the azimuthal momentum 
space coordinate 
$\varphi$, we do not consider the problem of recovering integrals over this variable here.
We note that a suitable quadrature was proposed by Mysovskikh \cite{mysovskikh88} and 
applications of this quadrature to LB modeling can be found in 
Refs.~\cite{watari03,romatschke11,ambrus12}.

In Sec.~\ref{sec:LB:quad}, the quadrature problem is stated. Sections~\ref{sec:LB:qp} and 
\ref{sec:LB:qxi} discuss the quadrature method for the radial and angular coordinates $p$ and 
$\xi$, respectively, as well as the construction of the momentum space derivatives occuring in 
Eq.~\eqref{eq:boltz_Milne} for the one-dimensional boost-invariant expansion.
In Sec.~\ref{sec:LB:feq}, the expansion of $\feq$ 
is presented, while Sec.~\ref{sec:LB:discrete} summarizes the model construction.
The finite-difference scheme employed in this paper is summarized in Sec.~\ref{sec:LB:num}.

\subsection{Quadratures}\label{sec:LB:quad}

In the following, we will focus on constructing a quadrature procedure suitable for the recovery of
the following type of moments of $f$:
\begin{equation}
 M_{s,r} = \int \frac{d^3p}{p^{\hat{0}}} f\, p^{s+1} \xi^r.
 \label{eq:momsr_def}
\end{equation}
Since $f$ is independent of $\varphi$, the integral with respect to $\varphi$ can be 
performed analytically:
\begin{align}
 M_{s,r} = 2\pi \int_0^{\infty} dp\, p^{s+2}\, \int_{-1}^1 d\xi\, f\, \xi^r.\label{eq:momsr}
\end{align}
With respect to the above notation, the non-vanishing components of $N^\halpha$ and 
$T^{\halpha\hbeta}$ that appear in 
Eqs.~\eqref{eq:tmunu_comp} in the context of the Sod shock tube problem 
can be written as:
\begin{gather}
 N^{\hatt} = M_{0,0}, \qquad N^{\hat{z}} = M_{0,1}, \nn \\ 
 T^{\hatt\hatt} = M_{1,0}, \qquad T^{\hatt\hatz} = M_{1,1}, \qquad 
 T^{\hatz\hatz} = M_{1,2},\label{eq:tmunu_comp_momsr}
\end{gather}
while $T^{\hat{x}\hat{x}} = T^{\hat{y}\hat{y}} = \frac{1}{2}(M_{1,0} - M_{1,2})$.
Similar relations hold in the case of the Bjorken flow, when 
$N^{\hat{\eta}_M} = T^{\htau\hat{\eta}_M} = 0$.
The recovery of the moments \eqref{eq:momsr} requires quadrature methods 
which are discussed below.

\subsection{Radial quadrature}\label{sec:LB:qp}

\begin{table}
\begin{center}
\begin{ruledtabular}
\begin{tabular}{l|l|l}
 $k$ & $\wp_k$ & $w_k^L$ \\\hline\hline
 $1$ & $2$ & $1.5$\\
 $2$ & $6$ & $0.5$ \\
\end{tabular}
\end{ruledtabular}
\end{center}
\caption{The quadrature roots and weights employed for the
radial quadrature.\label{tab:quad_lag}
}
\end{table}

In order to tackle the integral with respect to $p$ in Eq.~\eqref{eq:momsr},
$f$ can be expanded as follows:
\begin{equation}
 f = \frac{1}{T_0^3} e^{-\wp} \sum_{\ell = 0}^\infty \frac{1}{(\ell + 1)(\ell + 2)} 
 \mathcal{F}_\ell L_\ell^{(2)}(\wp),
 \label{eq:fseriesp}
\end{equation}
where $\wp = p / T_0$ is the ratio between the magnitude $p$ of the particle momentum and some reference
temperature $T_0$ \cite{romatschke11}, $L_\ell^{(2)}$ is the generalized Laguerre polynomial of
type $2$ and order $\ell$, given explicitly in Eq.~\eqref{eq:lag_expl}, 
while the expansion coefficients $\mathcal{F}_\ell$ can be obtained as follows:
\begin{equation}
 \mathcal{F}_\ell = \int_0^\infty dp\, p^2\, f\, L^{(2)}_{\ell}(\wp).
 \label{eq:F}
\end{equation}
The compatibility between Eqs.~\eqref{eq:fseriesp} and \eqref{eq:F} is ensured by the
orthogonality and completeness relations satisfied by the generalized Laguerre polynomials:
\begin{subequations}\label{eq:lag_ortho_compl}
\begin{align}
 \int_0^\infty d\wp\, e^{-\wp} \wp^2\, L^{(2)}_{\ell}(\wp) L^{(2)}_{\ell'}(\wp) = 
 (\ell + 1)(\ell + 2) \delta_{\ell,\ell'},
 \label{eq:lag_ortho}\\
 \sum_{\ell = 0}^\infty \frac{L_\ell^{(2)}(\wp) L_\ell^{(2)}(\wp')}{(\ell + 1)(\ell + 2)}
 = \delta(\wp'-\wp) \frac{e^{(\wp + \wp')/2}}{\wp\,\wp'}.\label{eq:compl}
\end{align}
\end{subequations}
An expansion similar to the one in Eq.~\eqref{eq:fseriesp} can be performed for 
the term involving the $p$ derivative of $f$ in Eq.~\eqref{eq:boltz_Milne}:
\begin{equation}
 \frac{1}{p^2} \frac{\partial(f p^3)}{\partial p} = 
 \frac{e^{-\wp}}{T_0^3} \sum_{\ell = 0}^\infty \frac{1}{(\ell + 1)(\ell + 2)} \mathcal{F}^{p}_\ell L_\ell^{(2)}(\wp),
 \label{eq:f_lag_force}
\end{equation}
where the coefficients $\mathcal{F}^{p}_\ell$ can be obtained starting from Eq.~\eqref{eq:F},
by applying integration by parts:
\begin{equation}
 \mathcal{F}^{p}_{\ell} = -\int_0^\infty dp\,p^2\,f\, p \frac{\partial L_\ell^{(2)}(\wp)}{\partial p}.
 \label{eq:f_lag_force_coeff_aux}
\end{equation}
Combining the following properties of the generalized Laguerre polynomials \cite{olver10}:
\begin{align}
 \frac{dL_\ell^{(\alpha)}(\wp)}{d\wp} =& - L_{\ell - 1}^{(\alpha + 1)}(\wp),\nonumber\\
 \wp L_{\ell-1}^{(\alpha + 1)}(\wp) =& (\ell + \alpha) L_{\ell-1}^{(\alpha)}(\wp) - 
 \ell L_\ell^{(\alpha)}(\wp),
\end{align}
the term $p [d L_\ell^{(2)}(\wp) / dp]$ can be expressed using the generalized Laguerre polynomials as follows:
\begin{equation}
 p \frac{\partial L_\ell^{(2)}(\wp)}{\partial p} = \ell L_\ell^{(2)}(\wp) - (\ell + 2) L_{\ell - 1}^{(2)}(\wp),
\end{equation}
such that Eq.~\eqref{eq:f_lag_force_coeff_aux} becomes:
\begin{equation}
 \mathcal{F}^{p}_\ell = 
 \begin{cases}
  (\ell + 2) \mathcal{F}_{\ell - 1} - \ell \mathcal{F}_\ell,& \ell > 0,\\
  0,& \ell = 0.
 \end{cases}
\end{equation}
Substituting the above result into Eq.~\eqref{eq:f_lag_force} leads to the following expression:
\begin{equation}
 \frac{1}{p^2} \frac{\partial(f p^3)}{\partial p} = 
 \frac{e^{-\wp}}{T_0^3} \sum_{\ell = 1}^\infty \frac{L_\ell^{(2)}(\wp)}{(\ell + 1)}
 \left[\mathcal{F}_{\ell - 1} - \frac{\ell}{\ell + 2} \mathcal{F}_\ell\right].
 \label{eq:f_lag_force_sum}
\end{equation}

For later convenience, Eq.~\eqref{eq:f_lag_force_sum} can be expressed 
by taking into account the definition \eqref{eq:F} of the coefficients $\mathcal{F}_{\ell}$ 
in the following integral form:
\begin{equation}
 \frac{1}{p^2} \frac{\partial(f p^3)}{\partial p} = \int_0^\infty dp'\, p'{}^2\,\mathcal{K}^\rtin{L}(p, p') f(p'),
 \label{eq:f_lag_force_K}
\end{equation}
where we have introduced a kernel $\mathcal{K}^\rtin{L}(p,p')$ following Refs.~\cite{ambrus17pof,busuioc17}, 
having the following expression:
\begin{multline}
 \mathcal{K}^\rtin{L}(p,p') = \frac{e^{-\wp}}{T_0^3} \sum_{\ell = 1}^\infty \frac{1}{(\ell + 1)(\ell + 2)} L_\ell^{(2)}(\wp) \\
 \times \left[(\ell + 2) L_{\ell - 1}^{(2)}(\wp') - \ell\, L_\ell^{(2)}(\wp')\right].
 \label{eq:K_lag}
\end{multline}

We now consider the truncation $f^{(L)}$ of $f$ at order $L$ with respect to $\ell$:
\begin{equation}
 f^{(L)} = \frac{1}{T_0^3} e^{-\wp} \sum_{\ell = 0}^L \frac{1}{(\ell + 1)(\ell + 2)}
 \mathcal{F}_\ell L_\ell^{(2)}(\wp).\label{eq:fseriespL}
\end{equation}
After the replacement of $f$ with $f^{(L)}$ in Eq.~\eqref{eq:momsr}, the
moments $M_{s,r}$ become:
\begin{multline}
 M_{s,r} =  2\pi T_0^s \sum_{\ell =0}^L \frac{1}{(\ell + 1)(\ell + 2)} \\ 
 \times \left[\int_0^\infty d\wp\,e^{-\wp}
 \wp^{s + 2} L_{\ell}^{(2)}(\wp) \right]
 \left[\int_{-1}^1 d\xi \, \mathcal{F}_{\ell} \xi^r\right].\label{eq:momsr_aux}
\end{multline}
The integrand in
the integral with respect to $\wp$ can be written as a polynomial in $\wp$ multiplied by the weight
function $\omega(\wp) = e^{-\wp} \wp^2$. The recovery of such integrals can be performed using
the Gauss-Laguerre quadrature method, which can be summarized as:
\begin{equation}
 \int_0^\infty d\wp \, e^{-\wp} \wp^2\, L_\ell^{(2)}(\wp) \wp^s \simeq
 \sum_{k = 1}^{Q_\rtin{L}} w_k^L L^{(2)}_\ell(\wp_k) \wp_k^s, \label{eq:quadL}
\end{equation}
where the $Q_\rtin{L}$ quadrature points $\wp_k$ are
the roots of $L_{Q_\rtin{L}}^{(2)}$ and $w_k^L$ are the corresponding quadrature weights,
given by \cite{hildebrand87,shizgal15}:
\begin{equation}
 w_k^L = \frac{\wp_k (Q_\rtin{L} + 2)}{(Q_\rtin{L}+1) [L_{Q_\rtin{L}+1}^{(2)}(\wp_k)]^2}.
 \label{eq:wL}
\end{equation}
We note that the Laguerre polynomials $L_{\ell}^{(\alpha)}(x)$ defined 
in Ref.~\cite{hildebrand87} differ from the ones employed in this paper by a factor of $\ell!$, giving rise to
a different expression for the quadrature weights.
The equality in Eq.~\eqref{eq:quadL} is exact if $2Q_\rtin{L} > \ell + s$. We conclude that
for the recovery of the moments $M_{s,r}$ of $f$ with $s = L$, a number of $Q_\rtin{L} > L$ quadrature
points must be employed.

We now show that the exact recovery of the evolution of 
$N^\halpha$ and $T^{\halpha\hbeta}$ is ensured by taking 
only the terms with $\ell = 0$ and $\ell = 1$ in Eq.~\eqref{eq:fseriesp},
i.e., the terms with $\ell > 1$ do not contribute towards the 
evolution of these two quantities. 

In the context of the Sod shock tube problem, 
substituting the expansion \eqref{eq:fseriesp} into Eq.~\eqref{eq:boltz_Sod}
and taking the coefficient with respect to $L^{(2)}_{\ell}$ gives:
\begin{equation}
 \partial_t \mathcal{F}_\ell + \xi \partial_z \mathcal{F}_\ell =
 -\frac{u^0 - u^z \xi}{\tau_{\rtin{A-W}}} (\mathcal{F}_\ell - \mathcal{F}^{\rtin{(eq)}}_\ell).
 \label{eq:Sod_F}
\end{equation}
Similarly, substituting Eqs.~\eqref{eq:fseriesp} and \eqref{eq:f_lag_force_sum} into 
Eq.~\eqref{eq:boltz_Milne} yields:
\begin{multline}
 \frac{1}{\tau} \partial_\tau (\tau \mathcal{F}_\ell) - \frac{\xi^2}{\tau} 
 [(\ell + 2) \mathcal{F}_{\ell - 1} - \ell \mathcal{F}_\ell] 
 - \frac{1}{\tau} \partial_\xi[\xi(1 - \xi^2) \mathcal{F}_\ell] \\
= -\frac{1}{\tau_{\rtin{A-W}}} [\mathcal{F}_\ell - \mathcal{F}_{\ell}^{\rtin{(eq)}}], 
 \label{eq:Milne_F}
\end{multline}
where the second term has non-vanishing values only when $\ell >0$. 
The left-hand sides of Eqs.~\eqref{eq:Sod_F} and \eqref{eq:Milne_F} are 
independent of the coefficients $\mathcal{F}_{\ell'}$ 
with $\ell' > \ell$. There is an indirect dependence on 
$\mathcal{F}_0$ and $\mathcal{F}_1$ on the right-hand sides of 
Eqs.~\eqref{eq:Sod_F} and \eqref{eq:Milne_F} through the coefficients
$\mathcal{F}_\ell^{\rtin{(eq)}}$, since the Landau frame ($u$) and 
$\feq$ are constructed using information from $N^\halpha$ and $T^{\halpha\hbeta}$.
No higher-order terms are required for the construction of $\mathcal{F}_\ell^{\rtin{(eq)}}$,
since Eq.~\eqref{eq:tmunu_comp_momsr} shows that $N^\halpha$ and $T^{\halpha\hbeta}$ 
can be obtained as moments $M_{s,r}$ of $f$ of orders $s = 0$ and $s = 1$.
Thus, the evolution of $N^\halpha$ and $T^{\halpha\hbeta}$ can be tracked 
exactly by considering the evolution of only $f^{(L)}$ \eqref{eq:fseriespL}
with $L = 1$. The exact recovery of $N^\halpha$ and $T^{\halpha\hbeta}$ from 
this truncated expansion is ensured by using $Q_\rtin{L} = 2$ quadrature 
points on the radial direction, which are obtained as solutions of
the equation:
\begin{equation}
 L_2^{(2)}(\wp_k) \equiv 6 - 4\wp_k + \frac{1}{2}\wp_k^2 = 0
\end{equation}
The corresponding roots and quadrature weights are given in Table~\ref{tab:quad_lag}.

\subsection{Quadrature with respect to $\xi$}\label{sec:LB:qxi}

Let us now consider the expansion of $f$
with respect to $\xi = \cos\theta$:
\begin{equation}
 f = \sum_{m = 0}^\infty \frac{2m + 1}{2} \mathcal{P}_m P_m(\xi),
 \label{eq:fseriesxi}
\end{equation}
where $P_m(\xi)$ is the Legendre polynomial of order $m$. Using the following orthogonality
relation:
\begin{equation}
 \int_{-1}^1 d\xi \, P_m(\xi) P_{m'}(\xi) = \frac{2}{2m + 1} \delta_{m,m'},
 \label{eq:leg_ortho}
\end{equation}
the coefficients $\mathcal{P}_{m}$ can be obtained as
\begin{equation}
 \mathcal{P}_{m} = \int_{-1}^1 d\xi\, f\, P_m(\xi).
 \label{eq:P}
\end{equation}

We now turn to the derivative of $f$ with respect to $\xi$ in Eq.~\eqref{eq:boltz_Milne}.
Writing
\begin{equation}
 \frac{\partial[\xi (1 - \xi^2) f]}{\partial \xi} = 
 \sum_{m = 0}^\infty \frac{2m + 1}{2} \mathcal{P}^{\xi}_{m} P_m(\xi),
 \label{eq:f_leg_force_sum}
\end{equation}
the expansion coefficients $\mathcal{P}^{\xi}_{m}$ can be obtained using integration by parts as
\begin{equation}
 \mathcal{P}^{\xi}_{m} = -\int_{-1}^1 d\xi\, f\, \xi (1 - \xi^2) \frac{d P_m(\xi)}{d\xi}.
 \label{eq:f_leg_force_coeff_int}
\end{equation}
The following relation can be used to simplify the derivative of $P_m(\xi)$
\cite{gradshteyn14}:
\begin{equation}
 (1 - \xi^2) \frac{d P_m(\xi)}{d\xi} = -\frac{m(m + 1)}{2m + 1} \left[
 P_{m+1}(\xi) - P_{m - 1}(\xi)\right], \label{eq:legendre_1mxi2dP}
\end{equation}
while the following recurrence relation can be used to express $\xi P_m(\xi)$ in terms of the 
Legendre polynomials:
\begin{equation}
 \xi P_m(\xi) = \frac{m+1}{2m+1} P_{m+1}(\xi) + \frac{m}{2m+1} P_{m-1}(\xi).\label{eq:legendre_xiP}
\end{equation}
Using Eqs.~\eqref{eq:legendre_1mxi2dP} and \eqref{eq:legendre_xiP} into Eq.~\eqref{eq:f_leg_force_coeff_int}
yields:
\begin{multline}
 \frac{\partial[\xi (1 - \xi^2) f]}{\partial \xi} = 
 \sum_{m = 0}^\infty \frac{m(m + 1)}{2} P_m(\xi) 
 \left[ 
 \frac{m+2}{2m+3} \mathcal{P}_{m+2} \right.\\
 \left.-\left(\frac{m}{2m-1} - \frac{m+1}{2m+3}\right) \mathcal{P}_{m} - 
 \frac{m - 1}{2m - 1}\mathcal{P}_{m - 2}\right].\label{eq:F_leg_force_sum}
\end{multline}
For later convenience, Eq.~\eqref{eq:fseriesxi} can be used to 
express Eq.~\eqref{eq:F_leg_force_sum} in integral form:
\begin{equation}
 \frac{\partial[\xi (1 - \xi^2) f]}{\partial \xi} = \int_{-1}^1 d\xi'\, \mathcal{K}^P(\xi, \xi') f(\xi'),
 \label{eq:f_leg_force_K}
\end{equation}
where the integration kernel $\mathcal{K}^P(\xi, \xi')$ is
\begin{multline}
 \mathcal{K}^P(\xi, \xi') = \sum_{m = 1}^\infty \frac{m(m + 1)}{2} P_m(\xi) \left[
 \frac{m+2}{2m+3} P_{m+2}(\xi')\right.\\
 \left.- \left(\frac{m}{2m-1} - \frac{m+1}{2m+3}\right) P_m(\xi') - 
 \frac{m-1}{2m-1} P_{m-2}(\xi')\right].\label{eq:K_leg}
\end{multline}

In order to recover the integral with respect to $\xi$ in Eq.~\eqref{eq:momsr_aux},
the sum over $m$ in Eq.~\eqref{eq:fseriesxi} must be truncated at an order $P$:
\begin{equation}
 f^{(P)} = \sum_{m = 0}^{P} \frac{2m + 1}{2} \mathcal{P}_{m} P_m(\xi).
 \label{eq:fseriesxiP}
\end{equation}
The above truncation ensures that the integrand inside the integral with respect to $\xi$ in Eq.~\eqref{eq:momsr_aux}
is a polynomial of order $P + r$. Such integrals can be recovered using the Gauss-Legendre
quadrature:
\begin{equation}
 \int_{-1}^1 d\xi\, f^{(P)}(\xi) \xi^r \simeq
 \sum_{j = 1}^{Q_\xi} w_j^\xi f^{(P)}(\xi_j) \xi_j^r,\label{eq:quad_xi}
\end{equation}
where the $Q_\xi$ quadrature points
$\xi_j$ are the roots of the Legendre polynomial $P_{Q_\xi}(\xi)$ of order $Q_\xi$.
The quadrature weights $w_j^{\xi}$ can be obtained using the following formula \cite{hildebrand87}:
\begin{equation}
 w_j^{\xi} = \frac{2(1 - \xi_j^2)}{[(Q_{\xi} + 1) P_{Q_\xi + 1}(\xi_j)]^2}.
\end{equation}
The equality in Eq.~\eqref{eq:quad_xi} is exact if $2Q_\xi > P + r$. In order to ensure the
exact recovery of the moments $M_{s,r}$ \eqref{eq:momsr_aux} of orders $0 \le r \le P$, the
above restriction becomes $Q_\xi > P$. 

\subsection{Expanding $\feq$}\label{sec:LB:feq}

The expansion \eqref{eq:fseriesp} can be employed for $\feq$ to yield:
\begin{equation}
 \feq = \frac{1}{T_0^3} e^{-\wp} \sum_{\ell = 0}^\infty \frac{1}{(\ell + 1)(\ell + 2)}
 \mathcal{F}_\ell^{\rtin{(eq)}} L_\ell^{(2)}(\wp),
 \label{eq:feq_l}
\end{equation}
The coefficients $\mathcal{F}_\ell^{\rtin{(eq)}}$ can be calculated analytically:
\begin{align}
 \mathcal{F}^{\rtin{(eq)}}_\ell =& \frac{n}{8\pi T^3} \int_0^\infty dp\, p^2\,
 e^{-\left(\wp/\theta\right)(u^{\hat{0}} - u\cos \gamma_u)} L_\ell^{(2)}(\wp) \nonumber\\
 =& \frac{n}{8\pi} \frac{(\ell + 1)(\ell + 2)}{(u^{\hat{0}} - u \cos\gamma_u)^3}
 \left(1 - \frac{\theta}{u^{\hat{0}} - u\cos\gamma_u}\right)^\ell,
 \label{eq:Feq}
\end{align}
where $\theta = T/T_0$ (not to be confused with the polar angle)
and $\cos\gamma_u = \vu \cdot \vv / u$ represents the angle between
$\vu$ and $\vv = \vp / p$, such that
\begin{equation}
 u \cos\gamma_u = \frac{\vu \cdot \vp}{p} = \beta \gamma \xi,
\end{equation}
where the latter equality holds for the case when the macroscopic velocity 
$\vu$ is aligned along the third axis ($\beta = u^{\hat{3}} / u^{\hat{0}}$).
In deriving Eq.~\eqref{eq:Feq}, the following explicit expression for
the Laguerre polynomials was used:
\begin{equation}
 L_\ell^{(2)}(\wp) = \sum_{s = 0}^\ell \frac{(-\wp)^s (\ell + 2)!}{s!(s+2)!(\ell - s)!}.
 \label{eq:lag_expl}
\end{equation}

Since the coefficients $\mathcal{F}^{\rtin{(eq)}}_\ell$ \eqref{eq:Feq} now depend on 
$\xi$ through $\cos\gamma_u$, it is convenient to expand them with respect to 
the Legendre polynomials $P_m(\cos\gamma_u)$. The connection with an expansion 
with respect to $P_m(\xi)$ discussed in Sec.~\ref{sec:LB:qxi} will be 
presented in Eq.~\eqref{eq:leg_product} at the end of this subsection.
The expansion of $\mathcal{F}^{\rtin{(eq)}}_\ell$ thus reads:
\begin{equation}
 \mathcal{F}^{\rtin{(eq)}}_\ell = \sum_{m = 0}^{\infty} \frac{2m + 1}{2} a_{\ell,m}^{\rtin{(eq)}}
 P_m(\cos\gamma_u).
 \label{eq:aeq_def}
\end{equation}
Using Eq.~\eqref{eq:leg_ortho}, the coefficients $a_{\ell,m}^{\rtin{(eq)}}$ can be obtained as follows:
\begin{multline}
 a_{\ell,m}^{\rtin{(eq)}} = \frac{n(\ell + 1)(\ell + 2)}{8\pi}
 \int_{-1}^1 \frac{d(\cos\gamma_u) }{(u^{\hat{0}} - u\cos\gamma_u)^3} \\
 \times \left(1 - \frac{\theta}{u^{\hat{0}} - u \cos\gamma_u}\right)^\ell P_m(\cos\gamma_u).
 \label{eq:aeq_aux}
\end{multline}
Evaluating the above integrals is a rather tedious task, amenable to
algebraic computing techniques. The exact expressions
for $a_{\ell,m}$ for $0 \le \ell \le 1$ and $0 \le m \le 6$ are
presented in Appendix~\ref{app:feq}.

In order to construct the collision term in Eq.~\eqref{eq:boltz_Sod}, the expansion of
$\feq$ with respect to the Laguerre and Legendre polynomials must be truncated at
a finite order. The orders up to which $\feq$ is expanded with respect to $\wp$
and $\cos\gamma_u$ will be denoted $N_\rtin{L}$ and $N_\Omega$, respectively:
\begin{multline}
 \feq_{N_\rtin{L},N_\Omega} = \frac{1}{T_0^3} e^{-\wp} \sum_{\ell = 0}^{N_\rtin{L}} \frac{1}{(\ell + 1)(\ell+2)} L_\ell^{(2)}(\wp) \\
 \times \sum_{m = 0}^{N_\Omega} \frac{2m + 1}{2} a_{\ell,m}^{\rtin{(eq)}}
 P_m(\cos\gamma_u).
 \label{eq:feq_trunc}
\end{multline}
The analysis in Sec.~\ref{sec:LB:qp} revealed that setting $N_\rtin{L} = 1$ is sufficient to
exactly recover the evolution of $N^\mu$ and $T^{\mu\nu}$.
The truncation with respect to $N_\Omega$ is less obvious, since
according to the analysis in Sec.~\ref{sec:LB:qxi} and in Appendix~\ref{app:hyp},
a truncation at a finite value of $N_{\Omega}$ invariably introduces
errors which make the overall simulation results approximate.
In what follows, an analysis of the relative importance
of the terms of higher $m$ appearing in Eq.~\eqref{eq:aeq_def} is presented.

Let us begin by making the following expansion in the integrand in
Eq.~\eqref{eq:aeq_aux}:
\begin{gather}
 \frac{1}{(u^{\hat{0}} - u\cos\gamma_u)^3} \left(1 - \frac{\theta}{u^{\hat{0}} - u \cos\gamma_u}\right)^\ell \nn\\
 = \frac{1}{(u^{\hat{0}})^3} \sum_{i = 0}^\ell \frac{\ell!}{i! (\ell - i)!} \left(-\frac{\theta}{u^{\hat{0}}}\right)^i
 \left(1 - \frac{u}{u^{\hat{0}}} \cos\gamma_u\right)^{-i - 3}.
\end{gather}
Since $u / u^{\hat{0}} < 1$,
an expansion in powers of $u/u^{\hat{0}}$ can be performed:
\begin{multline}
 \frac{1}{(u^{\hat{0}} - u\cos\gamma_u)^3} \left(1 - \frac{\theta}{u^{\hat{0}} - u \cos\gamma_u}\right)^\ell 
 = \frac{1}{(u^{\hat{0}})^3} \\
 \times \sum_{j = 0}^\infty \frac{1}{j!} \left(\frac{u}{u^{\hat{0}}} \cos\gamma_u\right)^j
 \sum_{i = 0}^\ell \frac{\ell! (i + j + 2)!}{i! (i+2)! (\ell - i)!} \left(-\frac{\theta}{u^{\hat{0}}}\right)^i.
\end{multline}
Since $P_m(\cos\gamma_u)$ is orthogonal to all terms in the above sum for
which $0 \le j < m$, it is now clear that the leading term in $a_{\ell, m}^{\rtin{(eq)}}$ 
is of order $(u/u^{\hat{0}})^m$. This is also confirmed in the explicit series expansions given in 
Appendix~\ref{app:feq}.

From the above analysis, we conclude that the order $N_\Omega$ at which the expansion of $\feq$ with
respect to the angular coordinates can be truncated must be correlated to the 
maximum expected value of $u$ in the flow. We defer the analysis of the value of $N_\Omega$ suitable
for the simulations presented in this paper until Sec.~\ref{sec:Sod:conv:NO}.

Before ending this section, it is worth noting that for the unidirectional
flow considered in this paper, $\cos\gamma_u = {\rm sgn}(u^{\hat{3}}) \xi$,
implying that the expansion of $\feq$ with respect to $\cos\gamma_u$ is effectively
an expansion with respect to $\xi = p^{\hat{3}} / p^{\hat{0}}$. In the general case when 
the flow is not necessarily
aligned along the $z$ axis, the expansion of $\feq$ can be performed in a similar manner, since
the addition theorem can be employed to split $P_m(\cos\gamma_u)$ in terms of a series in
$\cos\theta$ and $\sin\theta$ \cite{abramowitz72,olver10}:
\begin{align}
 P_m(\cos\gamma_u) =& \frac{4\pi}{2m + 1} \sum_{q = -m}^m
 Y_{m,q}(\theta, \varphi) Y^*_{m, q}(\theta_u, \varphi_u)\nonumber\\
 =& P_m(\xi) P_m(\cos\theta_u)\nonumber
 + 2\sum_{q = 1}^m \frac{(m - q)!}{(m+q)!} \\ 
 &\times P_m^q(\xi) P_m^q(\cos\theta_u)
 \cos[q(\varphi - \varphi_u)].\label{eq:leg_product}
\end{align}
where $Y_{s,m}(\theta, \varphi)$ are the spherical harmonics and $P_m^q(z)$ are the associated
Legendre polynomials. The term containing $\varphi$ can be
expressed as $\cos[q(\varphi - \varphi_u)] = \cos q\varphi \cos q \varphi_u +
\sin q\varphi \sin q\varphi_u$, being automatically in the form required for the
Mysovskikh trigonometric quadrature presented in Refs.~\cite{mysovskikh88,romatschke11,ambrus12}.

\subsection{Discretization of the momentum space}\label{sec:LB:discrete}

In the previous subsections, we have shown that the recovery of the integrals with respect
to $\xi = p^{\hat{3}} / p^{\hat{0}}$ and $p = p^{\hat{0}}$ can be performed using the Gauss-Legendre and 
Gauss-Laguerre quadrature methods.
The main conclusion of Sec.~\ref{sec:LB:qp} is that $Q_\rtin{L} = 2$ is sufficient
for the recovery of the evolution of $T^{\mu\nu}$. Similarly, the expansion order 
$N_\rtin{L}$ of $\feq$ with respect to the generalized Laguerre polynomials is always 
set to $N_{\rtin{L}} = 1$. The number of quadrature points $Q_\xi$
on the $\xi$ coordinate of the momentum space remains a free parameter
and the effect of varying $Q_\xi$ on the accuracy of our LB simulations will 
be investigated in the numerical results sections (Secs.~\ref{sec:Sod} and \ref{sec:bjorken}).
The expansion order $N_\Omega$ of $\feq$ with respect to the Legendre polynomials 
$P_m(\cos\gamma_u)$ is also left as a free parameter.

After applying the above-mentioned quadrature methods, the momentum space is
discretized according to
\begin{gather}
 p^{\hat{\alpha}} = p (1, \cos\varphi \sqrt{1 - \xi^2}, \sin\varphi \sqrt{1 - \xi^2}, \xi)  \nn\\
 \rightarrow p^{\hat{\alpha}}_{jk} = p_k (1, \sqrt{1 - \xi_j^2}, 0, \xi_j),
 \label{eq:pjk}
\end{gather}
where we have implicitly selected one quadrature point $\varphi = 0$ along the azimuthal coordinate.
The radial components $p_k = T_0 \wp_k$ are related to the roots of $L_2^{(2)}(\wp)$ and 
are listed in Table~\ref{tab:quad_lag}, while
$\xi_j$ ($j = 1, 2, \dots Q_\xi$) are the roots of $P_{Q_\xi}(\xi)$.
Since at high values of $Q_\xi$, high-precision algebraic software is required 
in order to generate the roots $\xi_j$ and their corresponding weights $w_j^\xi$,
we provide in the Supplemental Material their values for $Q_\xi$ between 
$1$ and $1000$ [90].

Thus, the number of velocities required to discretize the momentum space is:
\begin{equation}
 N_{\rtin{vel}} = Q_L \times Q_\xi = 2Q_\xi,\label{eq:nvel}
\end{equation}
while the corresponding model is denoted $\text{R-SLB}(N_\Omega; Q_\xi)$. In these 
models, we always consider $N_\rtin{L} = 1$ and $Q_\rtin{L} = 2$, while $Q_\xi > N_\Omega \ge 0$.

The recovery of the moments $M_{s,r}$ \eqref{eq:momsr} is ensured through the
use of the Gauss-Laguerre and Gauss-Legendre quadratures, as follows:
\begin{equation}
 M_{s,r} = \sum_{j = 1}^{Q_{\xi}} \sum_{k = 1}^{Q_\rtin{L}} f_{jk} \xi_j^r p_k^s,
\end{equation}
where $f_{jk}$ are related to the distribution function $f(p, \xi, \varphi)$ through:
\begin{equation}
 f_{jk} = 2\pi T_0^3 e^{\wp_k} w_j^\xi w_k^L f(p_k, \xi_j, \varphi = 0),
 \label{eq:fjk}
\end{equation}
where the factor $2\pi$ corresponds to the azimuthal quadrature with a single point,
as explained in Ref.~\cite{ambrus12}.
In particular, $N^{\hat{\alpha}}$ and $T^{\hat{\alpha}\hat{\sigma}}$ can be recovered as follows:
\begin{gather}
 \begin{pmatrix}
  N^{\hat{0}}\\
  N^{\hat{3}}
 \end{pmatrix}
 = \sum_{k = 1}^{Q_\rtin{L}} \sum_{j = 1}^{Q_\xi} f_{jk} 
 \begin{pmatrix}
  1\\
  \xi_j
 \end{pmatrix},  \nn\\
 \begin{pmatrix}
  T^{\hat{0}\hat{0}} \\
  T^{\hat{0}\hat{3}} \\
  T^{\hat{3}\hat{3}}
 \end{pmatrix}
 = \sum_{k = 1}^{Q_\rtin{L}} \sum_{j = 1}^{Q_\xi} f_{jk} p_k
 \begin{pmatrix}
  1 \\
  \xi_j \\
  \xi_j^2
 \end{pmatrix},
 \label{eq:macro_jk}
\end{gather}
while $T^{\hat{1}\hat{1}} = T^{\hat{2}\hat{2}} = \frac{1}{2}(T^{\hat{0}\hat{0}} - T^{\hat{3}\hat{3}})$.

Equation~\eqref{eq:fjk} can be employed to make the transition between 
Eq.~\eqref{eq:f_lag_force_K} and its equivalent after discretization: 
\begin{equation}
 \left[\frac{1}{p^2} \frac{\partial(f p^3)}{\partial p}\right]_{jk} = \sum_{k' = 1}^{Q_L} 
 \mathcal{K}^\rtin{L}_{k,k'} f_{jk'},
\end{equation}
where the $Q_L \times Q_L$ matrix $\mathcal{K}^\rtin{L}_{k,k'}$ has the following elements:
\begin{multline}
 \mathcal{K}^\rtin{L}_{k,k'} = 
 w_k^L \sum_{\ell = 1}^{Q_L - 1} \frac{1}{(\ell + 1)(\ell + 2)} L_\ell^{(2)}(\wp_k) \\
 \times \left[(\ell + 2) L^{(2)}_{\ell - 1}(\wp_{k'}) - \ell\, L^{(2)}_\ell(\wp_{k'})\right].
 \label{eq:K_lag_disc}
\end{multline}
Since $Q_\rtin{L} = 2$, the above expression reduces to
\begin{equation}
 \mathcal{K}^\rtin{L}_{k,k'} = \frac{1}{6} w_k^L (3 - \wp_k) \wp_{k'}.
 \label{eq:K_lag_disc1}
\end{equation}

Similarly, Eq.~\eqref{eq:f_leg_force_K} can be replaced using a quadrature sum, yielding
\begin{equation}
 \left[\frac{\partial[\xi (1 - \xi^2) f]}{\partial \xi}\right]_{jk} = \sum_{j' = 1}^{Q_\xi} \mathcal{K}^P_{j,j'} 
 f_{j',k},
\end{equation}
where the $Q_\xi \times Q_\xi$ matrix $\mathcal{K}^P_{j,j'}$ has the following entries:
\begin{multline}
 \mathcal{K}^{P}_{j,j'} = w_j \sum_{m = 1}^{Q_\xi - 3} \frac{m(m + 1)(m+2)}{2(2m+3)} 
 P_m(\xi_j) P_{m+2}(\xi_{j'})\\
 - w_j \sum_{m=1}^{Q_\xi-1} \frac{m(m+1)}{2}P_m(\xi_j) \left[
 \frac{(2m+1)P_m(\xi_{j'})}{(2m-1)(2m+3)} \right.\\
 \left.+ \frac{m-1}{2m-1} P_{m-2}(\xi_{j'})\right].\label{eq:K_leg_disc}
\end{multline}

\subsection{Spatial and temporal discretization}\label{sec:LB:num}

The time variable is discretized using equal time steps.
In the context of the Sod shock tube problem discussed in Sec.~\ref{sec:Sod}, the time 
step is $\delta t$ and the value of the time variable after $n$ steps is:
\begin{equation}
 t_n = n \delta t.
\end{equation}
In the context of the one-dimensional boost-invariant expansion 
considered in Sec.~\ref{sec:bjorken}, the 
initial time is $\tau_0 > 0$, such that after $n$ time steps
of duration $\delta \tau$, the time variable has the value
\begin{equation}
 \tau_n = \tau_0 + n \delta \tau.
\end{equation}

The spatial discretization is trivial for the Bjorken flow, since in this case 
the flow is considered to be homogeneous along all spatial directions.
For the Sod shock tube, the flow is homogeneous with respect to the $x$ and $y$ coordinates, 
such that the flow domain consists of $Z$ equallysized cubic cells. 
Cell $s$ ($1 \le s \le Z$) is centered on the point having coordinates:
\begin{equation}
 x_s = y_s = 0, \qquad z_s = \left(s - \frac{1}{2}\right) \delta z - \frac{L}{2},
\end{equation}
where $\delta z = L / Z$ is the lattice spacing.

Following the discretization of the momentum space presented in Sec.~\ref{sec:LB:discrete}, 
the distribution function $f(z, p, \xi, t)$ is replaced by the set of functions 
$f_{jk}(z, t)$ according to Eq.~\eqref{eq:fjk}. 
At time $t = t_n$, the function $f_{jk}(z,t)$ is replaced by a set of cell averages 
defined as:
\begin{equation}
 f_{n;s;jk} \approx \int_{z_{s-1/2}}^{z_{s+1/2}} dz\, f_{jk}(z, t = t_n).
\end{equation}

With the above conventions the relativistic Boltzmann equation for the Sod shock tube 
\eqref{eq:boltz_Sod} reads
\begin{multline}
 (\partial_t f)_{n;s;jk} +  (\xi \partial_z f)_{n;s;jk} \\
 = -\frac{\gamma_{n;s}(1 - \beta_{n;s} \xi_j)}{\tau_{\rtin{A-W};n;s}} 
 [f_{n;s;jk} - \feq_{n;s;jk}],
 \label{eq:boltz_Sod_disc}
\end{multline}
while in the case of the one-dimensional boost-invariant expansion, 
Eq.~\eqref{eq:boltz_Milne} reduces to:
\begin{multline}
 (\partial_\tau f)_{n;jk} -
 \frac{\xi_j^2}{\tau_n} \sum_{k' = 1}^{Q_\rtin{L}} \mathcal{K}^{\rtin{L}}_{k,k'} 
 f_{n;j,k'} 
 - \frac{1}{\tau_n} \sum_{j' = 1}^{Q_\xi} \mathcal{K}^{P}_{j,j'} 
 f_{n;j',k}\\
 = -\frac{1}{\tau_n} f_{n;jk}
 -\frac{1}{\tau_{\rtin{A-W};n}} [f_{n;jk} - \feq_{n;jk}].\label{eq:boltz_Milne_disc}
\end{multline}

The time stepping is performed using the explicit, total variation diminishing (TVD),
nonlinearly stable third-order Runge-Kutta 
(RK3) method \cite{shu88,gottlieb98,henrick05,trangenstein07,busuioc17}.
In order to employ the RK3 algorithm, the Boltzmann equations for the 
Sod shock tube \eqref{eq:boltz_Sod} and for the Bjorken flow 
\eqref{eq:boltz_Milne} can be put in the form:
\begin{align}
(\partial_t f)_{n;s;jk} =& (L_{\rtin{Sod}}[f])_{n;s;jk}, \nonumber\\
(\partial_\tau f)_{n;jk} =& (L_{\rtin{Bjorken}}[f])_{n;jk},
\end{align}
where $L(f)$ is given for the Sod shock tube problem discussed in Sec.~\ref{sec:Sod} by
\begin{multline}
 L_{\rtin{Sod}}[f_{n;s;jk}] = - (\xi \partial_z f)_{n;s;jk} \\ - 
 \frac{u^0_{n;s} - u^z_{n;s} \xi_j}{\tau_{\rtin{A-W};n;s}} 
 \left[ f_{n;s;jk} - \feq_{n;s;jk} \right],
 \label{eq:L_Sod}
\end{multline}
while in the case of the Bjorken flow
discussed in Sec.~\ref{sec:bjorken}, it has the following form:
\begin{multline}
 L_{\rtin{Bjorken}}[f_{n;jk}] = 
 - \frac{f_{n;jk}}{\tau_n} + 
 \frac{\xi_j^2}{\tau_n} \sum_{k' = 1}^{Q_\rtin{L}} 
 \mathcal{K}^\rtin{L}_{k,k'} f_{n;j,k'} \\ + \frac{1}{\tau_n} \sum_{j' = 1}^{Q_\xi}
 \mathcal{K}^P_{j,j'} f_{n;j',k}
 -\frac{1}{\tau_{\rtin{A-W};n}} [f_{n;jk} - \feq_{n;jk}].\label{eq:L_Bjorken}
\end{multline}

The third order Runge-Kutta algorithm consists of applying the following steps \cite{shu88,gottlieb98}:
\begin{align}
f^{(1)}_{n;s;jk} &= f_{n;s;jk} + \delta t\, L[f_{n;s;jk}], \nonumber\\
f^{(2)}_{n;s;jk} &= \frac{3}{4} f_{n;s;jk} + \frac{1}{4} f^{(1)}_{n;s;jk} + 
\frac{1}{4}\delta t \, L[f^{(1)}_{n;s;jk}], \nonumber\\
f_{n+1;s;jk} &= \frac{1}{3} f_{n;s;jk} + \frac{2}{3} f^{(2)}_{n;s;jk} + 
\frac{2}{3} \delta t \, L[f^{(2)}_{n;s;jk}].
\label{eq:rk3}
\end{align}
The Butcher tableau \cite{butcher08} corresponding to the above procedure is presented 
in Table~\ref{tab:rk3}.
\begin{table}
\begin{center}
\begin{ruledtabular}
\begin{tabular}{l|lll}
 $0$ & & & \\
 $1$ & $1$ & &\\
 $1/2$ & $1/4$ & $1/4$ & \\\hline
 & $1/6$ & $1/6$ & $2/3$
\end{tabular}
\end{ruledtabular}
\end{center}
\caption{Butcher tableau corresponding to the third-order total variation diminshing (TVD) 
Runge-Kutta (RK3) time-stepping procedure described in Eq.~\eqref{eq:rk3}. \label{tab:rk3}}
\end{table}

The maximum values of the time steps $\delta t$ and $\delta \tau$ which can be 
employed with the explicit RK3 algorithm considered in this paper suffer 
from two constraints. In particular, the time step is bounded from above by the 
relaxation time $\tau_{\rtin{A-W}}$. This is because when the time step is of the order 
of $\tau_{\rtin{A-W}}$, 
the equation becomes stiff. The second constraint applies only to advection problems 
(the Sod shock tube in this paper) and requires that 
the Courant-Friedrichs-Lewy (CFL) number is smaller than 1:
\begin{equation}
 \text{CFL} = \frac{v_{\rtin{max}} \delta t }{\delta z} < 1.\label{eq:CFL}
\end{equation}
In the context of the Sod shock tube, the reference length is taken to be the flow domain 
length, such that $\delta z = 1/Z$, while the reference speed is taken to be the speed of light,
such that $v_{\rtin{max}} < 1$. This implies that
\begin{equation}
 \delta t \lesssim \frac{1}{Z}.
\end{equation}

Finally, the advection term for the Sod shock tube is implemented using the 
fifth-order weighted essentially non-oscillatory (WENO) scheme \cite{rezzolla13,gan11}.
The WENO scheme is particularly good for reproducing flows around discontinuities,
like shock waves or contact discontinuities, effectively eliminating spurious oscillations
and heavily suppressing numerical viscosity \cite{rezzolla13,gan11}. The scheme produces remarkably
sharp profiles around the discontinuities.
The implementation of the scheme starts by writing the advection term in the Boltzmann
equation \eqref{eq:boltz_Sod_disc} as
\begin{equation}
 (\partial_z \xi f)_{n;s;jk} = \frac{1}{\delta z} \left(\mathcal{F}_{n;s+1/2;jk} - \mathcal{F}_{n;s-1/2;jk}\right).
\end{equation}
The numerical fluxes are defined as
\begin{multline}
\mathcal{F}_{n;s+1/2;jk} = \overline{\omega}_1\mathcal{F}^1_{n;s+1/2;jk} +
\overline{\omega}_2\mathcal{F}^2_{n;s+1/2;jk} \\ + \overline{\omega}_3\mathcal{F}^3_{n;s+1/2;jk}.
\label{eq:flux_weno}
\end{multline}
The interpolating functions $\mathcal{F}^q_{n;s+1/2;jk}$ are computed in an upwind-biased 
manner by taking into account the sign of $J_{n;s;jk} = \xi_j f_{n;s;jk}$.
For the case when $J_{n;s;jk} > 0$, the interpolating functions are given by
\begin{align}
\mathcal{F}^1_{s+1/2} =& \frac{1}{3}J_{s-2} - \frac{7}{6} J_{s-1} + \frac{11}{6} J_s, \nn \\
\mathcal{F}^2_{s+1/2} =& -\frac{1}{6}J_{s-1} + \frac{5}{6} J_{s} + \frac{1}{3} J_{s+1}, \nn \\
\mathcal{F}^3_{s+1/2} =& \frac{1}{3}J_{s} + \frac{5}{6} J_{s+1} - \frac{1}{6} J_{s+2},
\end{align}
where, for simplicity, the labels $n$, $j$, and $k$ were omitted.
The construction of the flux $\mathcal{F}_{s-1/2}$, as well as the associated interpolation 
functions $\mathcal{F}^q_{s-1/2}$, can be performed as described above, by replacing 
$s \rightarrow s-1$. The weighting factors $\overline{\omega}_q$ appearing in 
Eq.~\eqref{eq:flux_weno} are defined as:
\begin{equation}
\overline{\omega}_q = \frac{\widetilde{\omega}_q}{\widetilde{\omega}_1+\widetilde{\omega}_2+\widetilde{\omega}_3}, 
\quad \widetilde{\omega}_q = \frac{\delta_q}{\sigma^2_q}.\label{eq:weno5_omega}
\end{equation}
The ideal weights $\delta_q$ are $\delta_1 = 0.1$, $\delta_2 = 0.6$, and $\delta_3 = 0.3$, 
while the smoothness functions $\sigma_q$ are given by:
\begin{align}
\sigma_1 =& \frac{13}{12} \left(J_{s-2} -2J_{s-1} + J_s \right)^2 + \frac{1}{4} \left( J_{s-2} - 4J_{s-1} + 3J_s \right)^2, \nn \\
\sigma_2 =& \frac{13}{12} \left(J_{s-1} -2J_{s} + J_{s+1} \right)^2 + \frac{1}{4} \left( J_{s-1} - J_{s+1} \right)^2, \nn \\
\sigma_3 =& \frac{13}{12} \left(J_{s} -2J_{s+1} + J_{s+2} \right)^2 + \frac{1}{4} \left( 3J_{s} -4 J_{s+1} + J_{s+2} \right)^2.
\end{align}

The weighing factors $\overline{\omega}_q$ are not defined when the smoothness functions $\sigma_q$
vanish. In order to avoid division by zero errors, it is customary to add a small but nonzero
term to $\sigma_q$ in the denominators of $\widetilde{\omega}_q$ \cite{gan11}.
In our numerical experiments, we found that this introduces small but visible 
spurious effects, so that instead we preferred to employ the following
limiting values of $\overline{\omega}_q$ \cite{busuioc17}:
\begin{description}
 \item[(i)] If $\sigma_1 = \sigma_2 = \sigma_3$,
 \begin{equation}
  \overline{\omega}_q = \delta_q.
 \end{equation}
 \item[(ii)] If $\sigma_q = 0$ for all $q$, except when $q = p$,
 \begin{equation}
  \overline{\omega}_q =
\begin{cases}
 {\displaystyle \frac{\delta_q}{\sum_{q' \neq p} \delta_{q'}}}, & q \neq p,\\
 0, & q = p.
\end{cases}
 \end{equation}
 \item[(iii)] If $\sigma_q = 0$ only when $q = p$,
\begin{equation}
  \overline{\omega}_q =
\begin{cases}
 1, & q = p, \\
 0, & q \neq p.
\end{cases}
\end{equation}
\end{description}
With the above definitions, the WENO5 scheme can be used without 
altering the smoothness functions.

\section{Sod shock tube}\label{sec:Sod}

In this section, we consider the validation of the LB models introduced in 
Sec.~\ref{sec:LB} in the case of a particular instance of the Riemann problem, 
called Sod's shock tube, which we investigate throughout the full range of 
the degree of rarefaction (starting from the nearly inviscid regime up to the 
ballistic regime). The problem statement is given in Sec.~\ref{sec:Sod:setup}.

The first validation test is performed in Sec.~\ref{sec:Sod:inviscid}, where 
we consider a comparison between our numerical results and the analytic solution 
of the Sod shock tube problem in the inviscid regime. 
Since in the kinetic formulation, there is always some finite dissipation 
due to the nonvanishing value of the relaxation time $\tau_{\rtin{A-W}}$, 
we also consider in this subsection the properties of the integrated heat flux 
and shear stress around the flow discontinuities, thus validating the 
correct recovery of dissipative effects as predicted through the Chapman-Enskog 
expansion. 
In Sec.~\ref{sec:Sod:ballistic}, we consider a comparison between our numerical 
results and the analytic solution of the Sod shock tube problem in the ballistic 
(free-streaming) regime. The final validation test is presented in 
Sec.~\ref{sec:Sod:BAMPS}, where our numerical results are compared with 
those reported in Refs.\cite{bouras10,bouras09prl,bouras09nucl}, obtained using the
Boltzmann approach of multiparton scattering (BAMPS) model, which
implies solving the full Boltzmann equation for on-shell particles, with a stochastic microscopic
model for the collision term \cite{bouras09prl}. We also compare our results with those obtained using the
viscous sharp and smooth transport algorithm (vSHASTA), which are reported also
in Refs.~\cite{bouras10,bouras09prl,bouras09nucl}.

Throughout the validation sections (Secs.~\ref{sec:Sod:inviscid}, \ref{sec:Sod:ballistic}, and
\ref{sec:Sod:BAMPS}), our simulation results are obtained using so-called ``reference models,'' 
which are LB models constructed as presented in Sec.~\ref{sec:LB}, with a sufficiently high 
quadrature order $Q_\xi$ to ensure good agreement with the benchmark data.
After validation, these reference models are used in Sec.~\ref{sec:Sod:conv} as
benchmark results in a convergence test designed to obtain the minimum quadrature 
$Q_\xi^{\rtin{conv}}$, as well as the minimum expansion order $N_\Omega$ of 
$\feq$ required to reduce the relative error in the simulation profiles below a certain 
threshold. 

Section~\ref{sec:Sod:summary} ends this section with a summary of our results 
and conclusions.

\subsection{Numerical setup and nondimensionalization convention}\label{sec:Sod:setup}

\begin{figure}
\begin{tabular}{c}
\includegraphics[width=0.82\linewidth]{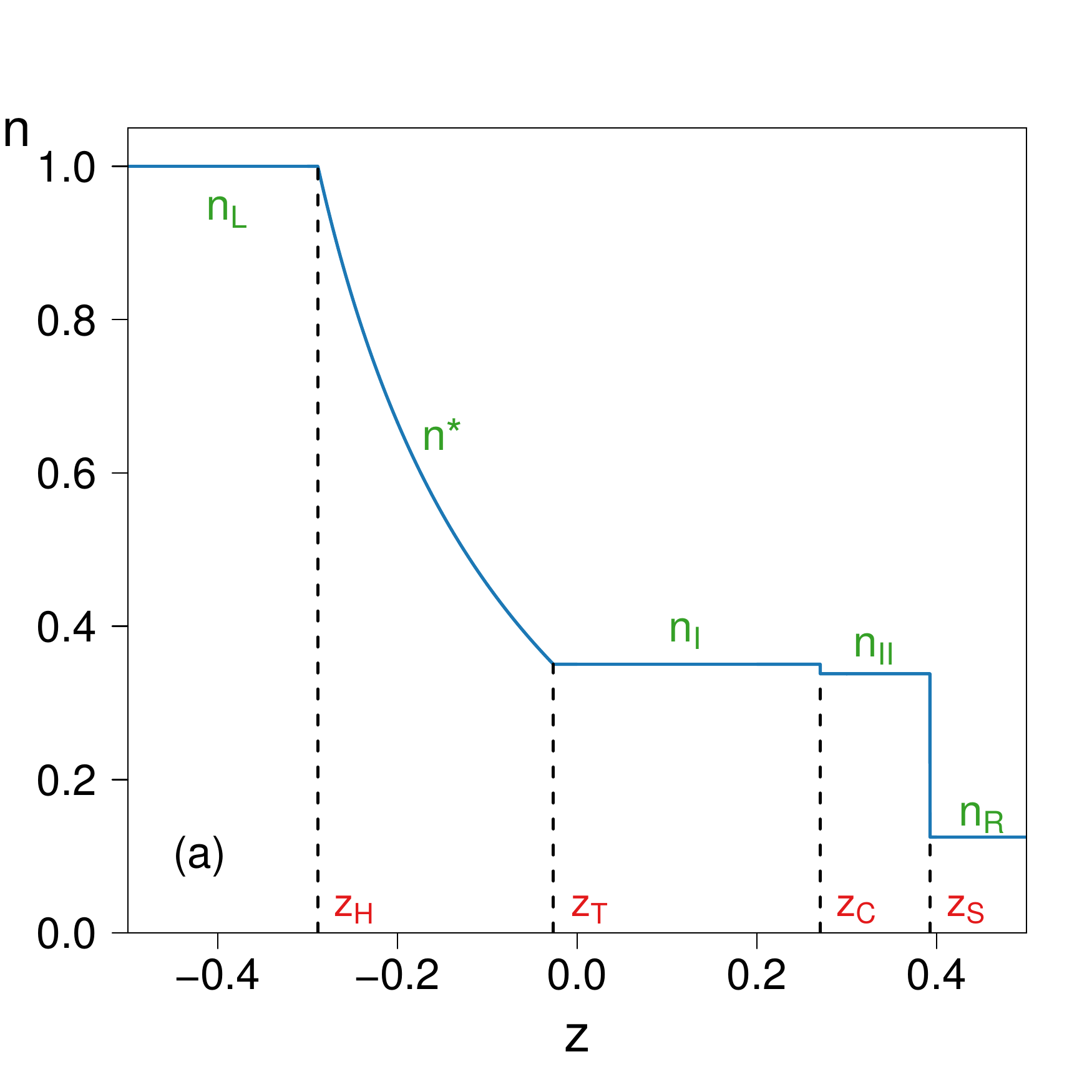} \\
\includegraphics[width=0.82\linewidth]{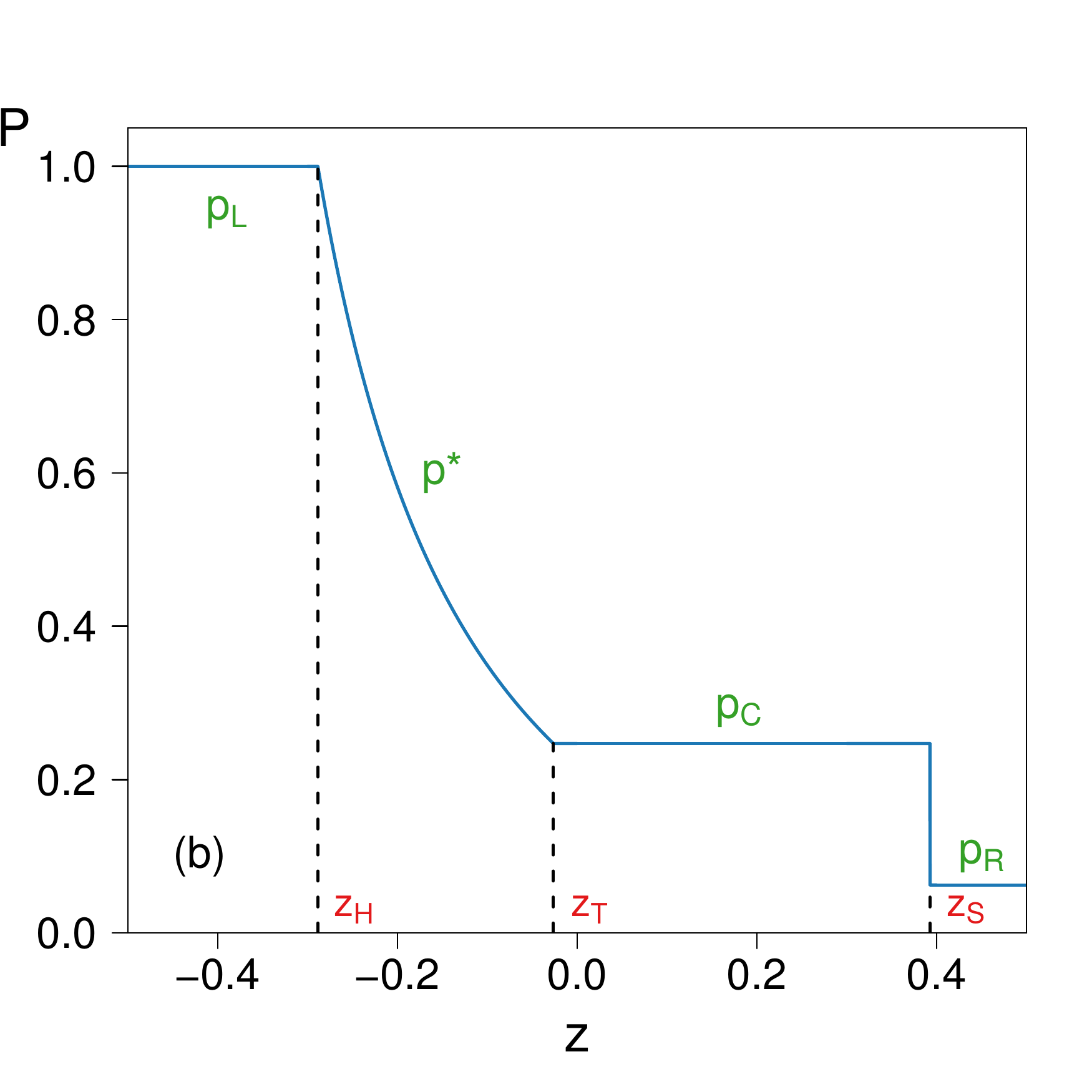} \\
\includegraphics[width=0.82\linewidth]{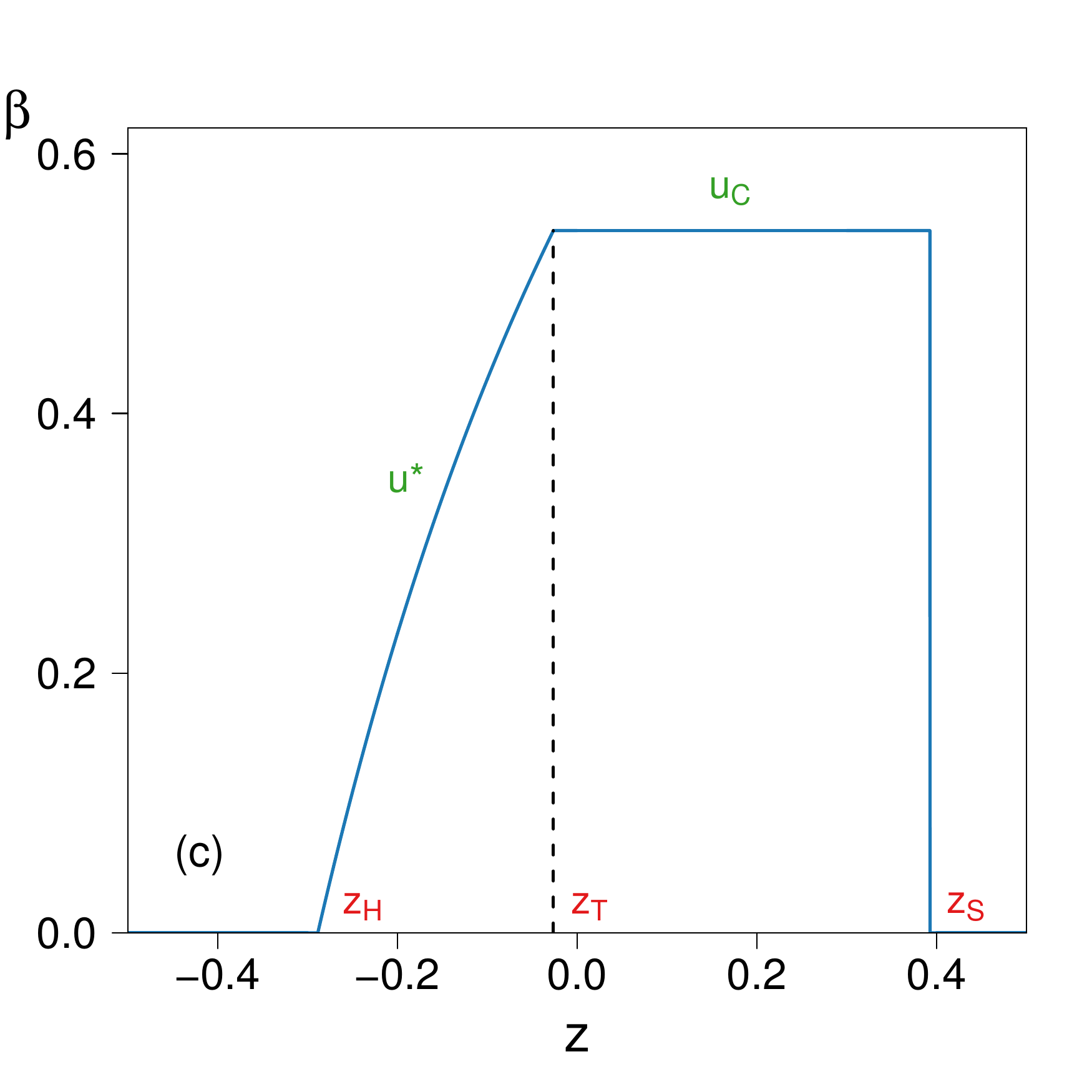} 
\end{tabular}
\caption{ The flow structure in the inviscid limit of the Sod shock tube problem.
\label{fig:inv_structure}}
\end{figure}

For the initial state in our simulations, we consider two chambers separated
by a thin membrane. The two parts of the fluid are in different states
such that the macroscopic fields describing the state of the fluid are
discontinuous at the interface. Although the setup is very simple, the presence
of large gradients make it a powerful test for numerical methods.
We assume that, before the membrane is removed, the fluid constituents in the two domains
are in local thermal equilibrium described by the following initial conditions:
\begin{equation}
(P,n,\beta) =
\begin{cases}
 (P_{\rtin{L}}, n_{\rtin{L}}, \beta_{\rtin{L}} = 0) & \text{if}\ \ z < 0,\\
 (P_{\rtin{R}}, n_{\rtin{R}}, \beta_{\rtin{R}} = 0) & \text{if}\ \ z > 0.
\end{cases}
 \label{eq:Sod:init}
\end{equation}
In the following, we take the reference pressure $\widetilde{P}_{\rm ref}$, 
temperature $\widetilde{T}_{\rm ref}$ and density $\widetilde{n}_{\rm ref}$ to be 
those of the left initial state, i.e.,
\begin{equation}
 \widetilde{T}_{\rtin{ref}} = \widetilde{T}_{\rtin{L}}, \qquad
 \widetilde{P}_{\rtin{ref}} = \widetilde{P}_{\rtin{L}}, \qquad
 \widetilde{n}_{\rtin{ref}} = \widetilde{n}_{\rtin{L}}.
\end{equation}
In the above, we employed the convention that dimensional quantities 
are denoted using a tilde $\widetilde{\phantom{T}}$.
The above choice implies that $P_{\rtin{L}} = n_{\rtin{L}} = T_{\rtin{L}} = 1$,
which is the convention that we will use throughout this section. 
Furthermore, the reference length $\widetilde{L}_{\rtin{ref}}$ is chosen to be
equal to the domain length, while the reference velocity
$\widetilde{v}_{\rtin{ref}} = \widetilde{c}$ is chosen to be the speed of light. 
This implies the following definition for the reference time:
\begin{equation}
 \widetilde{t}_{\rtin{ref}} = \frac{\widetilde{L}_{\rtin{ref}}}{\widetilde{c}}.
 \label{eq:Sod:tref}
\end{equation}

In the ideal (inviscid) case, as the system evolves in time, we see the formation of a rarefaction wave
(\textsl{R}) moving to the left, as well as a contact discontinuity (\textsl{C}) and shock wave (\textsl{S})
moving to the right. As shown in Fig.~\ref{fig:inv_structure}, these features divide the flow into 5 
regions: regions L and R represent the initial unperturbed states to the left of the rarefaction wave 
and to the right of the shock wave, regions I and II form the central plateau C to the left and right 
of the contact discontinuity, and the * region consists of the interval between the head and tail of 
the rarefaction wave. The region to which a 
particular macroscopic field refers will
be indicated via a subscript (e.g., $P_*$ refers to the pressure in the rarefaction wave).
A complete solution of the Riemann problem in the inviscid regime is represented by finding 
the values of the fields in the intermediate regions I, II, and * in terms of the initial conditions 
from the external unperturbed regions L and R. For completeness, we present in detail the procedure 
for obtaining such a solution in Sec.~\ref{sec:Sod:inviscid}.

As the viscosity is increased, dissipation causes the interfaces between the above mentioned regions
to become smooth, while in the ballistic regime, the fluid constituents stream freely, such that the above
regions become indistinguishable. In the viscous regime, we follow Ref.~\cite{bouras09prl} and characterize
the degree of rarefaction using the ratio $\etas$ (assumed to be constant) between the shear viscosity
$\eta$ and the entropy density $s$, expressed in Planck units. The connection between $\etas$ and the 
nondimensionalized Anderson-Witting relaxation time $\tau_{\rtin{A-W}}$ employed in this paper 
can be established as follows.

The nondimensional relaxation time
$\tau_{\rtin{A-W}} = \widetilde{\tau}_{\rtin{A-W}} / \widetilde{t}_{\rtin{ref}}$ 
can be related to the
ratio $\widetilde{\eta}/\widetilde{s}$ starting from \cite{cercignani02}:
\begin{equation}
 \frac{\widetilde{\eta}}{\widetilde{s}} =
 \frac{4 \widetilde{\tau}_{\rtin{A-W}} \widetilde{T}}{5 (4 - \ln \lambda)},
 \label{eq:etas_dim}
\end{equation}
where the fugacity $\lambda$ is defined as \cite{cercignani02}:
\begin{equation}
 \lambda = \frac{\widetilde{n} \pi^2}{g_s}
 \left(\frac{\widetilde{\hbar} \widetilde{c}}{\widetilde{k}_B \widetilde{T}}\right)^3.
 \label{eq:lambda_def}
\end{equation}
In the context of quark-gluon plasma systems, $g_s = 16$ is the number of degrees of freedom for 
gluons. It is convenient to express $\lambda = \overline{\lambda} \lambda_{\rtin{ref}}$
using the relative fugacity $\overline{\lambda}$ \eqref{eq:fugacity}, in terms of the fugacity
$\lambda_{\rtin{ref}} = \lambda_{\rtin{L}}$ in the left side of the channel:
\begin{equation}
 \overline{\lambda} = \frac{n}{T^3}, \qquad
 \lambda_{\rtin{ref}} = \frac{\widetilde{n}_{\rtin{ref}} \pi^2}{g_s}
 \left(\frac{\widetilde{\hbar} \widetilde{c}}{\widetilde{k}_B \widetilde{T}_{\rtin{ref}}}\right)^3.
 \label{eq:lambda}
\end{equation}
With the above notation, $\tau_{\rtin{A-W}}$ can be obtained from Eq.~\eqref{eq:etas_dim} as:
\begin{equation}
 \tau_{\rtin{A-W}} = \frac{\eta}{s} \frac{5}{4 T} [4 -\ln(\overline{\lambda} \lambda_{\rtin{ref}})],
\end{equation}
where $\eta / s$ is:
\begin{equation}
 \frac{\eta}{s} = \frac{\widetilde{\eta}}{\widetilde{s}}
 \frac{\widetilde{c}}{\widetilde{T}_{\rtin{ref}} \widetilde{L}_{\rtin{ref}}}.
\end{equation}
The connection between $\etas$ and 
the ratio $\eta / s$ obtained by applying the non-dimensionalization 
procedure discussed in this subsection can be established as follows:
\begin{equation}
 \etas \equiv
 \frac{\widetilde{k}_B}{\widetilde{\hbar}} \left(\frac{\widetilde{\eta}}{\widetilde{s}}\right)
 = \frac{\widetilde{k}_B \widetilde{T}_{\rtin{ref}} \widetilde{L}_{\rtin{ref}}}
 {\widetilde{\hbar} \widetilde{c}} \left(\frac{\eta}{s}\right).
\end{equation}
The above procedure gives the following expression for the non-dimensional 
relaxation time $\tau_{\rtin{A-W}}$:
\begin{align}
 \tau_{\rtin{A-W}} =& 
 \frac{\tau_{\rtin{A-W}; 0}}{T}\left[1 - \frac{1}{4} \ln(\overline{\lambda} \lambda_{\rtin{ref}})\right], \nonumber\\
 \tau_{\rtin{A-W}; 0} =& 5 \frac{\widetilde{\hbar} \widetilde{c} \etas}
 {\widetilde{k}_B \widetilde{T}_{\rtin{ref}} \widetilde{L}_{\rtin{ref}}}.\label{eq:Sod:tau}
\end{align}

As discussed at the beginning of this subsection, at $t = 0$, the system is initialized 
using 
\begin{equation}
 f(z, t = 0) = 
 \begin{cases}
  \feq_{\rtin{L}}, & z < 0,\\
  \feq_{\rtin{R}}, & z > 0,
 \end{cases}
\end{equation}
where 
\begin{equation}
 \feq_{\rtin{L}} = \frac{n_\rtin{L}}{8\pi T_\rtin{L}^3} e^{-p/T_\rtin{L}},\qquad
 \feq_{\rtin{R}} = \frac{n_\rtin{R}}{8\pi T_\rtin{R}^3} e^{-p/T_\rtin{R}}.
\end{equation}
In order to employ the numerical scheme described in Sec.~\ref{sec:LB:num}, boundary conditions 
must be employed to the left and right of the fluid domain. Since the WENO5 scheme requires 
information from three nodes in the upstream direction, three ghost nodes are required 
on both sides of the fluid domain. According to the discussion in Sec.~\ref{sec:LB:num}, 
the nodes to the left of the fluid domain have labels $s = 0,\ -1$ and $-2$, while the 
nodes to the right are given by $s = Z + 1$, $Z + 2$, and $Z + 3$.
In this paper, we follow Ref.~\cite{gan11} and set:
\begin{align}
 f_{n;0;jk} = f_{n;-1;jk} = f_{n;-2;jk} =& \feq_{\rtin{L}; jk}, \nonumber\\
 f_{n;Z+1;jk} = f_{n;Z+2;jk} = f_{n;Z+3;jk} =& \feq_{\rtin{R}; jk}.
\end{align}

\subsection{Inviscid regime}\label{sec:Sod:inviscid}

The analytic solution for the Sod shock tube problem is well known 
\cite{rezzolla13}. Since the solution is usually not 
particularized for the case of massless particles considered in this 
paper, we review the derivation of the analytic solution for
this case in Sec.~\ref{sec:Sod:inviscid:analytic}.
Our simulation results for the macroscopic profiles are presented in 
Sec.~\ref{sec:Sod:inviscid:macro}. 
In order to assess the capabilities of our models to recover dissipative 
effects, we present in Sec.~\ref{sec:Sod:inviscid:noneq} an analysis of 
the nonequilibrium quantities (heat flux and shear stress) 
induced near the flow discontinuities.

\subsubsection{Analytic profiles}\label{sec:Sod:inviscid:analytic}

In the case of the Riemann problem with the initial conditions \eqref{eq:Sod:init}, 
the density profile of the flow at a finite time contains three features: a rarefaction 
wave (R) traveling to the left and a shock wave (S) and contact discontinuity (C)
traveling to the right. The rarefaction wave is a simple wave. This means that the 
flow is isentropic and has a number of Riemann invariants, 
properties which can be exploited to obtain the values of the thermodynamic fields 
inside the wave \cite{rezzolla13}.
In the case of shock-waves either one or several of the fields present discontinuities. 
The macroscopic conservation equations then induce a number of junction conditions 
(the Rankine-Hugoniot junction conditions) involving 
the density, pressure, and velocity of the fluid on the two sides of the discontinuity. 

\paragraph{Rarefaction wave.}
In order to find relations that determine the fields inside the rarefaction wave, 
we start from the macroscopic conservation equations \cite{cercignani02}: 
\begin{align}
Dn =& -n\, \partial_\mu u^\mu, \nonumber\\
Du^\alpha =& - \frac{1}{4P} \Delta^{\alpha \mu} \partial_\mu P, \nonumber\\
DP =& -\frac{4P}{3} \partial_\mu u^\mu,\label{eq:Sod:euler}
\end{align}
where $D = u^\alpha \partial_\alpha = \gamma(\partial_t + \beta \partial_z)$
and $\Delta^{\alpha \beta} = \eta^{\alpha\beta} + u^\alpha u^\beta$.
The rarefaction wave, as any simple wave, is self-similar. This means that 
the fields depend on space and time only through the self-similarity variable $\zeta = \frac{z}{t}$,
such that Eq.~\eqref{eq:Sod:euler} becomes:
\begin{subequations}
\begin{align}
(\beta-\zeta)\frac{dn}{d\zeta}\ =& -n\gamma^2(1-\beta\zeta)\frac{d\beta}{d\zeta}, \label{eq:cons_n}\\
(\beta-\zeta)\frac{d\beta}{d\zeta} =& -\frac{1}{4P \gamma^2} (1 -\beta\zeta) \frac{dP}{d\zeta},  \label{eq:cons_v}\\
(\beta-\zeta)\frac{dP}{d\zeta}\ =& - \frac{4P}{3} \gamma^2(1-\beta\zeta) \frac{d\beta}{d\zeta}. \label{eq:cons_p}
\end{align}
\end{subequations}
Using Eq.~\eqref{eq:cons_p} to eliminate the pressure in Eq.~\eqref{eq:cons_v} gives the following constraint for $\zeta$:
\begin{equation} \label{eq:v(xi)}
\zeta = \frac{\beta \mp c_s}{1 \mp c_s \beta},
\end{equation}
where $c_s = 1/\sqrt{3} $ is the speed of sound for an ultrarelativistic fluid.
The lower sign refers to the case when the rarefaction wave moves to the right 
and is therefore irrelevant for the present case.
Plugging relation \eqref{eq:v(xi)} back into Eqs.~\eqref{eq:cons_n} and \eqref{eq:cons_p} gives:
\begin{align}
\frac{1}{n\sqrt{3}}\frac{dn}{d\zeta} + \gamma^2 \frac{d\beta}{d\zeta} =& 0, \nonumber\\
\frac{\sqrt{3}}{4P}\frac{dP}{d\zeta} + \gamma^2 \frac{d\beta}{d\zeta} =& 0.
\end{align}
Integrating these relations and with a bit of rearrangement, the expressions 
for the Riemann invariants can be obtained:
\begin{equation} \label{eq:R_inv}
n\left(\frac{1+\beta}{1-\beta}\right)^{\frac{\sqrt{3}}{2}} = \text{const}, \qquad 
P\left(\frac{1+\beta}{1-\beta}\right)^\frac{2}{\sqrt{3}} = \text{const}.
\end{equation}
The Riemann invariants are conserved along the rarefaction wave,thus allowing the 
quantities in this region (denoted using a star $*$) to be expressed as functions of the 
parameters in the unperturbed left state ($n_\rtin{L}, P_\rtin{L}, \beta_\rtin{L} = 0$):
\begin{subequations}
\begin{align}
\beta_* =& \frac{c_s + \zeta}{1 + c_s \zeta}, \label{eq:betastar}\\
P_* =& P_{\rtin{L}} \left[\frac{(1 - c_s)(1-\zeta)}{(1 + c_s)(1+\zeta)}\right]^{2/\sqrt{3}}, \label{eq:Pstar}\\
n_* =& n_{\rtin{L}} \left[\frac{(1 - c_s)(1-\zeta)}{(1 + c_s)(1+\zeta)}\right]^{\sqrt{3}/2}. \label{eq:nstar}
\end{align}
\end{subequations}
It can be seen that throughout the rarefaction wave, the relative fugacity 
\eqref{eq:fugacity} remains equal to the relative fugacity in the left state:
\begin{equation}
 \overline{\lambda}_* = \frac{n_*^4}{P_*^3} = \frac{n_\rtin{L}^4}{P_\rtin{L}^3}  = 
 \overline{\lambda}_\rtin{L}.\label{eq:inv_lambda_rar}
\end{equation}
Once the similarity variable $\zeta_T$ of the tail of the rarefaction wave is known, the 
values of $n_{\rtin{I}}$, $P_{\rtin C}$, and $\beta_{\rtin{C}}$ can be found. 
The problem is resolved by finding the velocity on the central plateau (from 
the conditions at the shock front) and matching with Eq.~(\ref{eq:v(xi)}).

\paragraph{Shock wave.}
In the hydrodynamic description, the shockwaves represent discontinuities in the values 
of the macroscopic fields of the fluid. The values of these quantities to the 
left and right of the shockwave are given by the Rankine-Hugoniot 
junction conditions \cite{rezzolla13}:
\begin{subequations}\label{eq:junction}
\begin{align} 
n_{\rtin{II}}\, \gamma_{\rtin{C,\textsl{S}}}\, \beta_{\rtin{C,\textsl{S}}} =&  
n_{\rtin{R}}\, \gamma_{\rtin{R,\textsl{S}}}\, \beta_{\rtin{R,\textsl{S}}}, \\
P_{\rtin{C}}\gamma_{\rtin{C,\textsl{S}}}^2\left(1 + 3\,  \beta_{\rtin{C,\textsl{S}}}^2 \right) =& 
P_{\rtin{R}} \gamma_{\rtin{R,\textsl{S}}}^2\, \left(1 + 3\, \beta_{\rtin{R,\textsl{S}}}^2 \right), \\
P_{\rtin{C}} \gamma_{\rtin{C,\textsl{S}}}^2 \beta_{\rtin{C,\textsl{S}}} =&  
P_{\rtin{R}} \gamma_{\rtin{R,\textsl{S}}}^2 \beta_{\rtin{R,\textsl{S}}},
\end{align}
\end{subequations}
where the subscript $s$ indicates that the velocities $\beta_{\rtin{C,\textsl{S}}}$ and 
$\beta_{\rtin{R,\textsl{S}}}$ are evaluated in the rest frame of the shock front. 
Manipulating Eq.~\eqref{eq:junction},
the velocities of the fluid on the two sides of the shock can be obtained:
\begin{equation}
\beta_{\rtin{C,\textsl{S}}} = -\sqrt{\frac{P_{\rtin{C}} + 3P_{\rtin{R}}}{3(P_{\rtin{R}}+3P_{\rtin{C}})}}, \qquad
\beta_{\rtin{R,\textsl{S}}} = - \sqrt{\frac{P_{\rtin{R}}+3P_{\rtin{C}}}{3(P_{\rtin{C}} + 3P_{\rtin{R}})}}.
\end{equation}
The velocities of the shock front and on the central plateau with respect to the Eulerian frame 
in which the unperturbed fluid is at rest can be obtained using the relativistic law of 
velocity composition:
\begin{gather}
 \beta_{\rtin{shock}} = \sqrt{\frac{P_{\rtin{R}}+3P_{\rtin{C}}}
 {3(P_{\rtin{C}} + 3P_{\rtin{R}})}}, \nn\\
\beta_{\rtin{C}} = \frac{\beta_{\rtin{C,\textsl{S}}} - \beta_{\rtin{R,\textsl{S}}}}
{ 1 - \beta_{\rtin{C,\textsl{S}}}\beta_{\rtin{R,\textsl{S}}}} 
  = \sqrt{\frac{3(P_{\rtin{C}}-P_{\rtin{R}})^2}{(P_{\rtin{C}}+3P_{\rtin{\textsl{R}}})(3P_{\rtin{C}}+P_{\rtin{\textsl{R}}})}}.
  \label{eq:Sod:uplateau}
\end{gather}
From the first junction condition \eqref{eq:junction}, the density in the left side of the shock 
$n_{\rtin{II}}$ can be obtained as:
\begin{equation}
n_{\rtin{II}} = n_{\rtin{R}} 
\frac{ \gamma_{\rtin{R,\textsl{S}}}\, \beta_{\rtin{R,\textsl{S}}} }
{ \gamma_{\rtin{C,\textsl{S}}}\, \beta_{\rtin{C,\textsl{S}}} } = 
n_R \sqrt{\frac{P_C(P_R + 3P_C)}{P_R(P_C + 3P_R)}}.
\label{eq:Sod:nII}
\end{equation}

\paragraph{Contact discontinuity and central plateau.}
The contact discontinuity is a particular kind of shockwave where there is no mass transport through 
the discontinuity surface with the pressure and velocity
of the fluid being constant throughout, such that 
$\beta_{\rtin{I}} = \beta_{\rtin{II}} \equiv \beta_{\rtin{C}}$, 
$P_{\rtin{I}} = P_{\rtin{II}} \equiv P_\rtin{C}$, and 
$n_{\rtin{I}} \neq n_{\rtin{II}}$.

The last parameter needed to find the complete solution of the Sod problem is the pressure on the plateau. 
This can be found by requiring that Eqs.~\eqref{eq:Sod:uplateau}, \eqref{eq:Pstar}, and \eqref{eq:v(xi)} 
are simultaneously satisfied. This yields the following expression for the coordinate of the 
rarefaction tail $\zeta_{\rtin{T}}$:
\begin{equation}
\zeta_{\rtin{T}} = \frac{\sqrt{3} - 1 - (\sqrt{3} + 1) (P_\rtin{C} / P_{\rtin L})^{\sqrt{3}/2}}
{\sqrt{3} - 1 + (\sqrt{3} + 1) (P_\rtin{C} / P_{\rtin L})^{\sqrt{3}/2}}.
\label{eq:Sod:xiT}
\end{equation}
The pressure on the central plateau $P_{\rtin{C}}$ can be found as a root of the equation
\begin{equation}
\frac{(P_{\rtin{L}}/P_\rtin{C})^{\sqrt{3}/2} - 1}{(P_{\rtin{L}}/P_\rtin{C})^{\sqrt{3}/2} + 1} - 
\sqrt{\frac{3(P_\rtin{C}-P_{\rtin{R}})^2}{(P_\rtin{C}+3P_{\rtin{R}})(3P_\rtin{C}+P_{\rtin{R}})}} = 0.
\label{eq:Sod:PC}
\end{equation}
The density $n_{\rtin{I}}$ between the rarefaction tail and the contact discontinuity 
can be found using Eq.~\eqref{eq:nstar} with $\zeta = \zeta_{\rtin{T}}$:
\begin{equation}
 n_{\rtin{I}} = n_{\rtin{L}} \left(\frac{P_{\rtin{C}}}{P_{\rtin{L}}}\right)^{3/4}.
 \label{eq:Sod:nI}
\end{equation}

\paragraph{Full solution.}
In order to obtain the full solution of the Sod shock tube problem, the following steps can 
be taken. First, the pressure on the central plateau $P_{\rtin{C}}$ can be found as a root 
of Eq.~\eqref{eq:Sod:PC}. Then, the shock front velocity $\beta_{\rtin{shock}}$, 
central plateau velocity $\beta_{\rtin{C}}$, and rarefaction tail similarity coordinate $\zeta_T$
can be found using Eqs.~\eqref{eq:Sod:uplateau} and \eqref{eq:Sod:xiT}.
The densities $n_{\rtin{I}}$ and $n_{\rtin{II}}$ to the left and right 
of the contact discontinuity can be found using Eqs.~\eqref{eq:Sod:nI} 
and \eqref{eq:Sod:nII}, respectively. The structure of the 
full solution is represented in Fig.~\ref{fig:inv_structure} and 
has the following mathematical formulation:
\begin{gather}
  \beta(z, t) = \begin{cases}
    \beta_{\rtin{L}}, &  z<z_{\rtin{H}},\\
    \beta_*, &  z_{\rtin{H}} < z < z_{\rtin{T}}, \\
    \beta_{\rtin{C}},& z_{\rtin{T}} <  z < z_{\rtin{S}},\\
    \beta_{\rtin{R}}, & z>z_{\rtin{S}},
  \end{cases}  	\nn \\
 P(z, t) = \begin{cases}
    P_{\rtin{L}}, & z<z_{\rtin{H}},\\
    P_*, &  z_{\rtin{H}} < z< z_{\rtin{T}}, \\
    P_{\rtin{C}},& z_{\rtin{T}} < z< z_{\rtin{S}},\\
    P_{\rtin{R}}, & z>z_{\rtin{S}},
  \end{cases}, \nonumber\\
  n(z, t) = \begin{cases}
    n_{\rtin{L}}, & z<z_{\rtin{H}},\\
    n_*,  & z_{\rtin{H}} < z < z_{\rtin{T}}, \\
    n_{\rtin{I}}, & z_{\rtin{T}} < z < z_{\rtin{C}},\\
    n_{\rtin{II}},& z_{\rtin{C}} < z < z_{\rtin{S}},\\
    n_{\rtin{R}}, & z>z_{\rtin{S}}.
  \end{cases}
\end{gather}
In the above, $z_{\rtin{H}}$ ($z_{\rtin{T}}$), $z_{\rtin{C}}$ and $z_{\rtin{S}}$ represent 
the locations of the head (tail) of the rarefaction wave, of the contact discontinuity and 
of the shock front, respectively, being given by
\begin{align}
 z_{\rtin{H}} =& -c_s t, \qquad 
 z_{\rtin{T}} = \left(\frac{\beta_{\rtin{C}} - c_s}{1- \beta_{\rtin{C}}c_s} \right) t, \qquad  \\
 z_{\rtin{C}} =&\, \beta_{\rtin{C}}t, \qquad \quad 
 z_{\rtin{S}} = \beta_{\rtin{shock}}t,
\end{align}
where $c_s = 1/\sqrt{3}$ is the sound speed and $\beta_{\rtin{C}}$ and 
$\beta_{\rtin{shock}}$ are given in Eq.~\eqref{eq:Sod:uplateau}.

\subsubsection{Macroscopic profiles}\label{sec:Sod:inviscid:macro}

\begin{figure*}
\begin{tabular}{cc}
\includegraphics[width=0.4\linewidth]{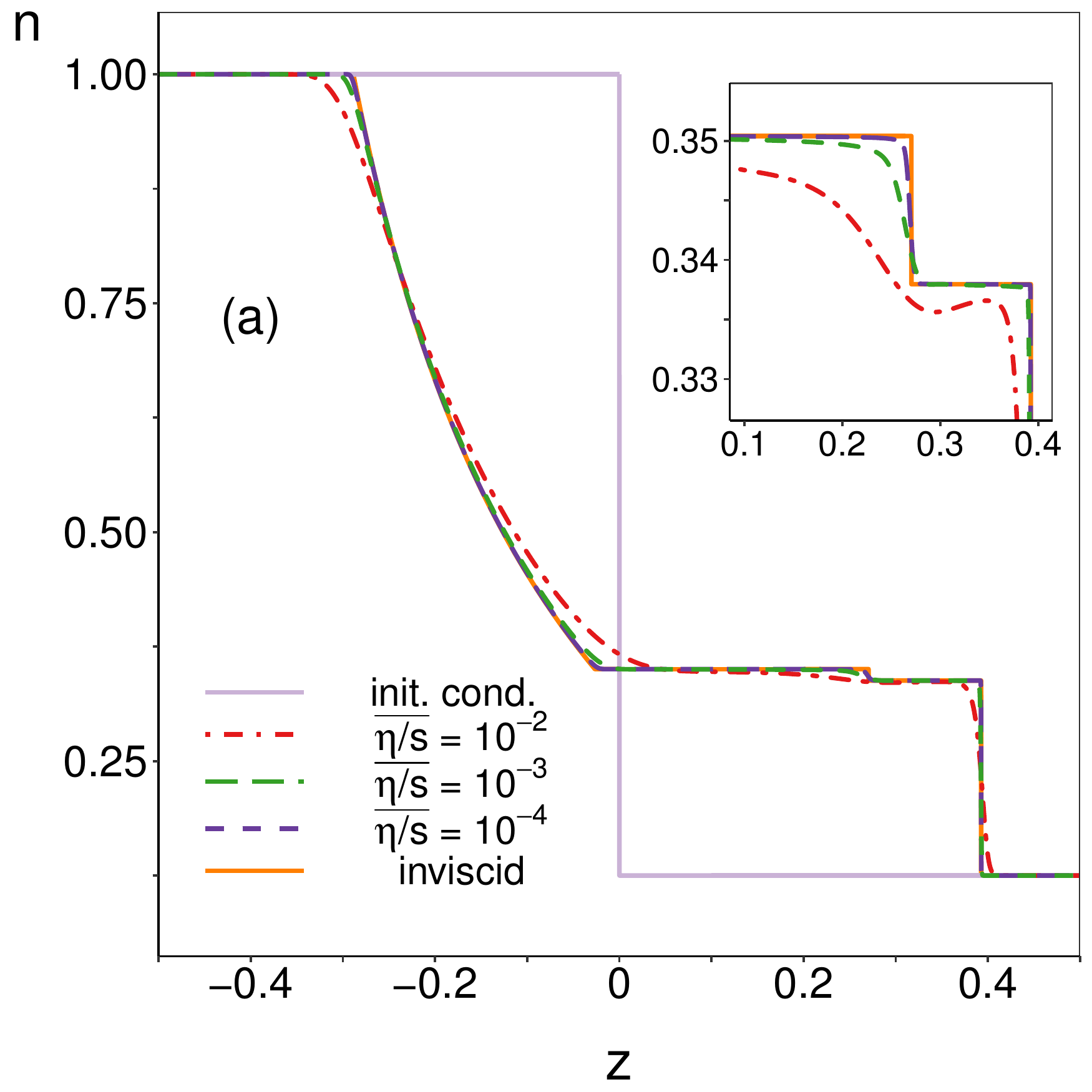} &
\includegraphics[width=0.4\linewidth]{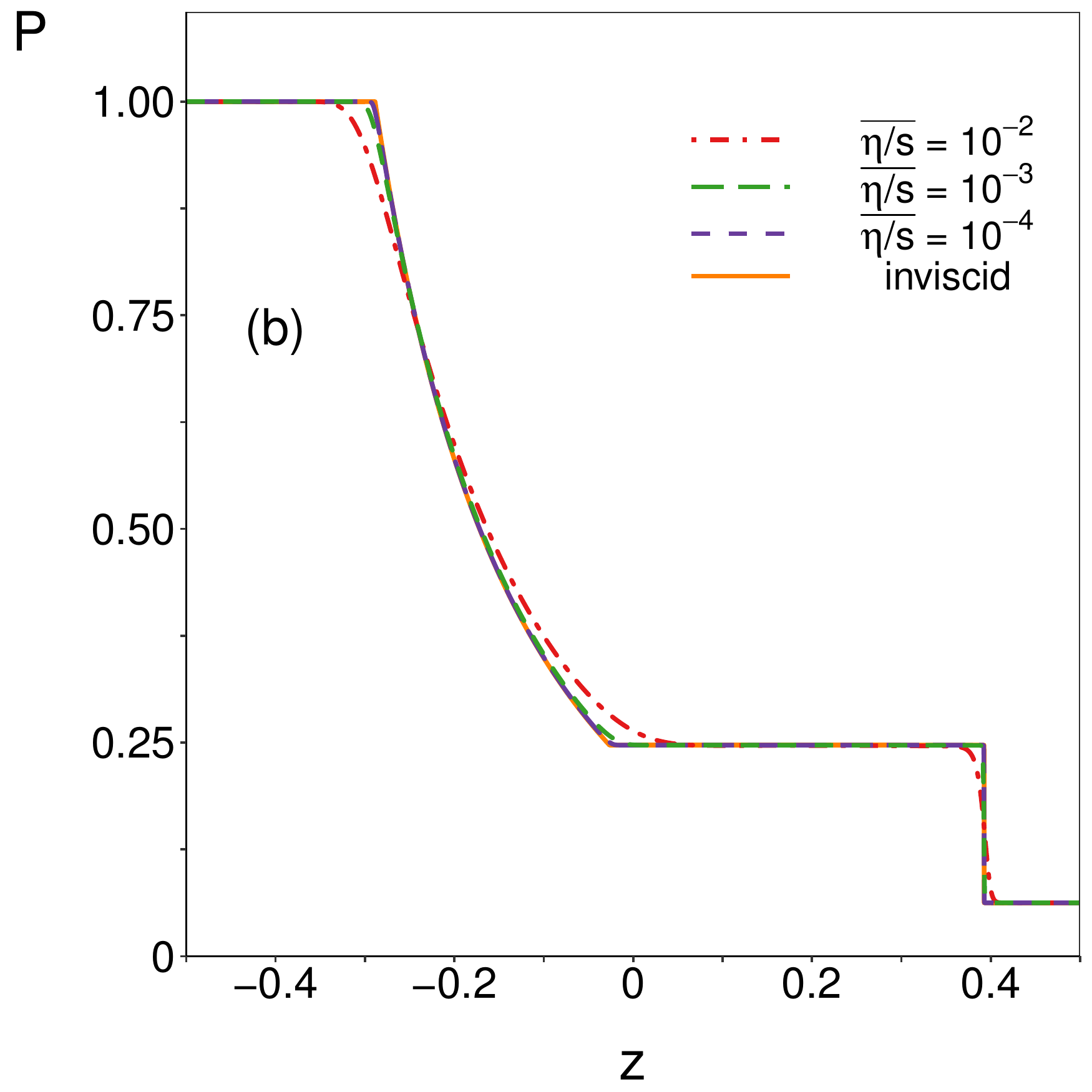} 
\\
\includegraphics[width=0.4\linewidth]{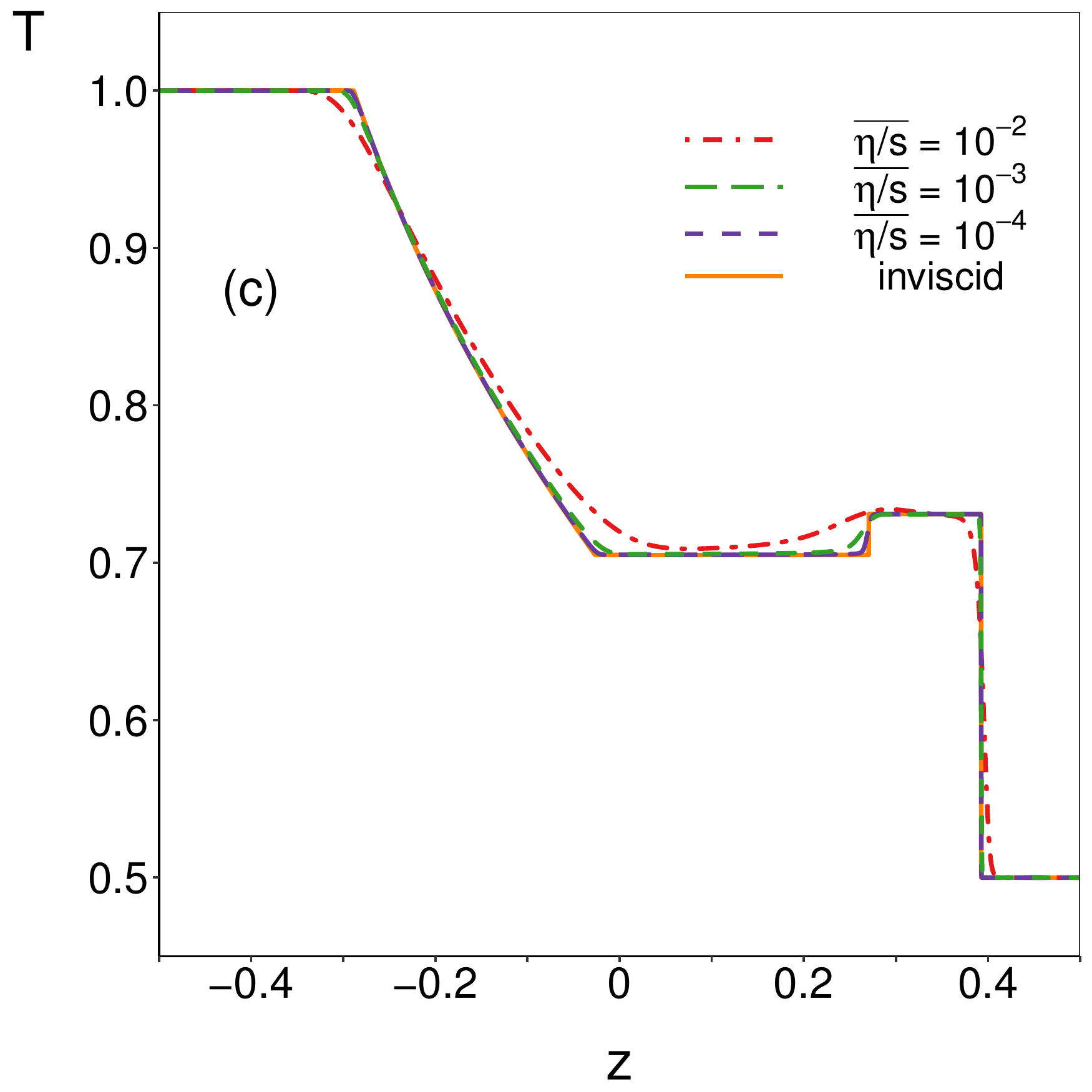} &
\includegraphics[width=0.4\linewidth]{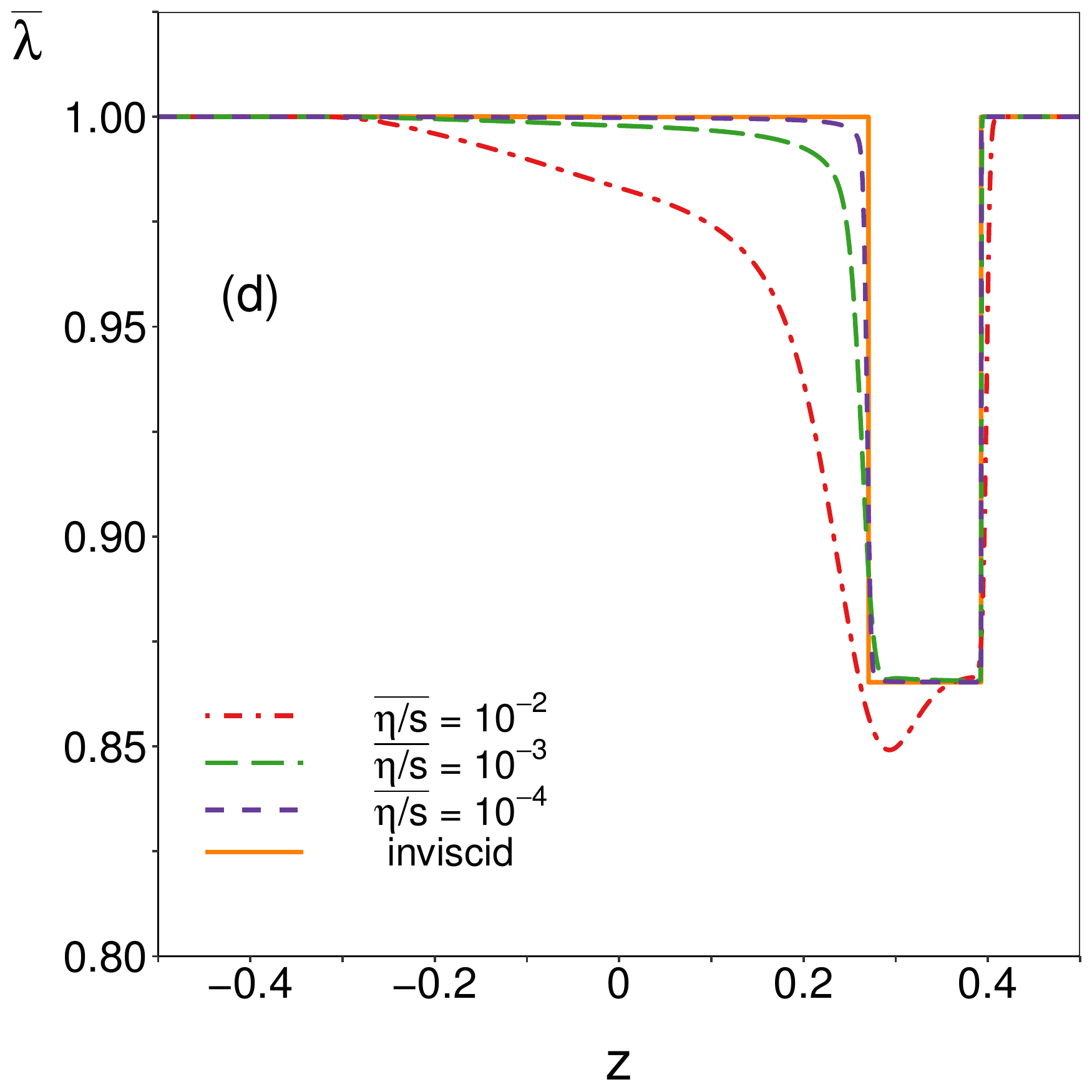} 
\end{tabular}
\begin{center}
\begin{tabular}{c}
\includegraphics[width=0.4\linewidth]{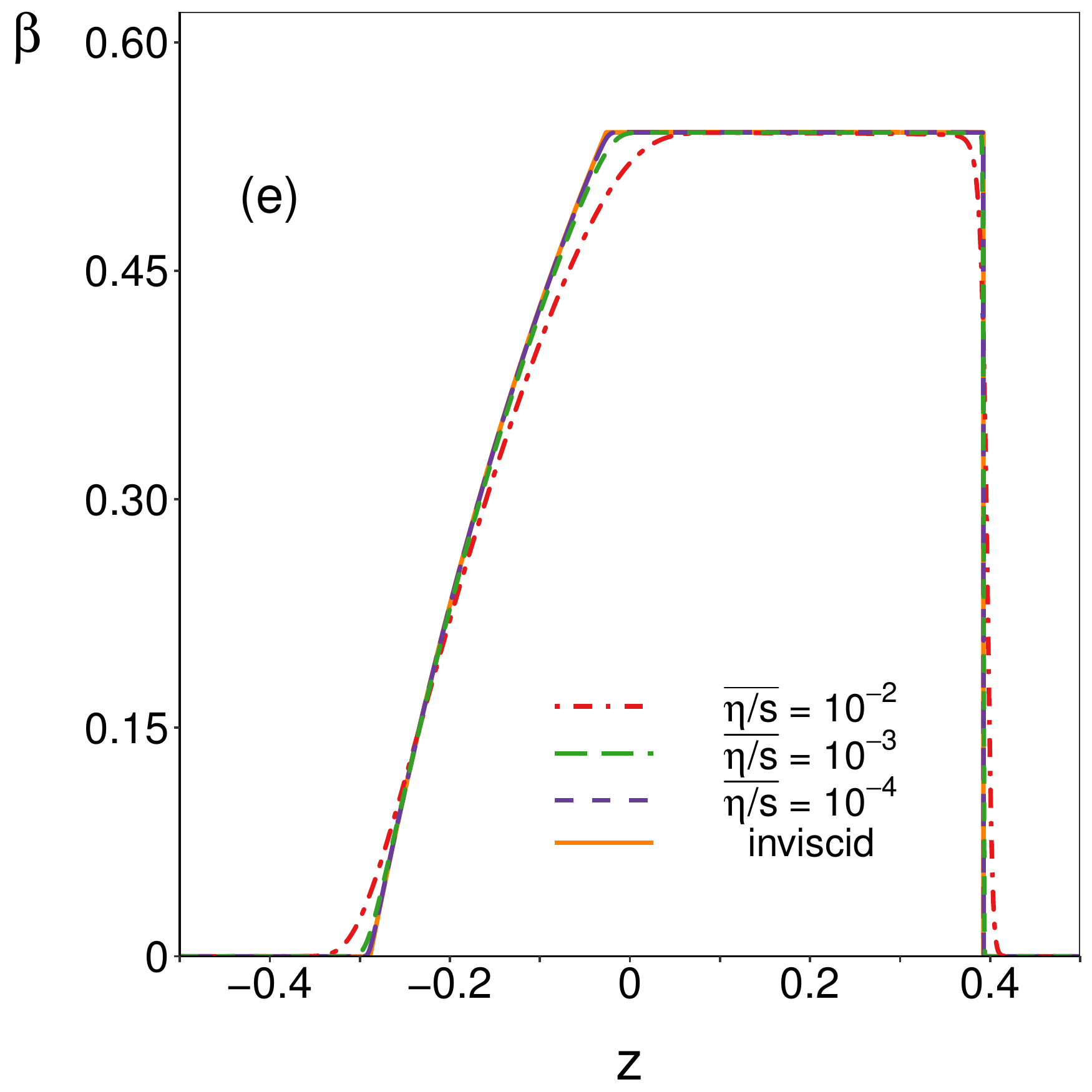} 
\end{tabular}
\end{center}
\caption{Density (a), pressure (b), temperature (c),
fugacity (d) and velocity (e) profiles
for decreasing values of the ratio $\etas$,
compared with the analytical results for the inviscid limit, presented
in Sec.~\ref{sec:Sod:inviscid:analytic}. The quadrature order is $Q_\xi = 6$, the number of points 
is $Z = 10\,000$, the time step was set to $5 \times 10^{-6}$ and the snapshot 
corresponds to $t = 0.5$ (i.e., $100\,000$ iterations were performed).
The initial conditions are $(n_\rtin{L}, T_\rtin{L}, P_\rtin{L}) = (1,1,1)$ and 
$(n_\rtin{R}, T_\rtin{R}, P_\rtin{R}) = (0.125, 0.5, 0.0625)$.
}
\label{fig:Rinviscid}
\end{figure*}

Let us now specialize the solution obtained in Sec.~\ref{sec:Sod:inviscid:analytic} 
to the case considered in 
Refs.~\cite{bouras10,bouras09nucl,hupp11,mohseni13}, where
\begin{gather}
 \widetilde{k}_B \widetilde{T}_{\rtin{ref}} = 0.4\ {\rm GeV}, \qquad
 \widetilde{P}_{\rtin{ref}} = 5.43\ {\rm GeV}/{\rm fm}^3, \nn\\
 \widetilde{n}_{\rtin{ref}} = 13.57\ {\rm fm}^{-3},
 \label{eq:Sod:left}
\end{gather}
while the reference length is $\widetilde{L}_{\rtin{ref}} = 6.4\ {\rm fm}$.
With the above quantities, $\lambda_{\rtin{ref}} \simeq 1$ and
\begin{align}
 \tau_{\rtin{A-W}} =& \frac{\tau_{\rtin{A-W};0}}{T}\left(1 - \frac{\ln \overline{\lambda}}{4} \right), 
 \nonumber\\
 \tau_{\rtin{A-W}; 0} \simeq& 0.3854 \etas.
 \label{eq:Sod:tau1}
\end{align}
The reference quantities given in Eq.~\eqref{eq:Sod:left} refer to the initial state
in the left half of the channel. In the right half, the following values are used:
\begin{gather}
 \widetilde{k}_B \widetilde{T}_{\rtin{R}} = 0.2\ {\rm GeV}, \qquad
 \widetilde{P}_{\rtin{R}} = 0.33\ {\rm GeV}/{\rm fm}^3, \nn\\
 \widetilde{n}_{\rtin{R}} = 1.66\ {\rm fm}^{-3},
 \label{eq:Sod:right}
\end{gather}
which correspond to:
\begin{equation}
 T_{\rtin R} = 0.5, \qquad
 P_{\rtin R} = 0.0625, \qquad
 n_{\rtin R} = 0.125.\label{eq:Sod:right_adim}
\end{equation}
With the above values, the relative fugacity \eqref{eq:fugacity} at initial 
time is constant throughout the channel:
\begin{equation}
 \overline{\lambda}_{\rtin L} = \overline{\lambda}_{\rtin R} = 1.\label{eq:lambda_LR}
\end{equation}
The relaxation times in the two halves of the channel are:
\begin{equation}
 \tau_{\rtin{A-W}; \rtin{L}} \simeq 0.3854 \etas, \qquad
 \tau_{\rtin{A-W}; \rtin{R}} \simeq 0.7708 \etas.\label{eq:tau_LR}
\end{equation}

Applying the analysis in Sec.~\ref{sec:Sod:inviscid:analytic}, the following numerical values 
are obtained:
\begin{gather}
 \beta_{\rtin{shock}} \simeq 0.785, \qquad 
 \beta_{\rtin{C}} \simeq 0.541, \nn\\
 P_{\rtin{C}} \simeq 0.247, \qquad 
 n_{\rtin{I}} \simeq 0.350, \qquad 
 n_{\rtin{II}} \simeq 0.338, 
\end{gather}
while the rarefaction head and tail are located at 
$z_{\rtin{H}} = -t/\sqrt{3} \simeq -0.577 t$ and 
$z_T = \zeta_T t \simeq -0.053 t$.
The contact discontinuity is located at $z_{\rtin{C}} = \beta_{\rtin{C}} t \simeq 0.541 t$.

Figure~\ref{fig:Rinviscid} shows the profiles of $n$, $P$, $T$, and $\overline{\lambda}$
obtained from our simulations at various values of $\etas$.
The convergence towards the analytical result can be clearly seen as 
$\etas \rightarrow 0$ (the inviscid limit). 
For these simulations, we used a grid of $Z = 10\,000$ nodes and a time step 
of $\delta t = 5 \times 10^{-6}$.
While the shock front in the particle number density profile is already well recovered at
$\etas = 10^{-3}$, at $\etas = 10^{-4}$ it spans around four to six grid points, which is equivalent to
$0.04$--$0.06\%$ of the size of the domain.

The orders of the quadratures are fixed at $Q_\rtin{L} = 2$, $Q_\varphi = 1$, 
and $Q_\xi = 6$ for the radial, azimuthal and polar quadratures, while 
the expansion of $\feq$ was truncated at $N_\rtin{L} = 1$ and 
$N_\Omega = 5$. We remind the reader that the values $Q_\xi = 6$ and 
$N_\Omega = 5$ are used here to validate a benchmark ``reference profile,'' while in
Sec.~\ref{sec:Sod:conv}, the capabilities of models with smaller values of $Q_\xi$ and 
$N_\Omega$ will be discussed. 

If $\etas$ is decreased below $10^{-4}$ while keeping $\delta t$
and $\delta z$ fixed at the values used above, the macroscopic profiles
start developing spurious oscillations. For the values of $\etas \ge 10^{-4}$,
our simulations are stable, mainly due to the increased stability ensured by the combination of
the fifth order WENO and third order Runge-Kutta schemes, employed for the implementation of the
spatial and temporal derivatives, respectively.

\subsubsection{Non-equilibrium quantities and dissipative effects} \label{sec:Sod:inviscid:noneq}

\begin{figure*}
\begin{center}
\begin{tabular}{cc}
\includegraphics[width=0.45\linewidth]{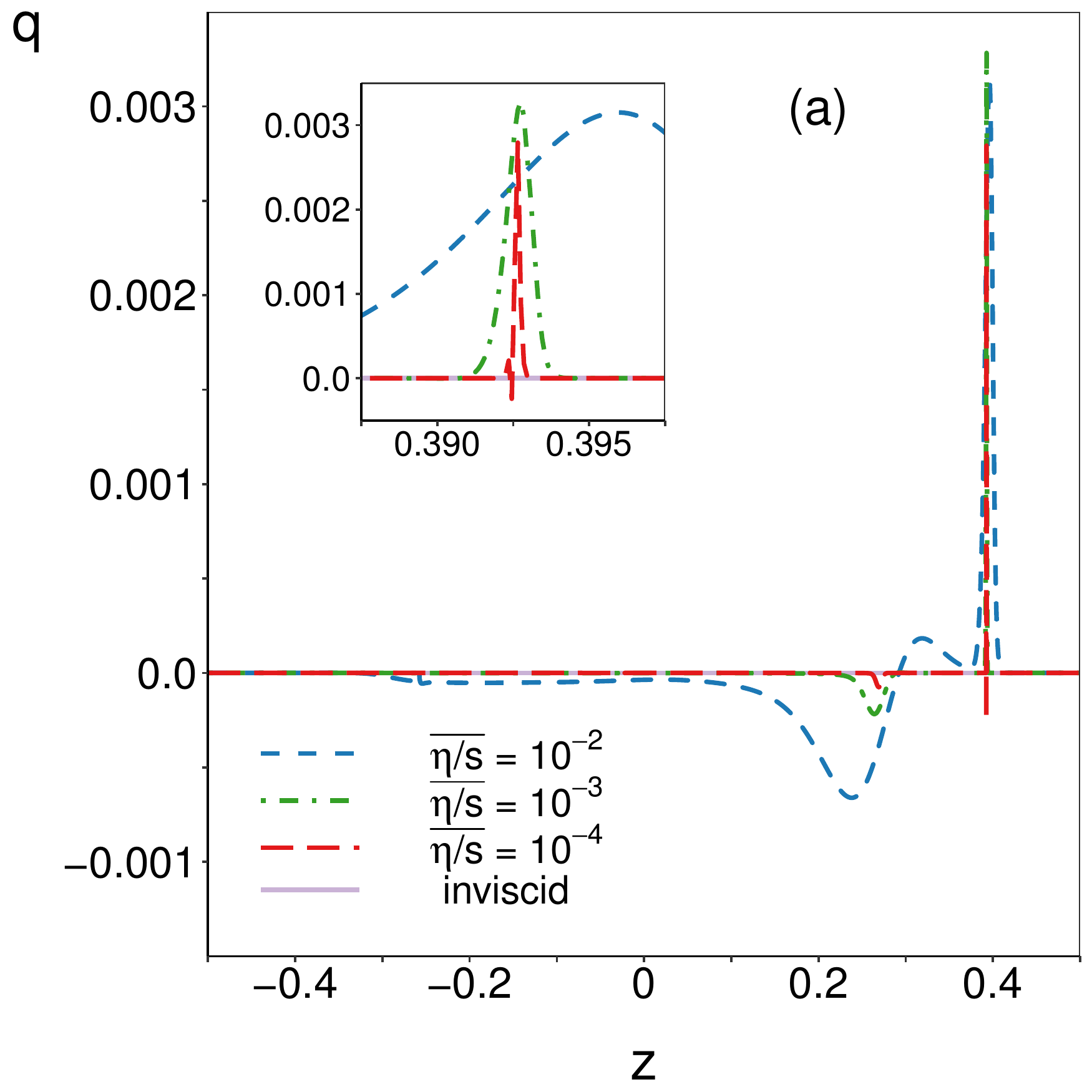} &
\includegraphics[width=0.45\linewidth]{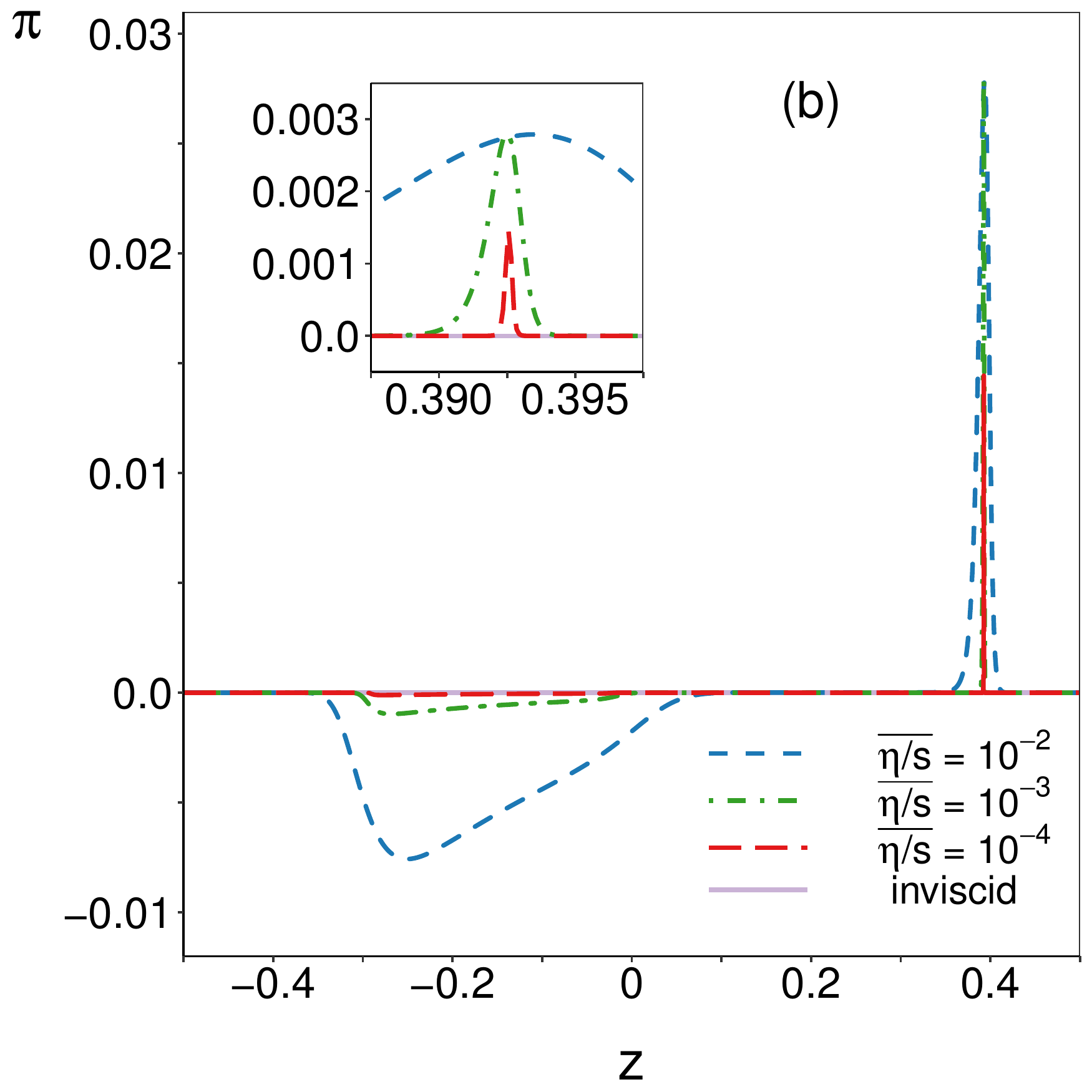} 
\end{tabular}
\begin{tabular}{c}
\includegraphics[width=0.45\linewidth]{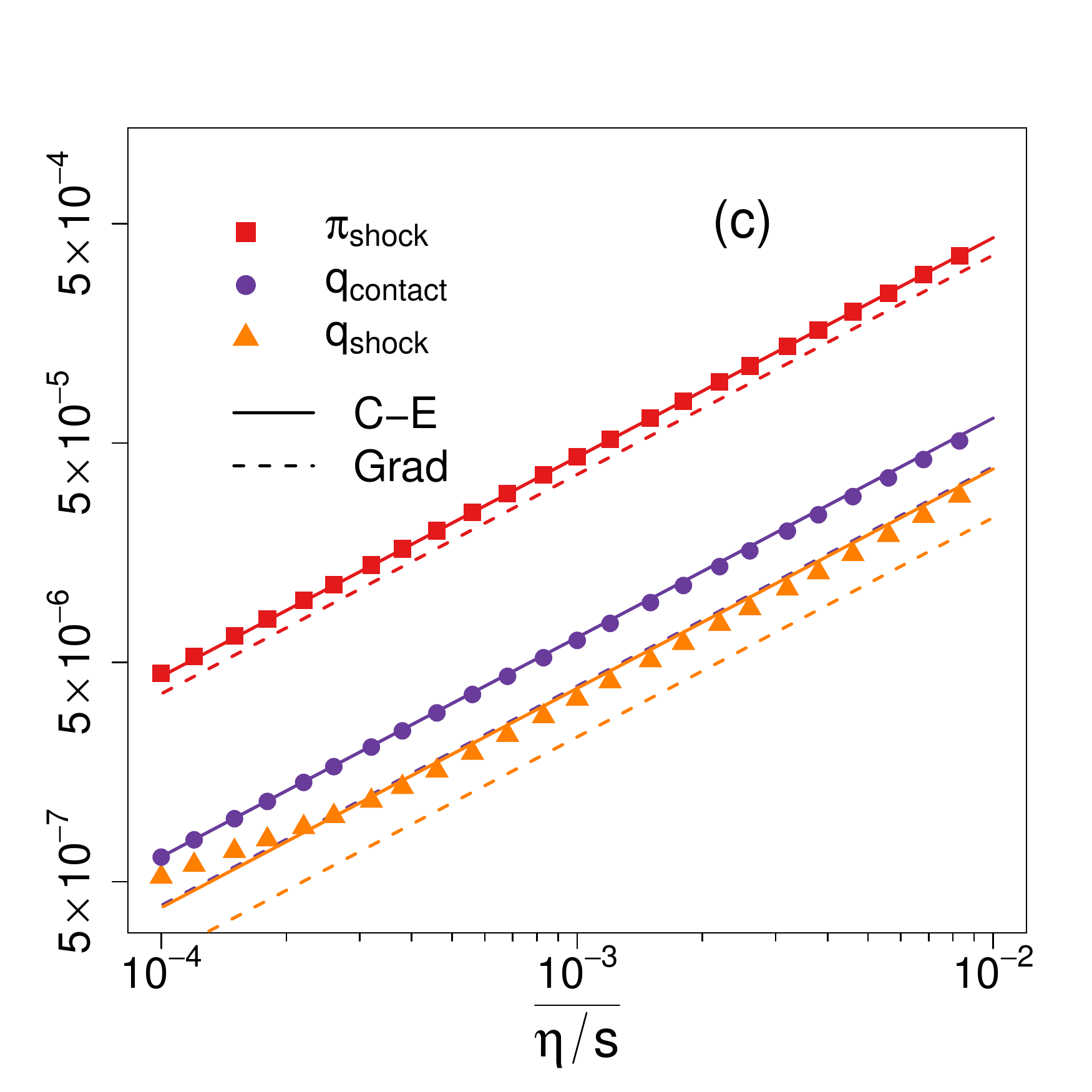} 
\end{tabular}
\end{center}
\caption{Profiles of $q$ (a) and $\Pi$ (b) for decreasing values of $\etas$.
(c) The integrated values of $q$ and $\Pi$ over small vicinities around the 
contact discontinuity and the shock front are shown in absolute value as functions of $\etas$ in log-log scale
(the integral of $\Pi$ around the contact 
discontinuity is not shown, since $\Pi$ vanishes in this region). The analytic results 
\eqref{eq:intq} and \eqref{eq:intpi} are also represented for
comparison, for values of the transport coefficients obtained with both the Chapman-Enskog (solid lines) 
and Grad (dashed lines) methods. 
The simulation parameters are presented in the caption of Fig.~\ref{fig:Rinviscid}.}
\label{fig:inv_noneqb}
\end{figure*}

In Subsec.~\ref{sec:Sod:inviscid:analytic}, we considered the flow of a perfect fluid. In theory, this corresponds
to taking the limit of vanishing relaxation time in the Anderson-Witting Boltzmann equation \eqref{eq:boltz_Sod}.
In our implementation of the Boltzmann equation, the relaxation time can never
be decreased to $0$, such that the transport coefficients defined in Eqs.~\eqref{eq:tcoeff_ce},
assumed to vanish in the inviscid limit presented above, will always be finite.
According to Eqs.~\eqref{eq:hydro}, the heat flux $q^\mu$ and pressure deviator $\Pi^{\mu\nu}$
can be obtained in the hydrodynamic regime (small values of $\tau_{\rtin{A-W}}$) by taking gradients
of $T$, $P$ and $u^\mu$. In the inviscid limit, these quantities are discontinuous
in the vicinity of the shock front and of the contact discontinuity (the pressure
and velocity are discontinuous only at the shock front). Thus, it is natural to 
expect that their gradients will be sharply peaked in these regions.

As mentioned in Sec.~\ref{sec:boltz:Sod}, $q^\mu$ and $\Pi^{\mu\nu}$ can be fully characterized  
by the scalar quantities $q$ \eqref{eq:q_def} and $\Pi$ \eqref{eq:pi_def}, 
which we represent in Fig.~\ref{fig:inv_noneqb}.
It can be seen in plot (a) that $q$ presents a strong fluctuation around the contact discontinuity,
while plot (b) demonstrates that $\Pi$ has nonvanishing values mostly around the
rarefaction wave, peaking towards its head.
Both $q$ and $\Pi$ exhibit a strong spike at the shock front, even when $\etas = 10^{-4}$, induced 
via Eqs.~\eqref{eq:hydro} due to the strong gradients of the macroscopic fields.
The width of this spike decreases as $\etas$ is decreased and the shock front becomes 
narrower. 

In this subsection, we consider a quantitative analysis of these nonequilibrium 
effects by evaluating the integrated values of $q$ and $\Pi$ over the regions where 
they are non-negligible. The value of these integrals can be estimated analytically 
by considering the flow to be close to the inviscid limit, where the solution for 
the macroscopic profiles is given in Sec.~\ref{sec:Sod:inviscid:analytic}.

Considering that the macroscopic fields depend only on the similarity variable 
$\zeta = z/t$, the action of the convective derivative $D = u^\mu \partial_\mu$ and
of $\Delta^{\mu\nu} \partial_\nu = \eta^{\mu\nu} \partial_\nu + u^\mu D$ on an arbitrary function 
$f(\zeta)$ can be reduced to:
\begin{subequations}
\begin{align}
 D f(\zeta) =& \frac{\gamma}{t} (\beta - \zeta) \partial_\zeta f,\\
  \Delta^{\mu\nu} \partial_\nu f(\zeta) =& \frac{\gamma^2}{t}
  \left( \delta^{\mu}{}_z + \beta \delta^{\mu}{}_0 \right) 
  (1 - \beta \zeta) \partial_\zeta f.
\end{align}
\end{subequations}
With the above formulas, Eqs.~\eqref{eq:hydro_q} and \eqref{eq:hydro_pi}
can be used to find $q$ and $\Pi$ as:
\begin{subequations}
\begin{align}
 q =& \frac{\lambda_\rtin{heat} T}{4} \Delta^{z\nu} \partial_\nu \ln \overline{\lambda} 
 = \frac{\lambda_\rtin{heat} T \gamma^2}{4t} (1 - \beta \zeta) \partial_\zeta \ln \overline{\lambda},\label{eq:q_inv}\\
 \Pi =& -\frac{4\eta}{3} \partial_\mu u^\mu 
 = -\frac{4\eta}{3t} 
 (1 - \beta \zeta) \partial_\zeta (\gamma \beta),\label{eq:pi_inv}
\end{align}
\end{subequations}
where the definition \eqref{eq:fugacity} for the relative fugacity $\overline{\lambda}$ was 
employed.

Since, according to Eq.~\eqref{eq:inv_lambda_rar}, $\overline{\lambda}$ is constant from 
the unperturbed region on the left
up to the contact discontinuity, $q$ vanishes in this region. This is confirmed in
Fig.~\ref{fig:inv_noneqb}(a). Similarly, $\Pi$ vanishes between the tail of the rarefaction and 
the shock front, since $\beta = \beta_{\rtin{C}}$ is constant in this region. 
Figure~\ref{fig:inv_noneqb}(b) confirms this prediction.

Close to the shock front, both $q$ and $\Pi$ can be put in the form:
\begin{equation}
 f_{\rtin{disc}} = \frac{1}{t} g(\zeta) \partial_\zeta h(\zeta),
\end{equation}
where both $g$ and $h$ may be discontinuous at $\zeta = \zeta_d$.
Using the Heaviside step function, $g(\zeta)$ and $\partial_{\zeta} h(\zeta)$ 
can be written as:
\begin{subequations}
\begin{align}
 g(\zeta) =& \theta(\zeta - \zeta_d) g_{>} + \theta(\zeta_d - \zeta) g_{<}, \\
 \partial_\zeta h(\zeta) =& \delta(\zeta - \zeta_d) (h_> - h_<),
\end{align}
\end{subequations}
where the subscript $>$ ($<$) denotes the value of the function $g(\zeta)$ or 
$h(\zeta)$ to the right (left) of the discontinuity.
The integral of $f_{\rtin{disc}}$ over a small region $z_d \pm \delta z$ around the
point $z_d = \zeta_d t$ is given by:
\begin{equation}
 \lim_{\delta z \rightarrow 0} \int_{z_d - \delta z}^{z_d + \delta z} f_{\rtin{disc}} dz =
 \frac{1}{2} (g_> + g_<) (h_> - h_<).\label{eq:int_disc}
\end{equation}

The value of the integrals of $q$ over the vicinity 
of the contact discontinuity and of the shock front can 
be estimated through
\begin{subequations} \label{eq:intq}
\begin{align}
 \text{contact:}& \int_{z_\rtin{C} - \delta z}^{z_\rtin{C} + \delta z} q\, dz =
 \frac{1}{8} (\lambda_{\rtin{heat},\rtin{I}} T_\rtin{I} + \lambda_{\rtin{heat},\rtin{II}} T_\rtin{II})
 \ln \frac{\overline{\lambda}_\rtin{II}}{\overline{\lambda}_\rtin{I}},\label{eq:intq_c}\\
 \text{shock:}\ \ \, & \int_{z_\rtin{S} - \delta z}^{z_\rtin{S} + \delta z} q\, dz = 
 \frac{1}{8} \left[\lambda_{\rtin{heat},R} T_R \right.\nonumber\\
 & \hspace{20pt} \left. + \lambda_{\rtin{heat},\rtin{II}} T_\rtin{II} 
 \gamma_\rtin{C}^2(1 - \beta_\rtin{C} \beta_\rtin{shock}) \right]
 \ln \frac{\overline{\lambda}_\rtin{R}}{\overline{\lambda}_\rtin{II}}.\label{eq:intq_s}
\end{align}
\end{subequations}
Similar expressions can be found for the integrals of $\Pi$:
\begin{subequations}\label{eq:intpi}
\begin{align}
 \text{contact:}& \int_{z_\rtin{C} - \delta z}^{z_\rtin{C} + \delta z} \Pi\, dz = 0,
 \label{eq:intpi_c}\\
 \text{shock:}\ \ \, & \int_{z_\rtin{S} - \delta z}^{z_\rtin{S} + \delta z} \Pi\, dz = 
 \frac{2 v_{\rtin{C}} \gamma_\rtin{C}}{3} \left[\eta_R +
 \eta_\rtin{II} (1 - v_\rtin{C} v_\rtin{S})\right].\label{eq:intpi_s}
\end{align}
\end{subequations}
In Fig.~\ref{fig:inv_noneqb}(c), the numerical estimate of the integrals calculated 
in Eqs.~\eqref{eq:intq} and \eqref{eq:intpi} are compared with the above analytic results
when the transport coefficients are obtained using either the Grad moment method \eqref{eq:tcoeff_grad}
or the Chapman-Enskog method \eqref{eq:tcoeff_ce}. 
The agreement between the numerical data and the analytic prediction 
when the Chapman-Enskog values of the transport coefficients are used
is remarkably good, even when $\etas \simeq 5 \times 10^{-3}$. 
This result confirms the analyses of 
Refs.~\cite{bhalerao14,ryblewski15,florkowski15,gabbana2017,ambrus18prc},
where further evidence was found supporting the validity of the 
Chapman-Enskog (as opposed to the Grad method) expressions for the transport coefficients,
even when large gradients are present.

\subsection{Ballistic regime}\label{sec:Sod:ballistic}

\begin{figure*}
\begin{tabular}{cc}
 \includegraphics[width=0.4\linewidth]{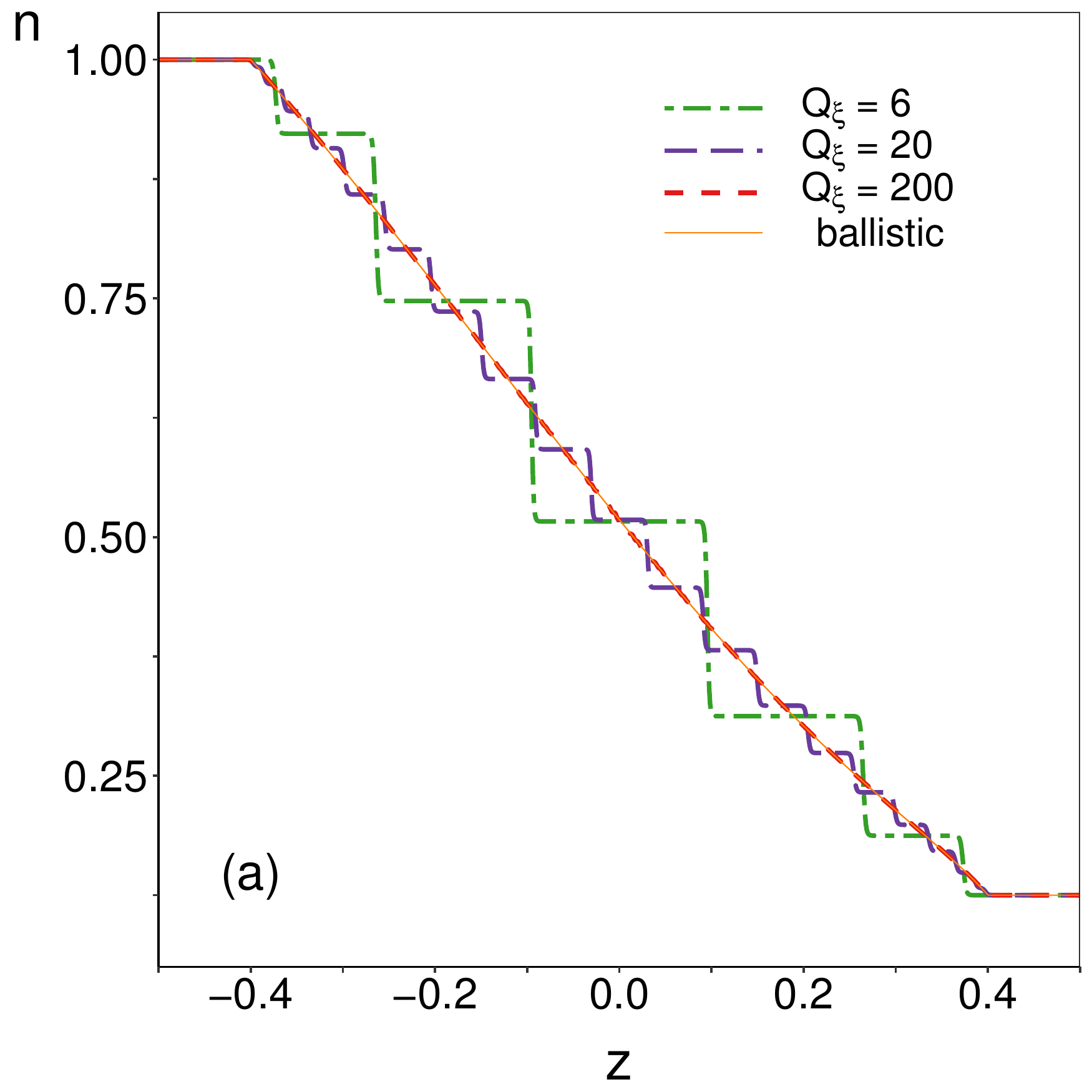} &
 \includegraphics[width=0.4\linewidth]{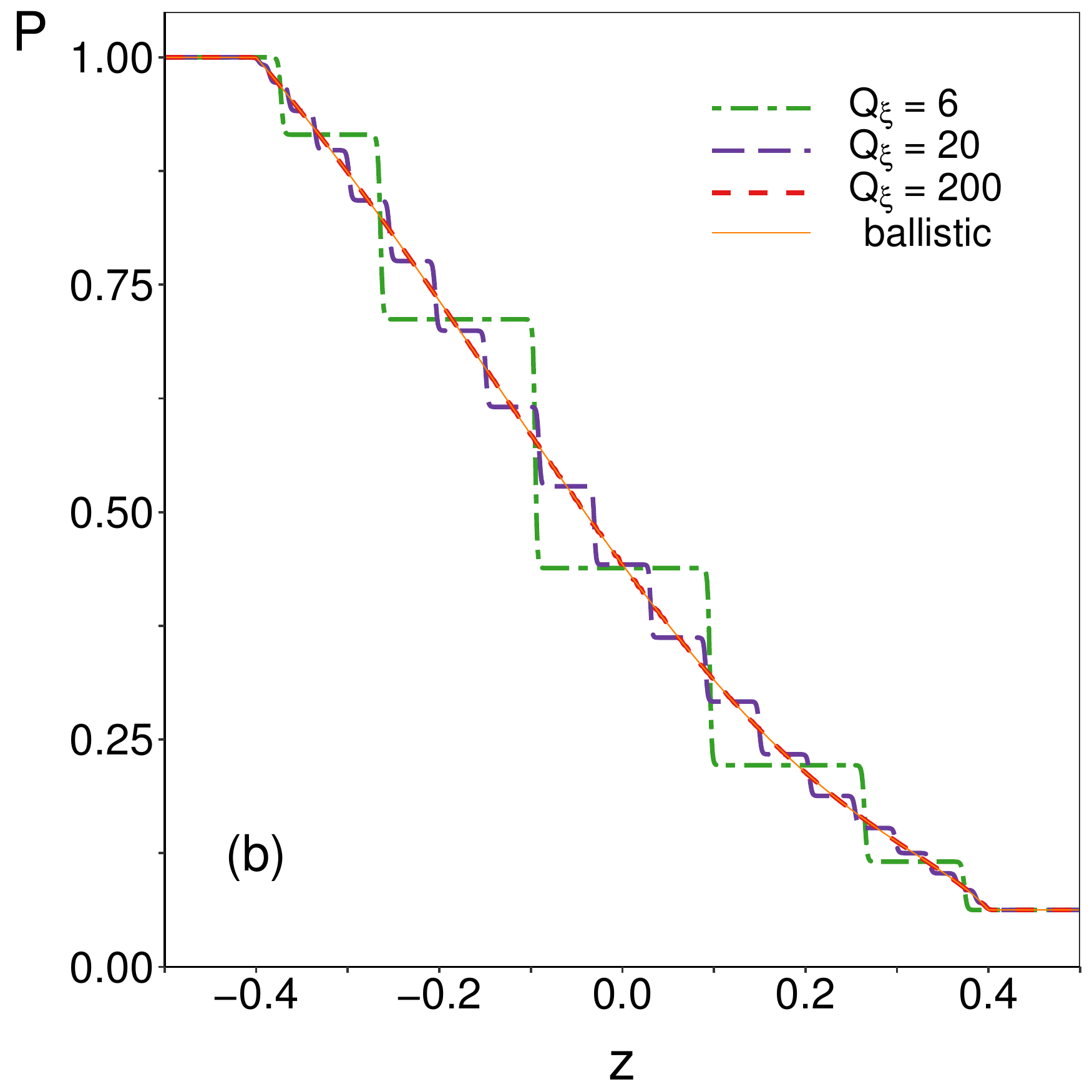}
 \\
 \includegraphics[width=0.4\linewidth]{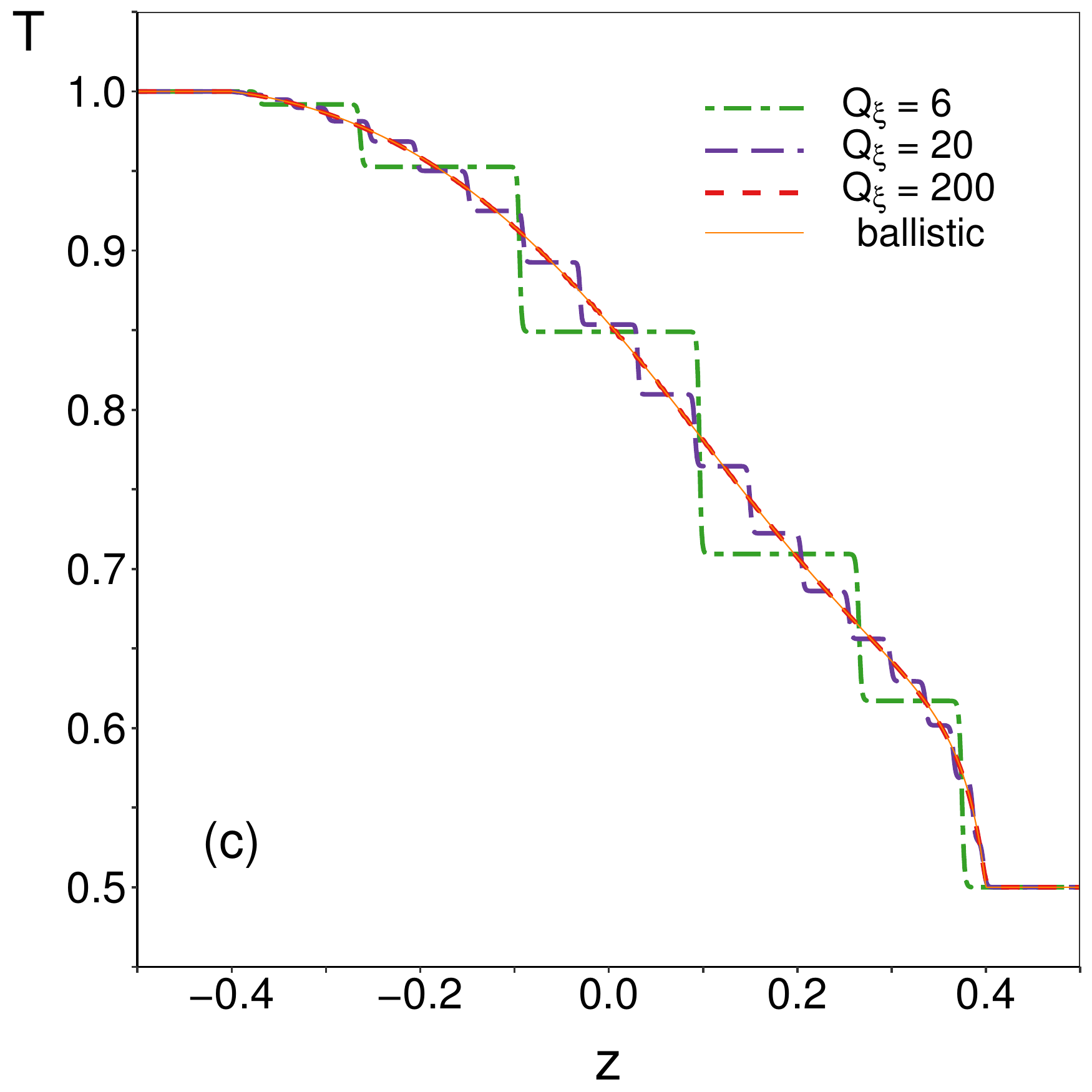} &
 \includegraphics[width=0.4\linewidth]{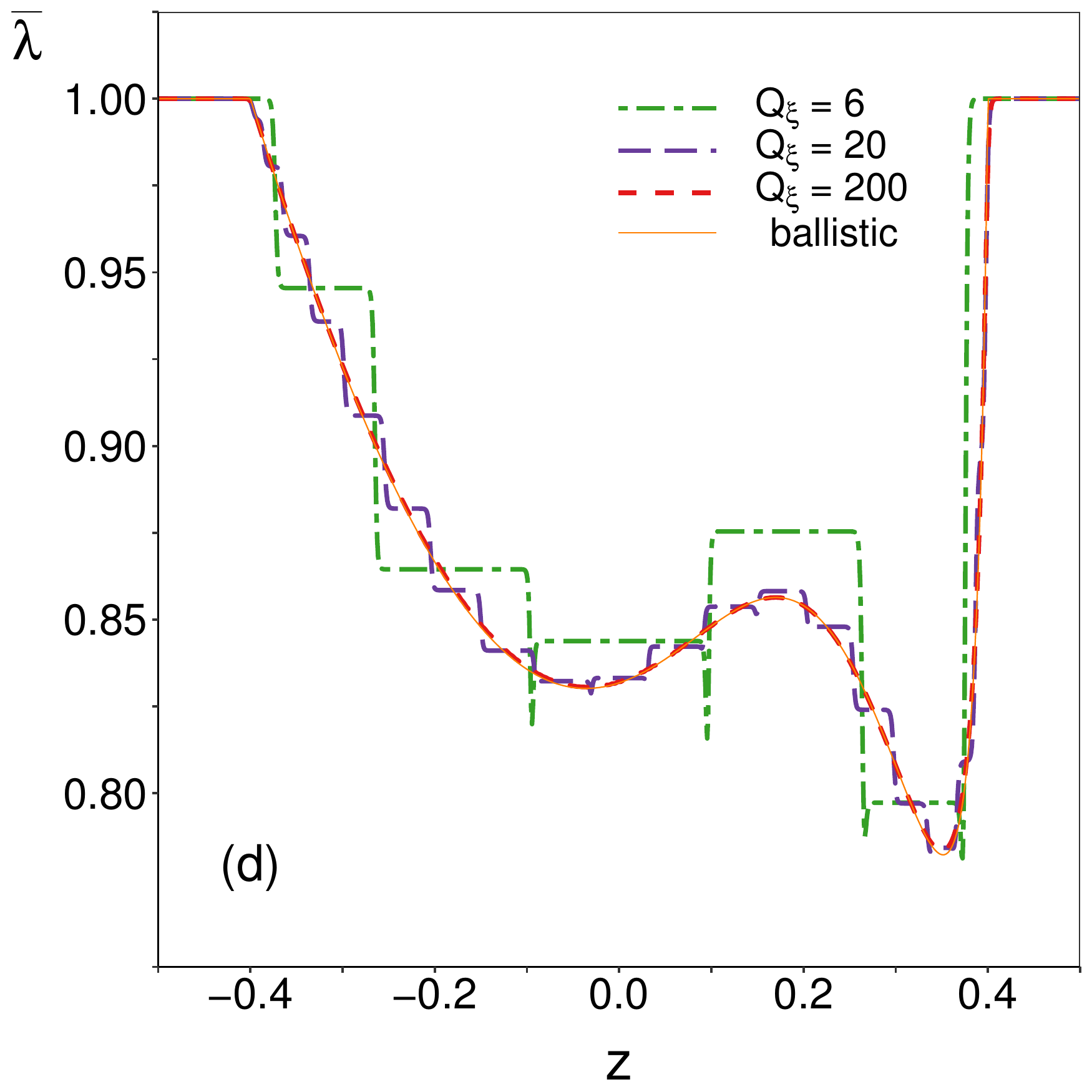} 
 \\
\end{tabular}
\begin{center}
 \begin{tabular}{c}
  \includegraphics[width=0.4\linewidth]{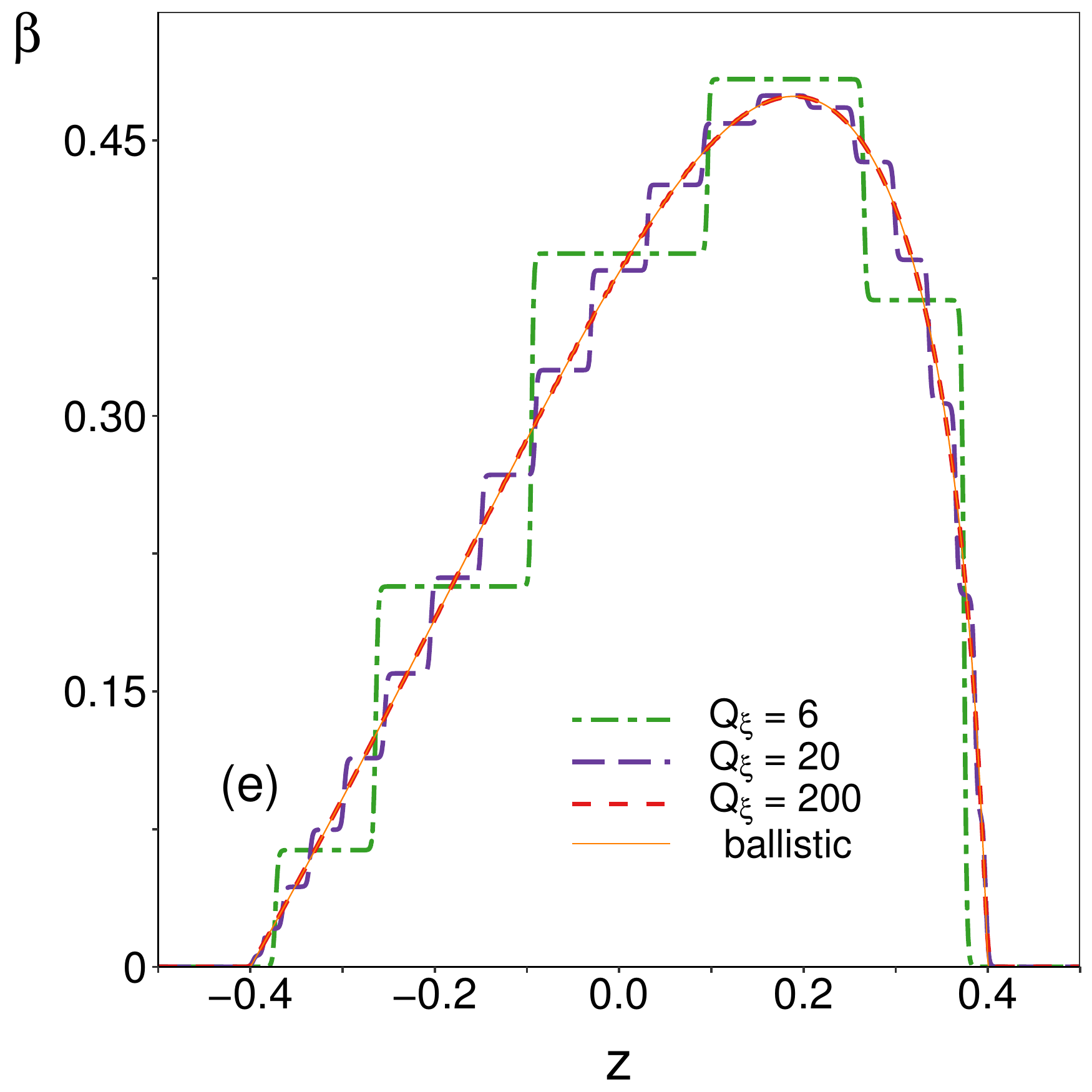} 
\end{tabular}
\end{center}
 \caption{Profiles of the density (a), relative fugacity (b), pressure (c), temperature (d),  and velocity (e), for different orders of the polar quadrature.
 The profiles corresponding to $Q_\xi = 200$ are overlapped with 
 the analytic solutions given in Sec.~\ref{sec:Sod:bal:analytic}. The profiles 
 represent snapshots at $t = 0.4$, obtained using a number of $Z = 1000$ nodes
 with a time step $\delta t = 5 \times 10^{-4}$. The initial state is described in 
 the caption of Fig.~\ref{fig:Rinviscid}.
}
\label{fig:ballistic}
\end{figure*}


\begin{figure}
  \begin{tabular}{cc}
 \includegraphics[width=0.82\linewidth]{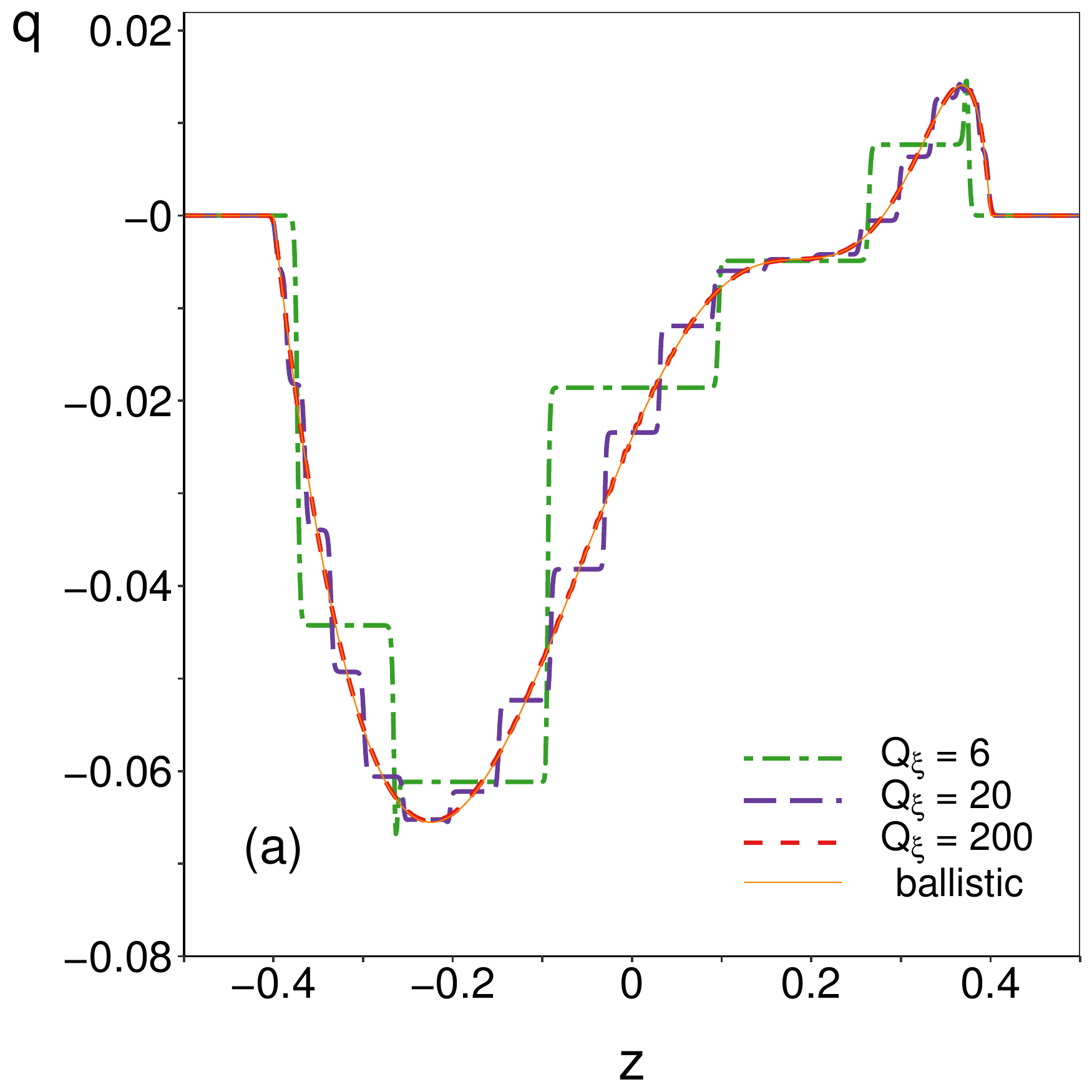} \\
 \includegraphics[width=0.82\linewidth]{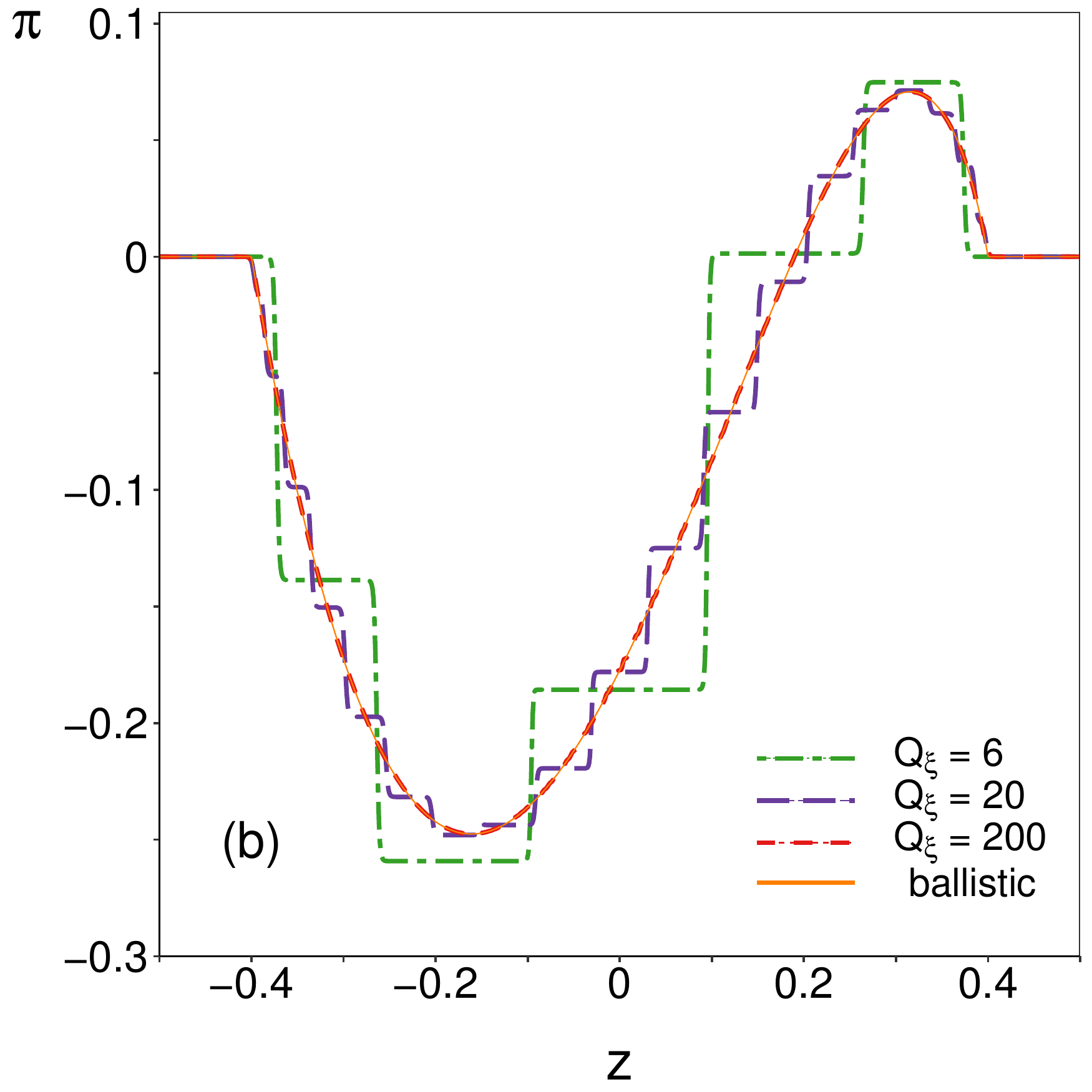} 
 \end{tabular}
  \caption{Ballistic regime profiles of the heatflux (a) 
  and pressure deviator (b), for different orders of the polar quadrature.
  The simulation parameters are described in the caption of Fig.\ref{fig:ballistic}.
  \label{fig:ballistic2}} 
\end{figure}

In the ballistic case, the collision term vanishes and the Boltzmann equation \eqref{eq:boltz_Sod} 
can be solved analytically (see, e.g., Refs.~\cite{greiner96} and \cite{guo15} 
for the analytic treatment of the ballistic regime of the 
sphericallysymmetric relativistic and 1D nonrelativistic cases of the Riemann problem).
For completeness, we present the free-streaming solution of the relativistic 
Sod shock tube problem in Sec.~\ref{sec:Sod:bal:analytic}. Our numerical results 
are discussed in Sec.~\ref{sec:Sod:bal:num}.

\subsubsection{Analytic solution}\label{sec:Sod:bal:analytic}

The collisionless version of the Boltzmann equation \eqref{eq:boltz_Sod} reduces to:
\begin{equation} \label{eq:ballistic}
\left(\partial_t + \xi \partial_z\right) f  = 0.
\end{equation}
At initial time, the flow is in thermal equilibrium characterized by the 
macroscopic fields given in Eq.~\eqref{eq:Sod:init}, such that
\begin{equation}
 f(z, \xi, t =0) = \theta(-z) \feq_{\rtin{L}} + \theta(z) \feq_{\rtin{R}},
 \label{eq:bal_ft0_theta}
\end{equation}
where 
\begin{equation}
 \feq_{\rtin{L}/\rtin{R}} = 
 \frac{n_{\rtin{L}/\rtin{R}}}{8\pi T^3_{\rtin{L}/\rtin{R}}} 
 \exp\left(-\frac{p}{T_{\rtin{L}/\rtin{R}}}\right).
\end{equation}
Equation~\eqref{eq:ballistic} is automatically satisfied if $f \equiv f( z - \xi t)$, which can be 
combined with the initial condition \eqref{eq:bal_ft0_theta} to yield:
\begin{equation}
 f(\zeta, \xi) = \theta(\xi - \zeta) \feq_{\rtin{L}} + 
 \theta(\zeta - \xi) \feq_{\rtin{R}},\label{eq:bal_f_theta_zeta}
\end{equation}
where $\zeta = z / t$ and $t > 0$ was assumed.
Noting that, due to causality, the regions $\zeta > 1$ and $\zeta < -1$ remain unperturbed, 
Eq.~\eqref{eq:bal_f_theta_zeta} can be put in the form
\begin{equation}
f(\zeta, \xi) = 
\begin{cases}
 \feq_{\rtin{L}}, & \zeta < -1, \\
 \theta(\xi - \zeta) \feq_{\rtin{L}} + \theta(\zeta - \xi) \feq_{\rtin{R}},  & |\zeta| < 1, \\
 \feq_{\rtin{R}}, & \zeta > 1.
\end{cases}
\end{equation}
In the two external regions $|\zeta| > 1$, the macroscopic fields 
are always at their initial equilibrium values. 
In the intermediate region, $N^\mu$ and $T^{\mu\nu}$
can be obtained through \eqref{eq:macro_def}.
Specifically, the nonvanishing components of $N^\mu$ are given by:
\begin{align}
N^{0}(\zeta)
=& \left(\frac{n_{\rtin{L}} + n_{\rtin{R}}}{2}\right) - \left(\frac{n_{\rtin{L}} - n_{\rtin{R}}}{2}\right)\zeta, \nn \\
N^{z}(\zeta)
=& \left(\frac{n_{\rtin{L}} - n_{\rtin{R}}}{4}\right)\left(1 - \zeta^2\right),
\label{eq:Sod:bal:Nmu}
\end{align}
while $N^x(t,z) =\ N^y(t,z) = 0$, as expected from Eq.~\eqref{eq:tmunu_comp}. Furthermore, 
the nonvanshing components of $T^{\mu\nu}$ are:
\begin{align}
T^{00} =& \frac{3}{2} (P_\rtin{L} + P_\rtin{R}) -  \frac{3}{2} (P_\rtin{L} - P_\rtin{R}) \zeta,   \nn \\
T^{0z} =& \frac{3}{4}(P_\rtin{L} - P_\rtin{R} ) \left(1 - \zeta^2 \right), \nn \\
T^{zz} =& \frac{1}{2}\left(P_\rtin{L} + P_\rtin{R}\right) - \frac{1}{2}\left(P_\rtin{L} - P_\rtin{R}\right)  \zeta^3,
\label{eq:Sod:bal:Tmunu}
\end{align}
while $T^{xx} = T^{yy} = \frac{1}{2}(T^{00} - T^{zz})$.
The energy density $E$, macroscopic velocity $u^\mu$, heat flux $q$, and shear pressure $\Pi$ can 
be computed using Eqs.~\eqref{eq:E_landau}--\eqref{eq:pi_val}. Because their analytic expressions
are cumbersome, we omit them here and mention that the profiles of these quantities can be represented
in a straightforward manner using the above solution. 

For completeness, we also include an analysis of the Eckart frame, in which the 
macroscopic velocity $u^\mu_e = \gamma_e(1, 0, 0, \beta_e)^T$ is defined 
as the unit vector which is parallel to the particle four-flow $N^\mu$:
\begin{equation}
 N^\mu = n_e u^\mu_e,\label{eq:eck_Nmu}
\end{equation}
While in the unperturbed regions, the macroscopic fields computed with respect to 
the Eckart and Landau frames coincide, in the perturbed regions they are generally different. 
The Eckart frame particle number density $n_e = -u^\mu_e N_\mu$ and velocity 
$\beta_e = u^z_e / u^t_e$ are given by:
\begin{align}
 n_{e} =&
 \left[\left(\frac{1+\zeta}{2} n_{\rtin{R}} + \frac{1-\zeta}{2} n_\rtin{L}\right)^2 \right.\nonumber\\
 & \left. - \left(\frac{n_\rtin{L} - n_\rtin{R}}{2}\right)^2\left(\frac{1-\zeta^2}{2}\right)^2\right]^{1/2}, \\
\beta_{e}  =& \frac{1}{2} 
    \frac{(n_\rtin{L} - n_\rtin{R})\left(1 - \zeta^2\right)}{\left(n_\rtin{L} + n_\rtin{R}\right) - 
    \left(n_\rtin{L}-n_\rtin{R}\right)\zeta}.
\end{align}
The other macroscopic quantities (energy, pressure, shear stress, and heat flux) can 
be obtained from Eqs.~\eqref{eq:Sod:bal:Tmunu}, as follows:
\begin{equation}
 E_e = u_e^\mu u_e^\nu T_{\mu\nu}, \quad  
 q_e^\mu = -\Delta^{\mu\nu}_{e} u_{e}^\lambda T_{\nu\lambda}, \quad 
 \Pi_e^{\mu\nu} = T^{\langle\mu\nu\rangle},
\end{equation}
while $P = \frac{1}{3} E$.
Since the final results are rather lengthy, 
their algebraic expressions are not reproduced here.
The maximum Eckart frame velocity $\beta_e^\rtin{max}$ 
of the flow occurs at $z = \beta_e^\rtin{max} t$ and 
is given by:
\begin{equation}
\beta_e^\rtin{max} = \frac{ 1  - \sqrt{n_\rtin{R} / n_\rtin{L}} }
{ 1 + \sqrt{n_\rtin{R} / n_\rtin{L}}}.
\end{equation}

\subsubsection{Numerical results}\label{sec:Sod:bal:num}

Figures~\ref{fig:ballistic} and \ref{fig:ballistic2} present our simulation results, obtained
using $Z = 1000$ nodes on the $z$ axis and a time step equal to 
$\delta t = 5 \times 10^{-4}$.
In choosing the polar quadrature order $Q_\xi$, we note that 
because the flow constituents stream freely, the populations corresponding to 
different values of $\xi_j$ ($1 \le j \le Q_\xi$) will travel along the $z$ 
axis with different velocities (see Figs.~\ref{fig:ballistic} and \ref{fig:ballistic2}). 
This gives rise to staircase-like profiles, with the number of steps being
equal to the quadrature order $Q_\xi$. We find that setting $Q_\xi = 200$ 
produces profiles which are well overlapped with the analytic solution 
presented in Sec.~\ref{sec:Sod:bal:analytic}. The effect of lowering 
$Q_\xi$ on the resulting profiles will be discussed in Sec.~\ref{sec:Sod:conv:Qxi}.
The distribution function was initialized using the expansion of $\feq$ truncated 
at $N_\rtin{L} = 1$ and $N_\Omega = 5$, while the radial quadrature 
was always kept at $Q_{\rtin{L}} = 2$.

\subsection{Viscous regime}\label{sec:Sod:BAMPS}

\begin{figure*}
\begin{tabular}{cc}
\includegraphics[width=0.4\linewidth]{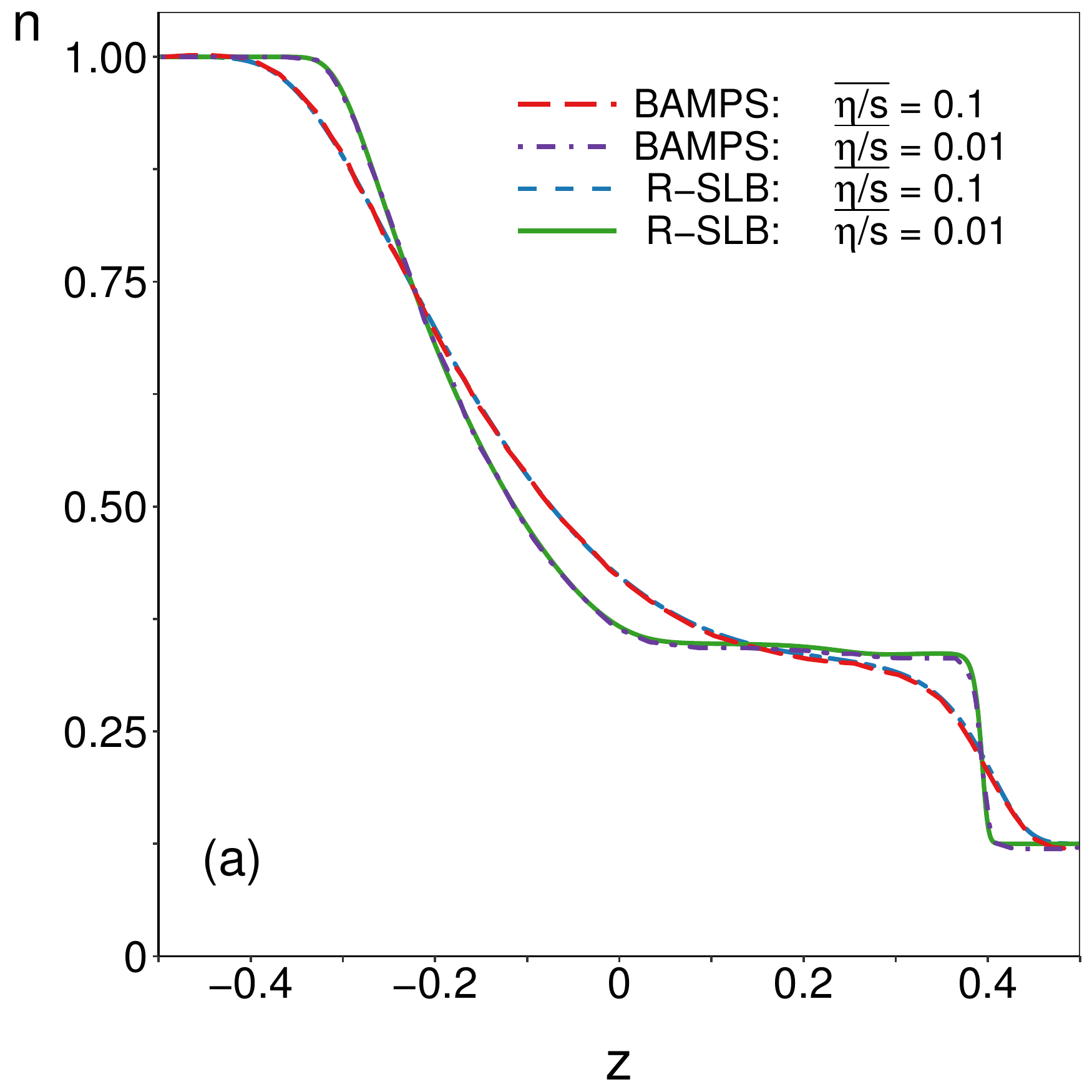} &
\includegraphics[width=0.4\linewidth]{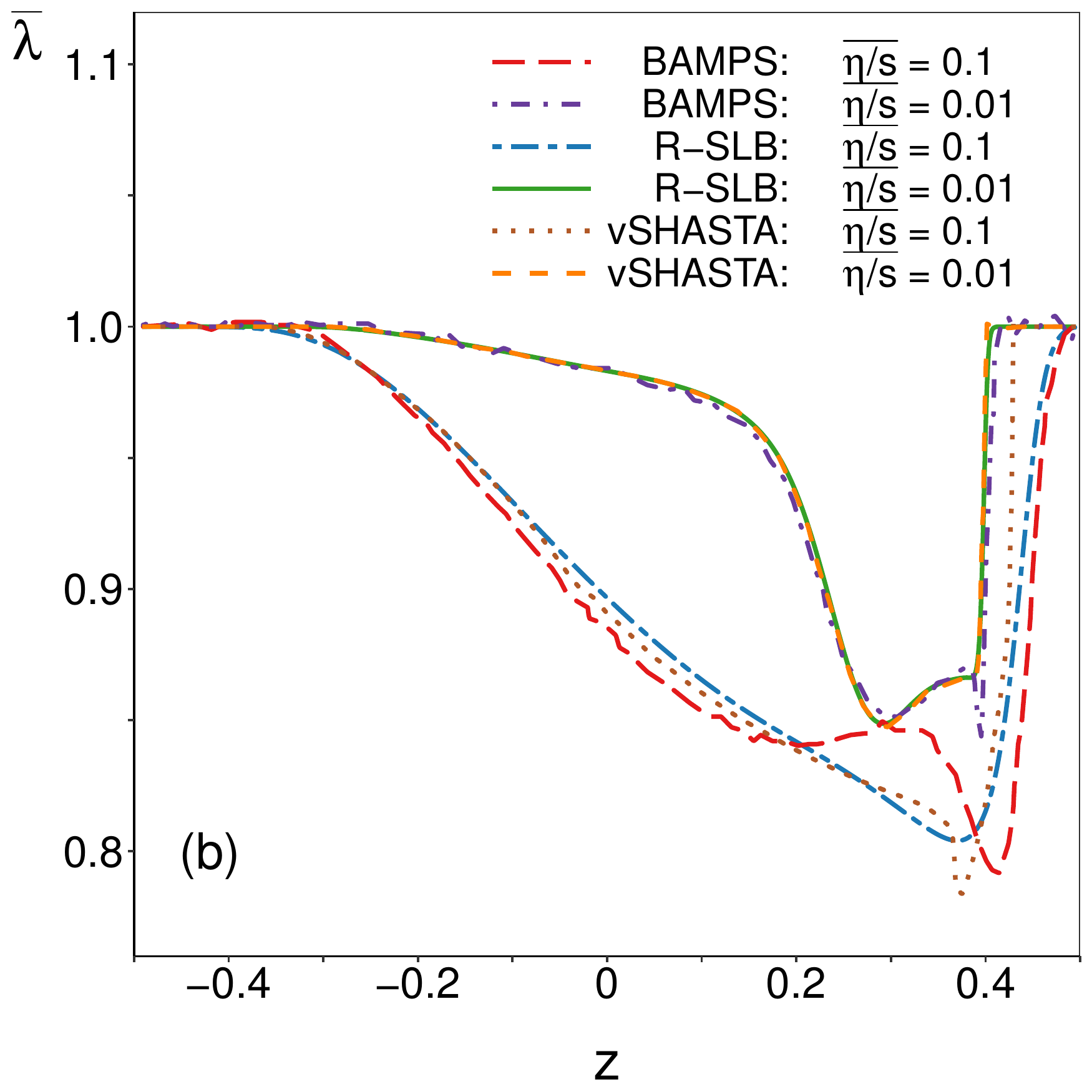} 
\\
\includegraphics[width=0.4\linewidth]{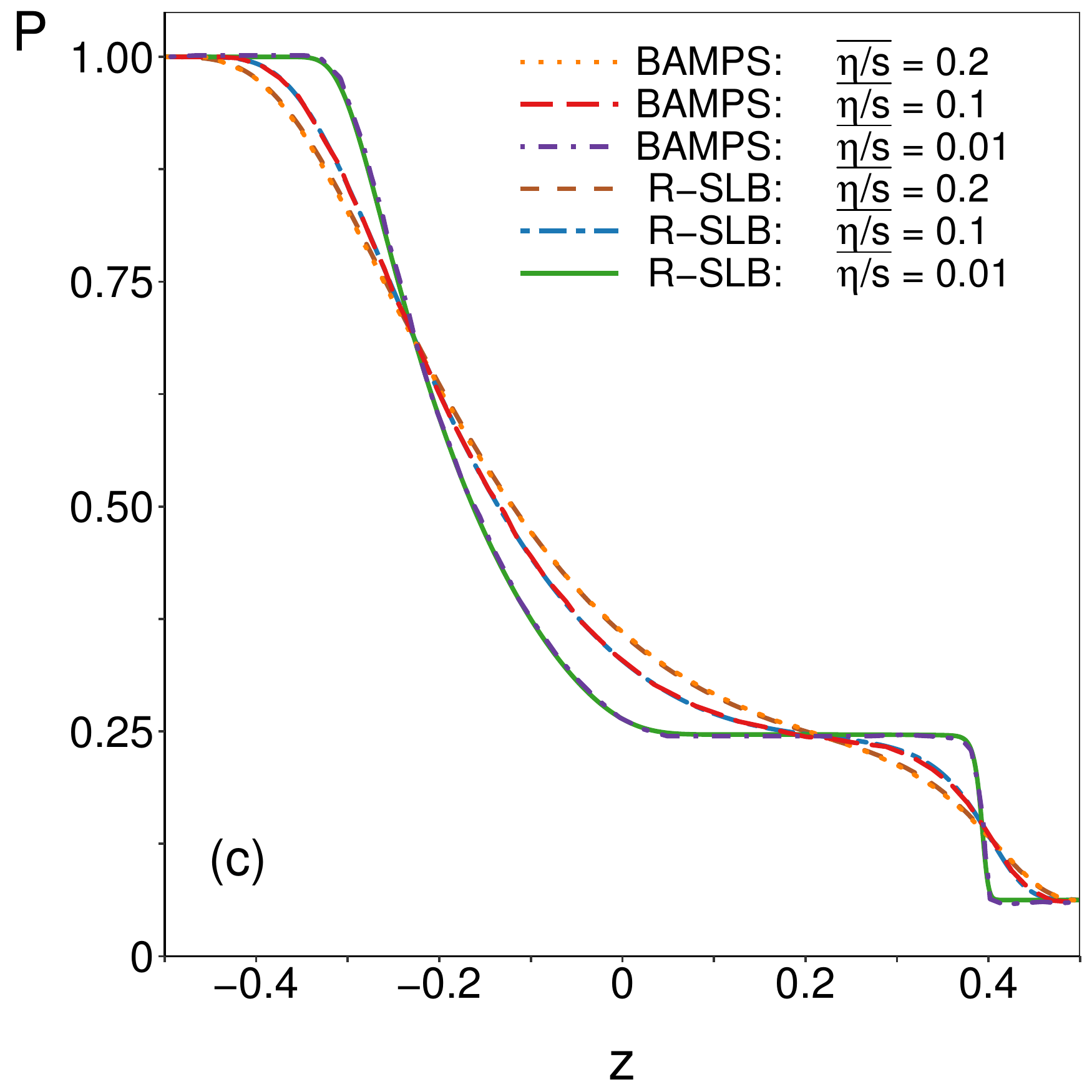} &
\includegraphics[width=0.4\linewidth]{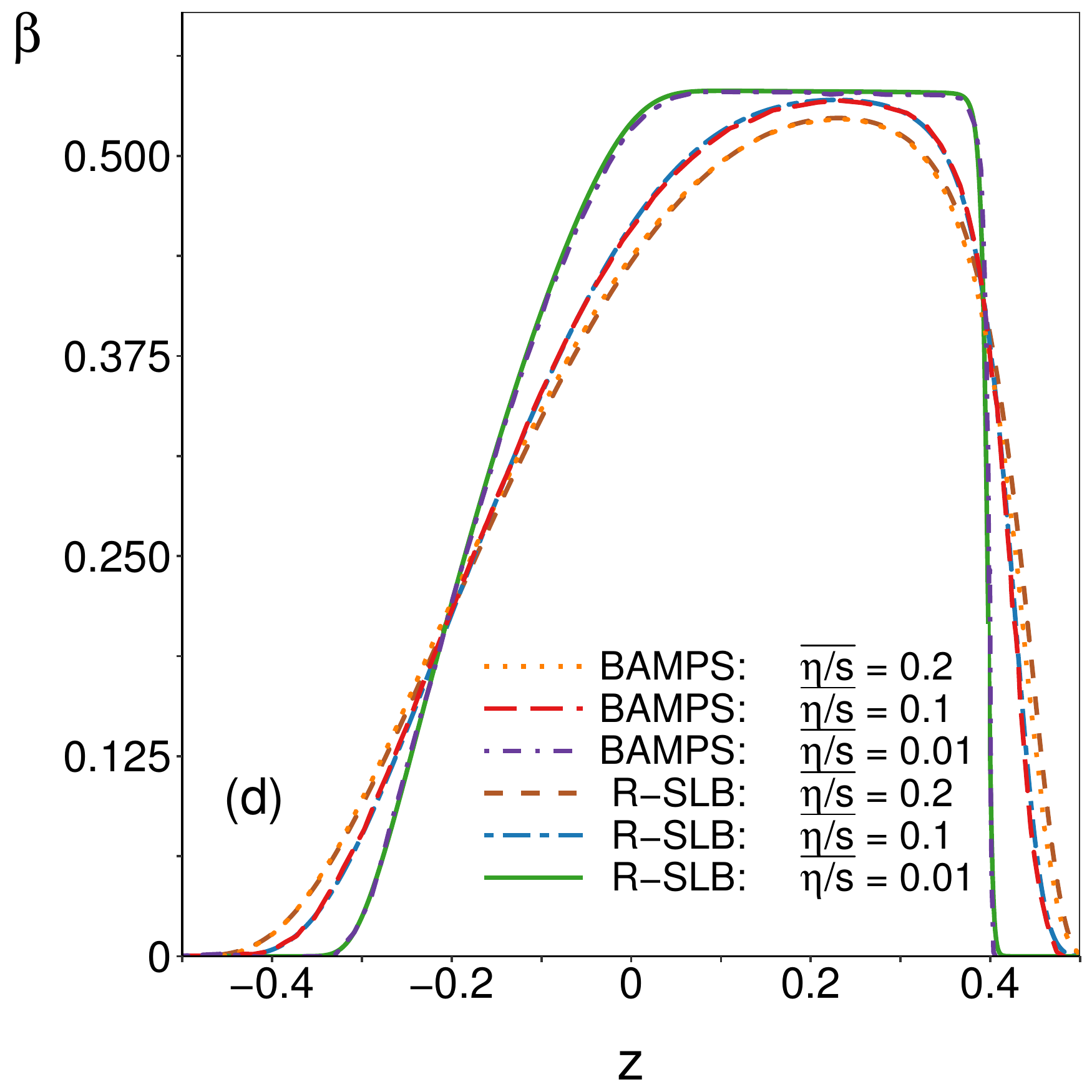} 
\\
\includegraphics[width=0.4\linewidth]{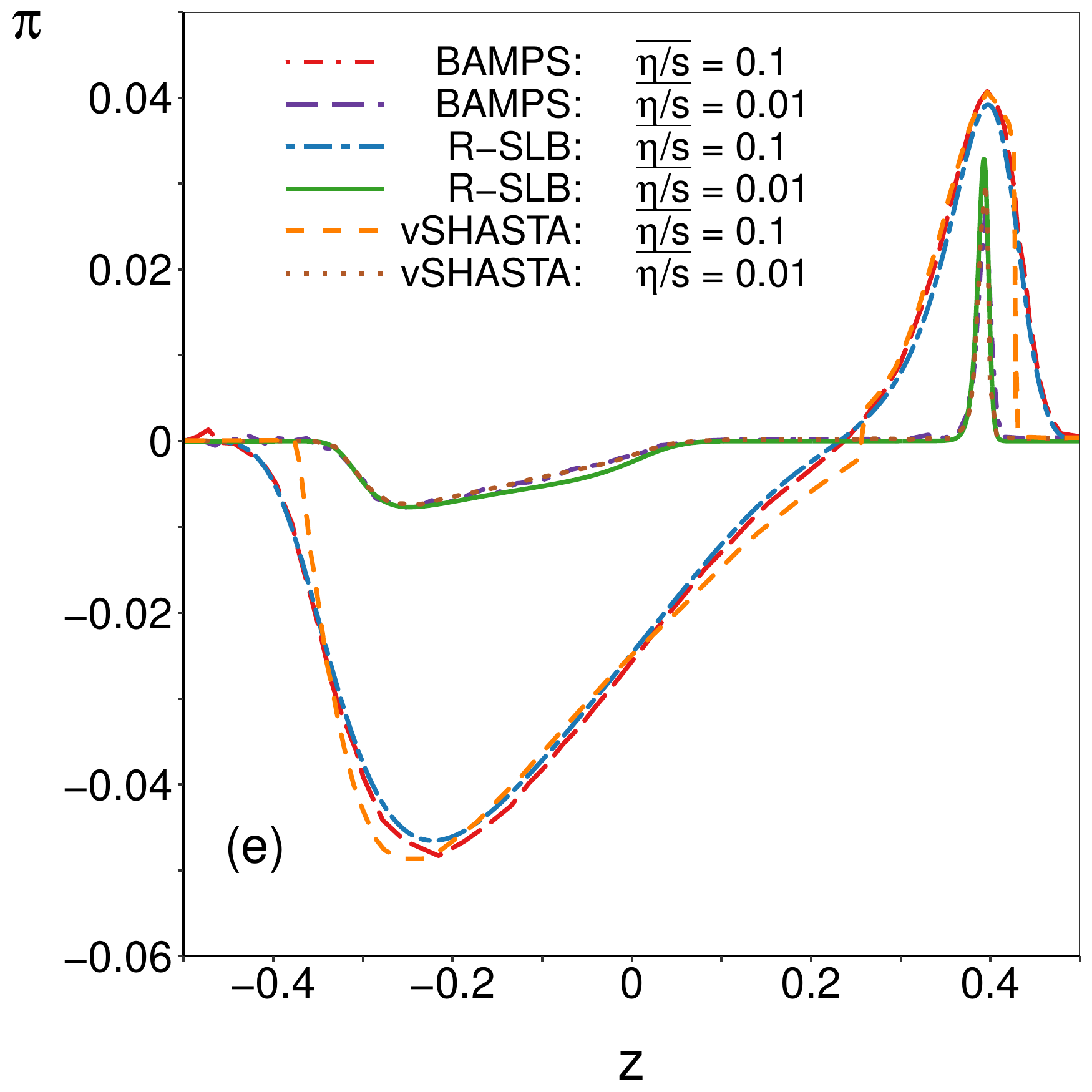} &
\includegraphics[width=0.4\linewidth]{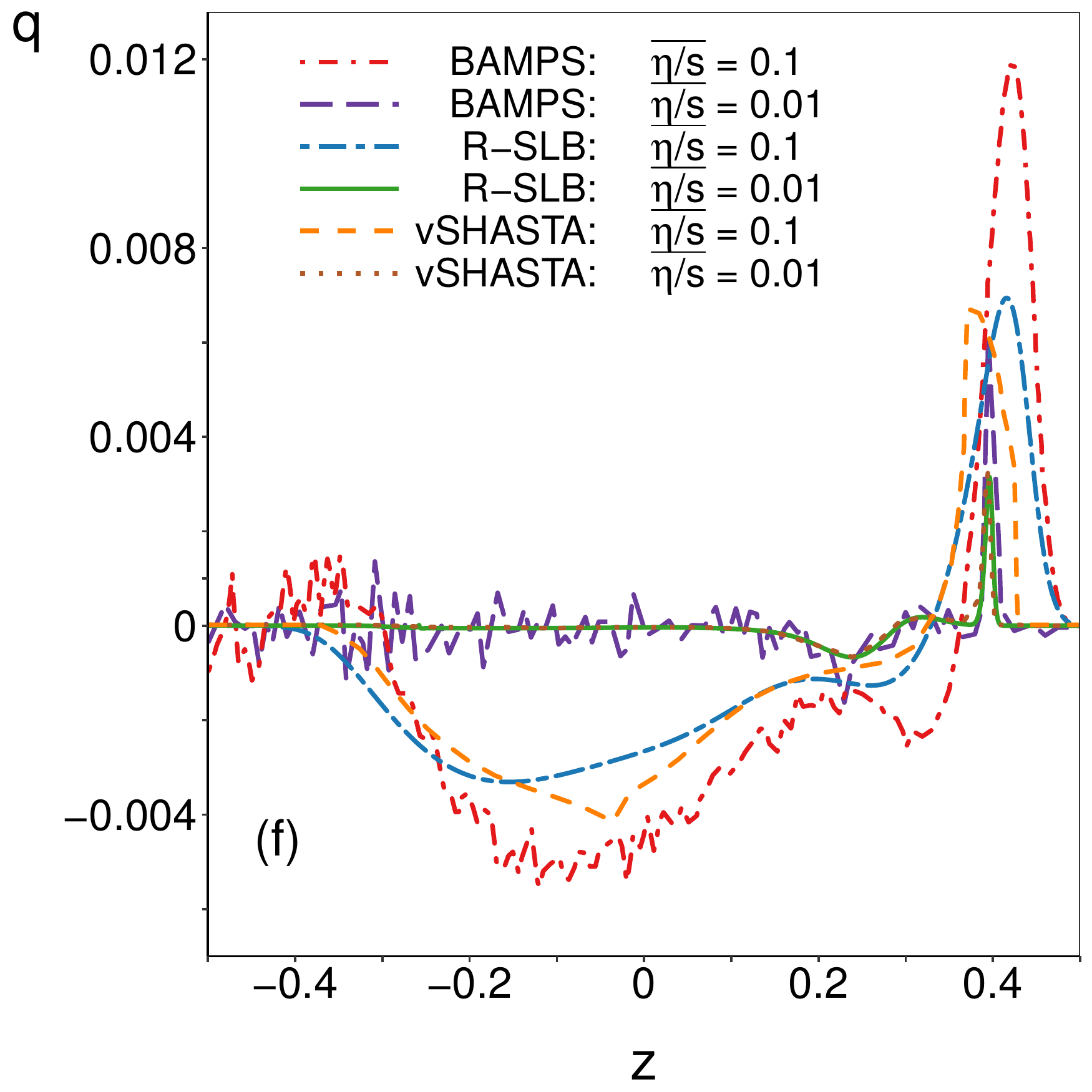} 
\end{tabular}
\caption{Comparison between the profiles obtained with the LB model with $Q_\xi = 500$ 
and $N_\Omega = 5$ and the results obtained using the BAMPS and vSHASTA methods, 
which were reported in Refs.~\cite{bouras10,bouras09nucl}, for various values of 
$\etas$. (a) Density $n$; (b) relative fugacity $\overline{\lambda}$; 
(c) Pressure $P$; (d) velocity $\beta = u^z / u^0$; (e) pressure deviator $\Pi$; 
(f) heat flux $q$. In (b), (e) and (f), the vSHASTA results are also included.
}
\label{fig:bamps1}
\end{figure*}

\begin{figure}
\centering
\includegraphics[width=0.9\linewidth]{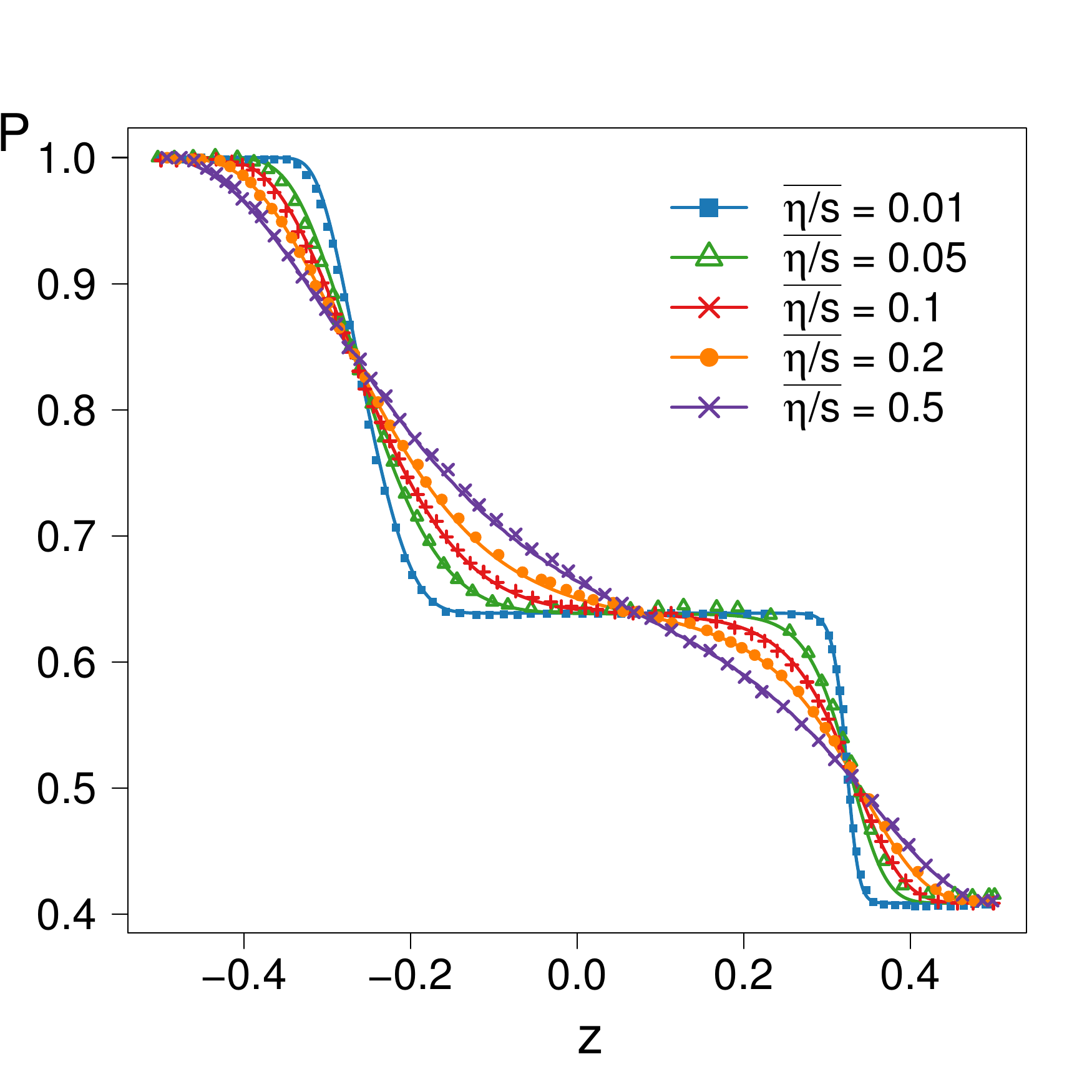}
\caption{Comparison between the profiles of the pressure $P$, obtained with the LB model 
with $Q_\xi = 500$ and $N_\Omega = 5$ (lines), and the results obtained using the BAMPS method
reported in Ref.~\cite{bouras09prl} (points), for various values of 
$\etas$.
}
\label{fig:bamps2}
\end{figure}

In this section, we discuss the validation of our models in the regime where 
dissipation becomes important.
We compare our simulation results with those obtained in
Refs.~\cite{bouras10,bouras09prl,bouras09nucl} using two methods:
the Boltzmann approach of multiparton scattering (BAMPS) model and 
the viscous sharp and smooth transport algorithm (vSHASTA).
For the simulations presented in this section, we used the ``reference model'' 
having quadrature orders $Q_\rtin{L} = 2$, $Q_\varphi = 1$, and $Q_\xi = 500$.
The expansion of $\feq$ was truncated at $N_\rtin{L} = 1$ and $N_\Omega = 5$.

In Fig.~\ref{fig:bamps1}, the profiles of $n$, $\overline{\lambda}$, $P$, $\beta$,
$\Pi$ and $q$
are represented for the initial conditions $(n_\rtin{L}, T_\rtin{L}, P_\rtin{L}) = (1,1,1)$ and 
$(n_\rtin{R}, T_\rtin{R}, P_\rtin{R}) = (0.125, 0.5, 0.0625)$, corresponding to the 
values used in Refs.~\cite{bouras10,hupp11,mohseni13,bouras09nucl}. Very good agreement 
between the results of our simulations and BAMPS is observed for the profiles of $n$, $P$ 
and $\beta$ when $\etas \in \{0.01, 0.1, 0.2\}$, where the connection between 
$\etas$, and the relaxation time $\tau_{\rtin{A-W}}$ is given in Eq.~\eqref{eq:Sod:tau1}.

Furthermore, since the relaxation time of the 
Anderson-Witting model is chosen such that $\etas$ matches the BAMPS value, 
we note that the shear stress $\Pi$ is also in good agreement with both the vSHASTA \cite{bouras10,bouras09nucl} 
and BAMPS data at $\etas = 0.01$. At $\etas = 0.1$, our simulation results for $\Pi$ are close to 
the BAMPS results, while the discrepancy with the vSHASTA data seems to indicate that the hydrodynamic 
description loses its validity. The situation is somewhat reversed 
for the relative fugacity $\overline{\lambda}$ \eqref{eq:fugacity} and the heat flux $q$. At 
$\etas = 0.01$, the agreement between our results and the vSHASTA data is very good, while the BAMPS data 
exhibits a spike in $\overline{\lambda}$ near the shock front. Since in the hydrodynamic limit, 
the heat flux is proportional to the derivatives of $\overline{\lambda}$ [see Eq.~\eqref{eq:hydro_q}],
the spike in the profile of $\overline{\lambda}$ induces a larger peak value of $q$ near the shock front.
Even though the vSHASTA method seems to be inaccurate at $\etas = 0.1$, our results are still closer to the 
vSHASTA data as compared to the BAMPS data.

The agreement between our simulation and the vSHASTA results is not surprising. 
According to Eq.~(53) in Ref.~\cite{bouras10}, the vSHASTA algorithm 
implements the heat conductivity $\lambda_\rtin{heat}$ (denoted $\kappa_q$ in Ref.~\cite{bouras10})
such that $\eta / \lambda_\rtin{heat} = 3 T / 5$, which is the same as that corresponding to 
the Chapman-Enskog expansion \eqref{eq:tcoeff_ce}. Since the value of $\etas$ is fixed in both 
our own and in the vSHASTA simulations to match the value of $\etas$ employed in the BAMPS simulations,
it follows that the value of $\lambda_\rtin{heat}$ arising in our simulations corresponds to that 
employed by vSHASTA. 

We attribute the strong fluctuations observed in the BAMPS result to the 
sensitivity of $\overline{\lambda} = n^4 / P^3$ to errors in $n$ or $P$, which are 
prone to be present in any stochastic numerical method. Indeed, allowing $n$ and 
$P$ to fluctuate by some small quantities $\delta n$ and $\delta P$, it can be seen that the 
fluctuation in $\overline{\lambda}$ is
\begin{equation}
 \frac{\delta \overline{\lambda}}{\overline{\lambda}} = \frac{4\delta n}{n} - \frac{3\delta P}{P}.
\end{equation}
In the inviscid regime $P_\rtin{C} \simeq 0.247$ around the contact discontinuity and 
$P_\rtin{R} = 0.0625$. This indicates that the fluctuation $\delta P$ in $P$ can be amplified 
by a factor between $12$ and $48$. 

In Fig.~\ref{fig:bamps2}, the pressure profile is compared with the BAMPS results 
reported in Ref.~\cite{bouras09prl} for higher values of $\etas$,
for the following reference values \cite{bouras09prl,mendoza10prl}:
\begin{gather}
 \widetilde{k}_B \widetilde{T}_{\rtin{ref}} = 0.35\ {\rm GeV}, \qquad
 \widetilde{P}_{\rtin{ref}} = 5.43\ {\rm GeV}/{\rm fm}^3, \nn\\
 \widetilde{n}_{\rtin{ref}} = 15.514\ {\rm fm}^{-3},
 \label{eq:Sod:left2}
\end{gather}
while the reference length is $\widetilde{L}_{\rtin{ref}} = 6.4\ {\rm fm}$.
With the above quantities, $\lambda_{\rtin{ref}} \simeq 1.715$ and
\begin{align}
 \tau_{\rtin{A-W}} \simeq& \frac{\tau_{\rtin{A-W};0}}{T}\left(0.865 - \frac{\ln \overline{\lambda}}{4} \right), 
 \nonumber\\
 \tau_{\rtin{A-W}; 0} \simeq& 0.4405 \etas.
 \label{eq:Sod:tau2}
\end{align}

In the right half of the domain, the system is initialized according to: 
\begin{gather}
 \widetilde{k}_B \widetilde{T}_{\rtin{R}} = 0.35\ {\rm GeV}, \qquad
 \widetilde{P}_{\rtin{R}} = 2.22\ {\rm GeV}/{\rm fm}^3, \nn\\
 \widetilde{n}_{\rtin{R}} = 6.34\ {\rm fm}^{-3},
 \label{eq:right2}
\end{gather}
which correspond to:
\begin{equation}
 T_{\rtin R} = 1, \qquad
 P_{\rtin R} = 0.409, \qquad
 n_{\rtin R} = 0.409.\label{eq:Sod:right2_adim}
\end{equation}
With the above values, the relative fugacity \eqref{eq:fugacity} takes the following values:
\begin{equation}
 \overline{\lambda}_{\rtin L} = 1, \qquad \overline{\lambda}_{\rtin R} = 0.409.\label{eq:lambda_LR2}
\end{equation}
The relaxation times in the two halves of the channel are:
\begin{equation}
 \tau_{\rtin{A-W}; \rtin L} \simeq 0.3811 \etas, \qquad
 \tau_{\rtin{A-W}; \rtin R} \simeq 0.4795 \etas.\label{eq:tau_LR2}
\end{equation}
The simulation conditions in 
this case are less challenging, but this choice of parameters allows us to validate our 
results for the pressure profile up to $\etas = 0.5$.

\subsection{Convergence test} \label{sec:Sod:conv}

To determine the adequate order for the quadratures at a given relaxation time, we 
follow Refs.~\cite{ambrus16jcp,ambrus16jocs} and employ a convergence test, which is 
described below.
In the inviscid and ballistic regimes, the profiles recovered with our LB 
models show a satisfactory convergence trend towards the analytic solutions. 
At a finite relaxation time however, as there 
is no analytic profile to which to compare, we can only check whether the procedure converges 
as we increase the order of the quadrature $Q_\xi$, as well as the truncation order $N_\Omega$ 
of $\feq$. To this end, we chose the model with $Q_\xi = 500$ and $N_\Omega = 6$ to be the 
reference model with which the reference profiles are obtained.
In order to test if a given model has achieved convergence, the following error measure is introduced:
\begin{equation}
\epsilon = \max_j (\epsilon_j), \qquad \epsilon_j =\left\vert \frac{ \max_z 
\left[ d_j(z) - d_{j,\rtin{\,ref}}(z) \right] }
{ \max_z [d_j(z)]  - \min_z [d_j(z)]}\right|,
\label{eq:eps_def}
\end{equation}
where the error is computed only for the macroscopic profiles 
$d_j \in \left\{ n, T, \gamma \right\}$ at time $t = 0.5$. The procedure selects, 
for a given quantity, the largest error in a single point on the whole spatial domain, for all 
tested profiles. We say that a model with a given $Q_\xi$ and $N_\Omega$ has 
achieved convergence when
\begin{equation}
 \epsilon < \epsilon_{\rtin{th}}
 \label{eq:eps_th}
\end{equation}
for a given threshold, which we set unless otherwise stated to
$\epsilon_{\rtin{th}} = 1\%$. In Sec.~\ref{sec:Sod:conv:NO}, we employ the above test to 
determine the dependence of $N_\Omega^{\rtin{conv}}$ required to achieve convergence 
for a wide range of $\etas$, while in Sec.~\ref{sec:Sod:conv:Qxi}, we determine the 
quadrature order $Q_\xi^{\rtin{conv}}$ required to achieve convergence 
at various values of $\epsilon_{\rtin{th}}$ from the inviscid regime to the ballistic regime.
These convergence tests are performed using the two initial states 
\eqref{eq:Sod:right_adim} and \eqref{eq:Sod:right2_adim}. All simulations 
are performed using $Z = 1000$ points and $\delta t = 5 \times 10^{-4}$.

\subsubsection{Convergence with respect to the expansion order of $\feq$}\label{sec:Sod:conv:NO}

\begin{figure}
\begin{tabular}{c}
\includegraphics[width=0.86\linewidth]{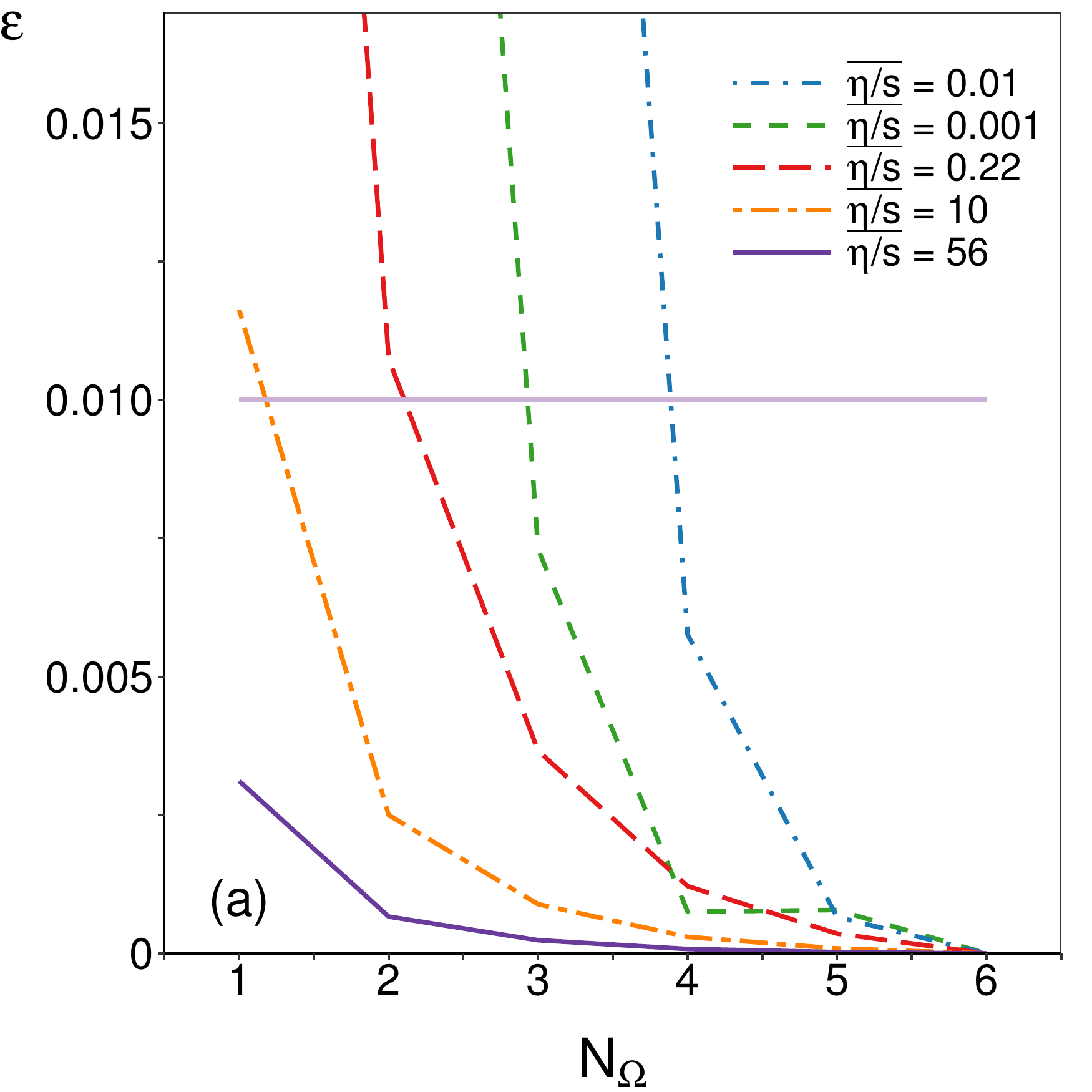} \ \ \\
\includegraphics[width=0.82\linewidth]{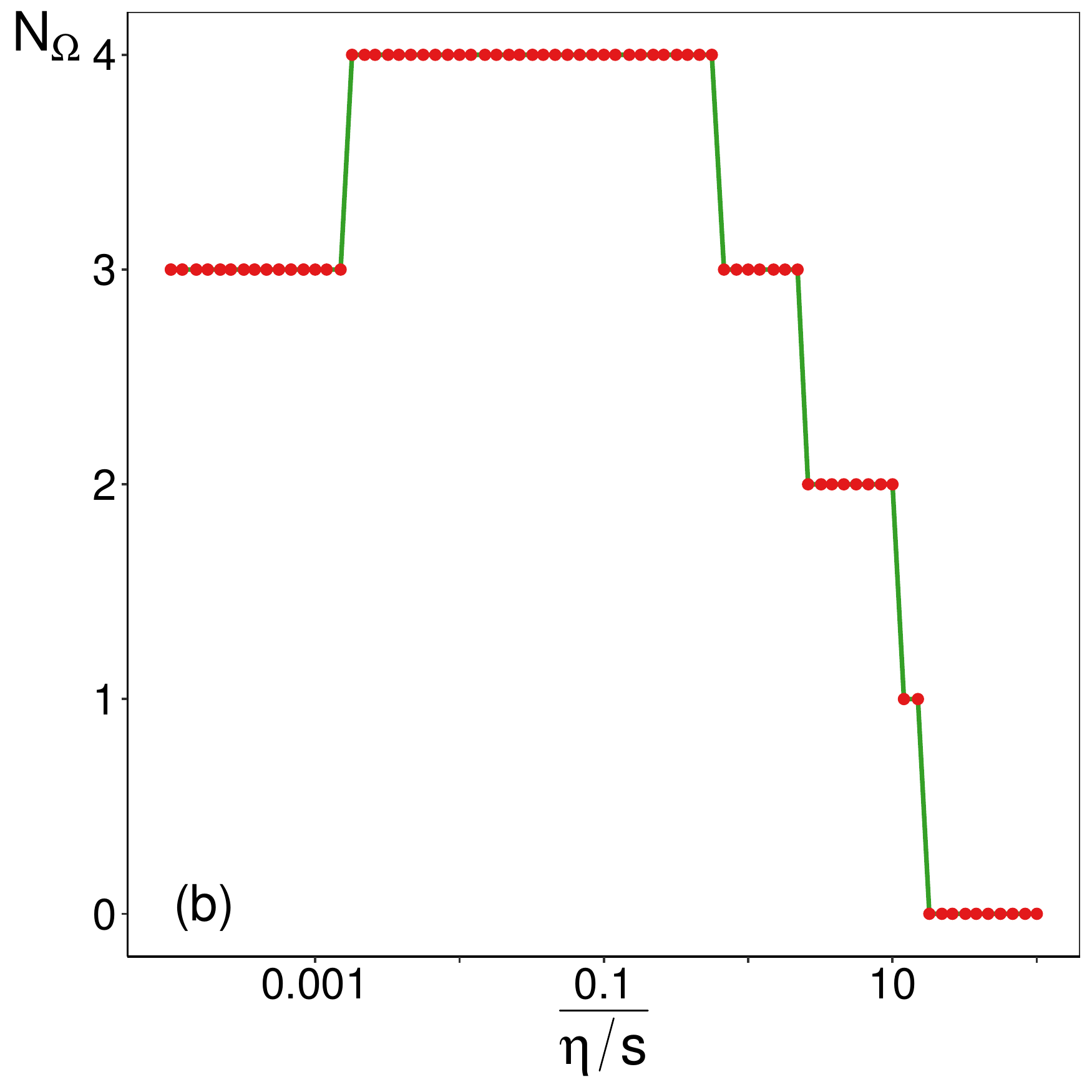} 
\end{tabular}
\caption{(a) The error $\epsilon$ \eqref{eq:eps_def} as a function of 
$N_\Omega$ for various values of $\etas$ ($Q_\xi = 500$ is kept constant). 
The reference profiles considered here were obtained using $Q_\xi = 500$ and $N_\Omega = 6$.
(b) The minimum expansion order $N_\Omega^{\rtin{conv}}$ required to reduce 
the error \eqref{eq:eps_def} of a model with $Q_\xi = 500$ below the threshold 
$\epsilon_{\rtin{th}} = 1\%$.}
\label{fig:NOconv}
\end{figure}

As mentioned in Sec.~\ref{sec:LB:feq}, the order $N_\Omega$ at which $\feq$ should be expanded with 
respect to the angular coordinates (in the case of the shock tube problem considered here,
the expansion involves only $\xi = \cos\theta$) cannot be determined {\it a priori}. 
The expansion coefficients required to construct $\feq$ up to $N_\Omega = 6$ are 
given in Appendix~\ref{app:feq}. All simulations presented in this subsection were performed
using the initial state considered in Fig.~\ref{fig:bamps1}.

Taking as a reference model the model with $Q_\xi = 500$ 
and $N_\Omega = 6$, Fig.~\ref{fig:NOconv}(a) shows the dependence of $\epsilon$ \eqref{eq:eps_def} 
on $N_\Omega$ at $Q_\xi = 500$ for various values of $\etas$. It can be seen that, for a fixed 
value of $\etas$, $\epsilon$ decreases as $N_\Omega$ is increased. Since the reference profiles 
correspond to $N_\Omega = 6$, the value of $\epsilon$ at $N_\Omega = 6$ is always $0$.

In Fig.~\ref{fig:NOconv}(b), the dependence with respect to $\etas$ 
of the order $N_\Omega^{\rtin{conv}}$ required to bring $\epsilon$ below the threshold 
$\epsilon_{\rtin{th}} = 1\%$ is shown. It can be seen that, for $\etas \lesssim 10^{-3}$,
$N_\Omega^{\rtin{conv}} = 3$. For $10^{-3} \lesssim \etas \lesssim 1$,
$N_\Omega^{\rtin{conv}}$ increases to $4$. At larger values of $\etas$, the flow enters the 
transition regime, where the effect of collisions becomes progressively small. The 
value of $N_\Omega^{\rtin{conv}}$ decreases with $\etas$ down to $N_\Omega^{\rtin{conv}} = 0$
as the free-streaming regime settles in at $\etas \gtrsim 18$. 

According to the analysis presented in this subsection, setting $N_\Omega = 4$ is sufficient to 
obtain results which are within $1\%$ error with respect to the reference profiles.
Before ending this section, it is important to note that the conclusions presented in this subsection 
hold for the flow parameters in Eq.~\eqref{eq:Sod:right_adim}. 
If the initial conditions are changed such that 
the flow develops a larger maximum velocity, the value of $N_\Omega^{\rtin{conv}}$ may 
have larger values. A similar analysis should be performed whenever the 
flow parameters are changed, in order to accurately determine the value of $N_\Omega$ 
necessary to satisfy the convergence test \eqref{eq:eps_th}.

\subsubsection{Convergence with respect to the quadrature order $Q_\xi$}\label{sec:Sod:conv:Qxi}


\begin{figure}
\begin{tabular}{c}
\includegraphics[width=0.82\linewidth]{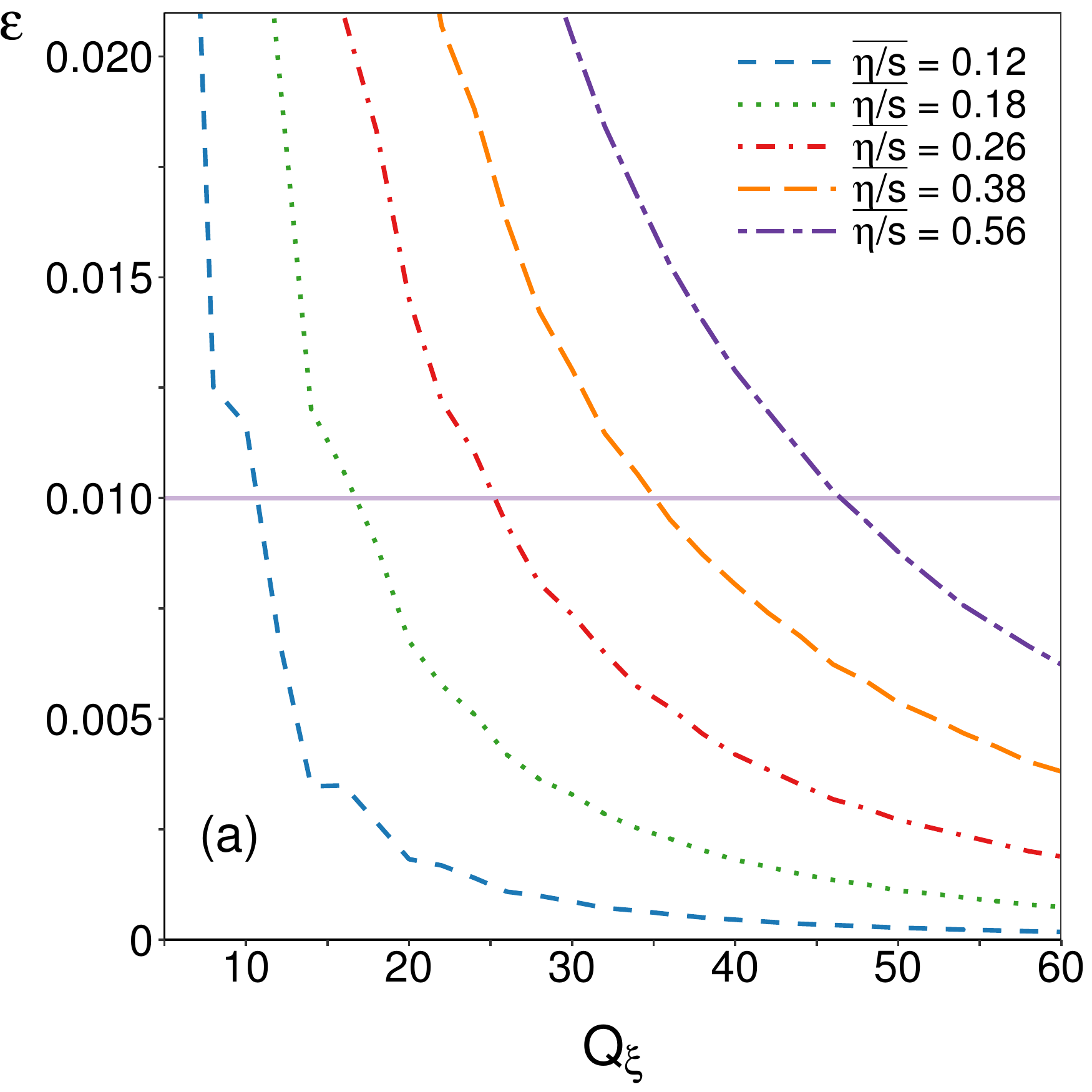} \\
\includegraphics[width=0.89\linewidth]{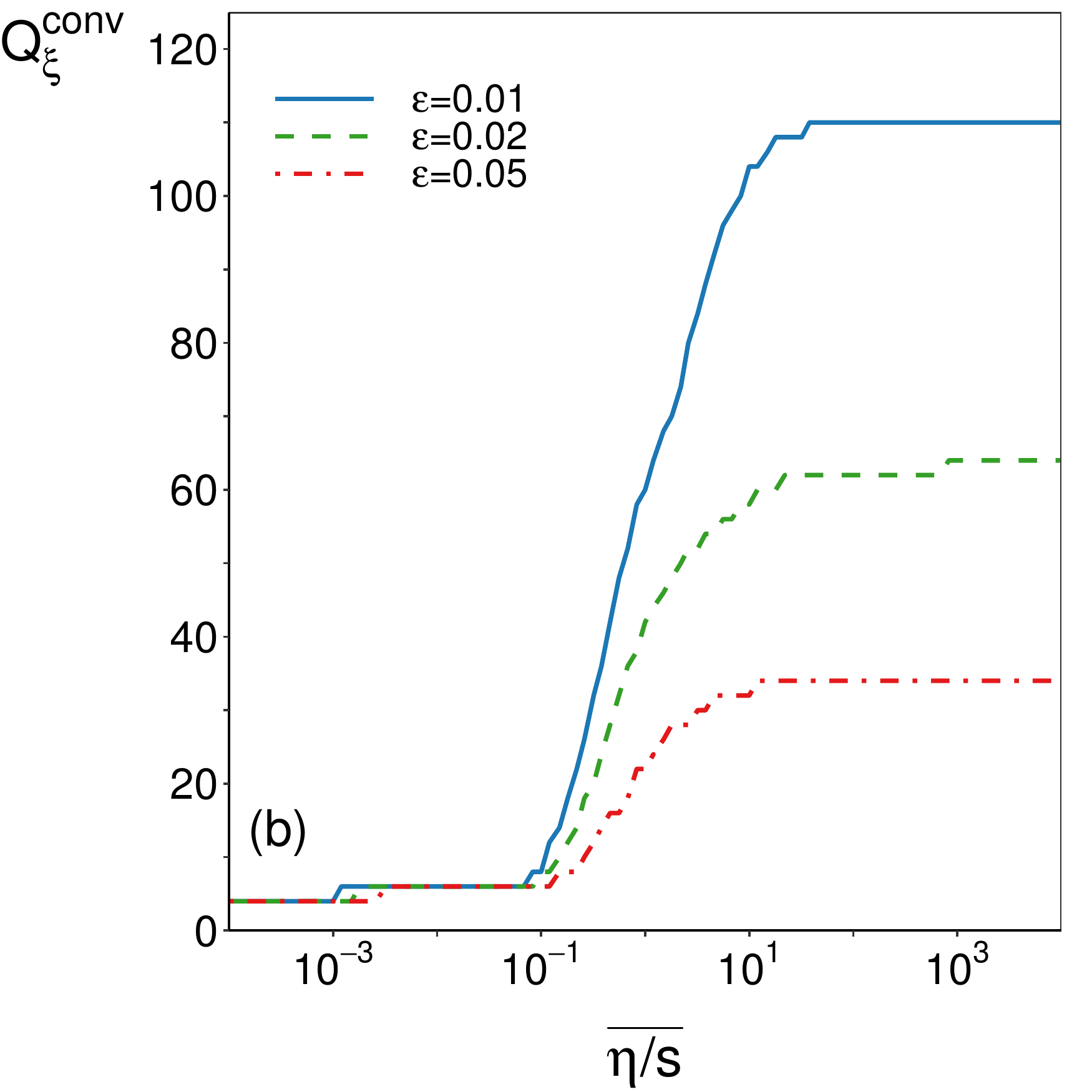} \ \ \ \ \ 
\end{tabular}
\caption{
(a) The error $\epsilon$ \eqref{eq:eps_def} 
as a function of $Q_\xi$ for various values of $\etas$. 
The reference profiles considered here were obtained using $Q_\xi = 500$ and $N_\Omega = 5$.
(b) The minimum quadrature order $Q_\xi^{\rtin{conv}}$ required to reduce 
the error \eqref{eq:eps_def} of a model with $N_\Omega = {\rm min}(Q_\xi - 1, 5)$ 
below various values of the threshold $\epsilon_{\rtin{th}}$.
The initial state for this analysis is given in Eq.~\eqref{eq:Sod:right_adim}.
}
\label{fig:Qxiconv}
\end{figure}

\begin{figure}
\begin{tabular}{c}
\includegraphics[width=0.89\linewidth]{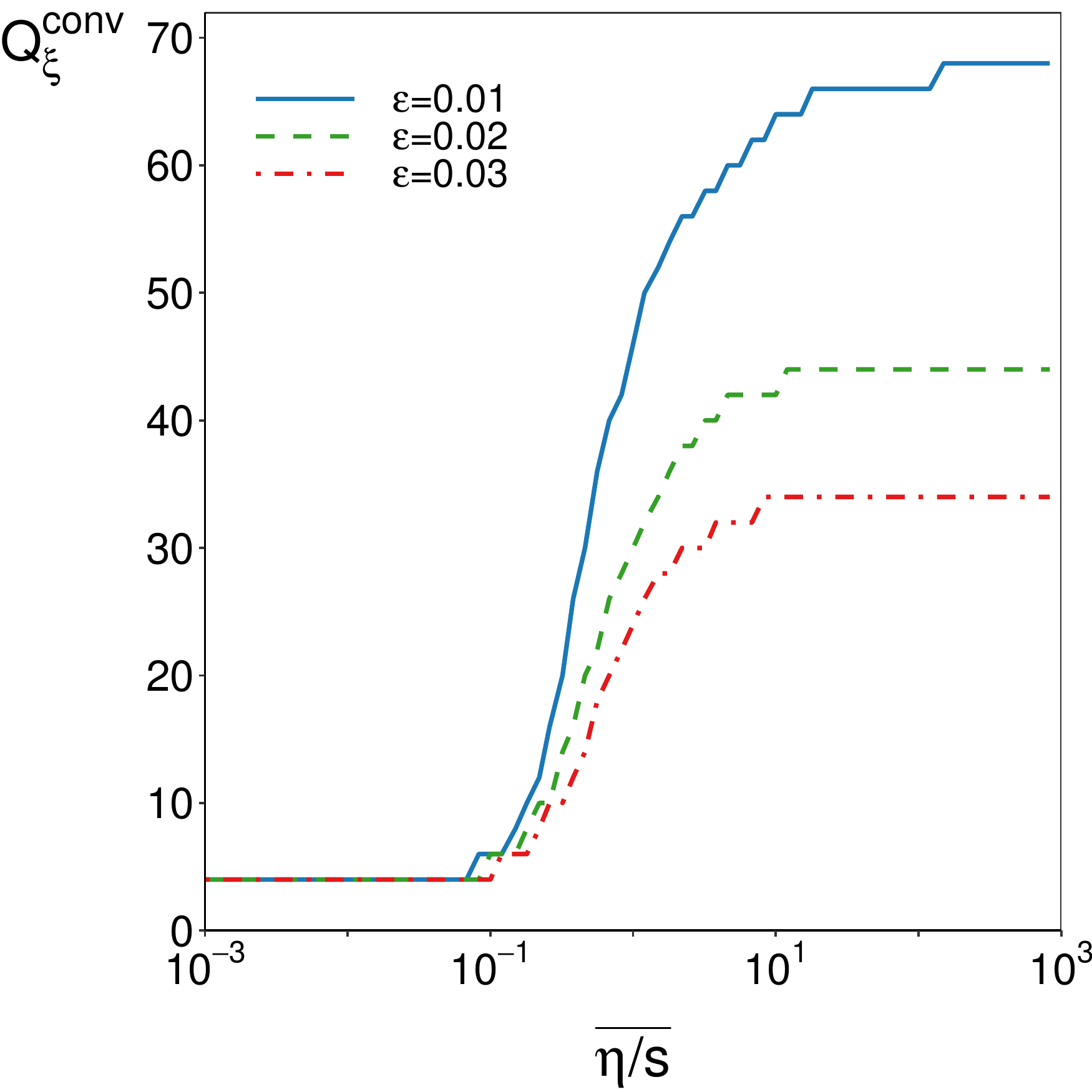} \ \ \ \ \ 
\end{tabular}
\caption{
The minimum quadrature order $Q_\xi^{\rtin{conv}}$ required to reduce 
the error \eqref{eq:eps_def} of a model with $N_\Omega = {\rm min}(Q_\xi - 1, 5)$ 
below various values of the threshold $\epsilon_{\rtin{th}}$ for the initial 
state given in Eq.~\eqref{eq:Sod:right2_adim}.
}
\label{fig:Qxiconv_etas05}
\end{figure}

\begin{table}
\begin{ruledtabular}
\begin{tabular}{l|l|r|r}
 Regime & Range of $\etas$ & $N_\rtin{vel} $ & $N_\Omega$ \\\hline\hline
Inviscid & $\etas \lesssim 10^{-3}$ & $ 8 $ & $ 3 $ \\ 
Hydrodynamic & $10^{-3} \lesssim \etas \lesssim 0.083$ & $ 10-12$ & $ 4$ \\ 
Transition  & $0.083 \lesssim \etas \lesssim 18$ & $ 12-200$ & $ 0-4$ \\ 
Ballistic & $\etas \gtrsim 18$ & $ 220$ & $0$
\end{tabular}
\end{ruledtabular}
\caption{Number of velocities $N_{\rtin{vel}}$ and expansion order 
$N_\Omega$ required in the simulation of the Sod shock tube problem 
with initial conditions \eqref{eq:Sod:right_adim}
to reduce the error $\epsilon$ \eqref{eq:eps_def} below
$\epsilon_{\rtin{th}} = 1\%$. The error is computed for the profiles of 
$n$, $T$, and $\gamma$ at $t = 0.5$ with respect to those obtained with 
$\text{R-SLB}(6;500)$ for various flow regimes.
\label{tab:Sod_conv}}
\end{table}

In this section, the effects of varying the quadrature order $Q_\xi$ on the profiles of $n$, $T$, and $\gamma$ 
are investigated. The convergence test \eqref{eq:eps_th} is performed with respect to the 
reference profiles obtained using the model with $Q_\xi = 500$ and $N_\Omega = 5$.
The initial states given by Eqs.~\eqref{eq:Sod:right_adim} and \eqref{eq:Sod:right2_adim} 
are discussed in Figs.~\ref{fig:Qxiconv} and \ref{fig:Qxiconv_etas05}, respectively.

Figure~\ref{fig:Qxiconv}(a) shows the dependence of the error $\epsilon$ \eqref{eq:eps_def} on the 
quadrature order $Q_\xi$ for various values of $\etas$. It can be seen that $\epsilon$ decreases smoothly 
as $Q_\xi$ is increased, demonstrating the stability of our quadrature procedure.

In Fig.~\ref{fig:Qxiconv}(b), the dependence on $\etas$ can be seen for 
the quadrature order $Q_\xi^\rtin{conv}$ required to reduce $\epsilon$ 
below the threshold error $\epsilon_\rtin{th} \in \{1\%, 2\%, 5\%\}$. 
For a fixed value of $\etas$, it can be seen that $Q_\xi^\rtin{conv}$ increases 
as $\epsilon_\rtin{th}$ is decreased. Similar features are observed 
in Fig.~\ref{fig:Qxiconv_etas05} for the initial state given in 
Eq.~\eqref{eq:Sod:right2_adim}. 
In both cases, for a fixed value of $\epsilon_\rtin{th}$,
three plateau regions and one transition region can be distinguished, which 
are described below for the most stringent case, namely 
the initial conditions in Eq.~\eqref{eq:Sod:right_adim} 
and for the threshold $\epsilon_\rtin{th} = 1\%$. 

The inviscid limit, for which $Q_\xi^\rtin{conv} = 4$ and $N_\Omega^\rtin{conv} = 3$,
occurs when $\etas \lesssim 10^{-3}$. In the hydrodynamic regime, 
[$10^{-3} \lesssim \etas \lesssim 0.083$], the non-equilibrium terms in $T^{\mu\nu}$ 
become important and the features of the shock become smooth, while 
$Q_\xi^\rtin{conv}$ increases to $6$ and $N_\Omega^\rtin{conv}$ increases to $4$.
During the transition regime [$0.083 \lesssim \etas \lesssim 18$], $Q_\xi$ increases sharply from 
$6$ to $\sim$100, while $N_\Omega^\rtin{conv}$ decreases from $4$ to $0$. 
Finally, in the free-streaming regime [$\etas \gtrsim 18$], $Q_\xi^\rtin{conv}$ reaches $110$,
while $N_\Omega^{\rtin{conv}} = 0$, confirming that in this regime, the particle 
collisions become insignificant. These results are summarized in Table~\ref{tab:Sod_conv}.

We conclude that, for the initial conditions \eqref{eq:Sod:right_adim} and 
\eqref{eq:Sod:right2_adim} and for $Z = 1000$ points, a number of $Q_\xi \simeq 100$ 
quadrature points are sufficient to reduce the errors in the profiles of $n$, $T$, and $\gamma$ below 
the threshold of $1\%$. At high values of $\etas$, $Q_\xi^{\rtin{conv}}$ shows a 
strong dependence on the number $Z$ of points used to discretize the flow domain. 
As discussed in Sec.~\ref{sec:Sod:bal:num}, in the free-streaming limit, the flow develops a 
staircaselike pattern, with the number of steps equal to $Q_\xi$. For a given number of fluid nodes, 
$Q_\xi$ must be increased such that the transition between these steps is smooth. If, for a fixed value of 
$Q_\xi$, the number of nodes (and hence the resolution) is increased, the staircase redevelops.
A similar argument holds for the case when $Z$ is decreased. 

\subsection{Summary}\label{sec:Sod:summary}

This section was devoted to validating the lattice Boltzmann (LB) model introduced in 
Sec.~\ref{sec:LB} in the context of the Sod shock tube problem. In order to compare 
our simulation results with the results obtained using the BAMPS stochastic approach 
and the vSHASTA hydrodynamic solver, we employed a constant shear viscosity to entropy 
ratio $\eta / s$. 

In the inviscid regime, our models were able to recover the analytic 
solution of the relativistic Euler equations to very good accuracy at $\etas = 10^{-4}$ 
[$\etas$ is the value of $\eta / s$ expressed in Planck units, as discussed in
Sec.~\ref{sec:Sod:setup}], where only $2Q_{\xi} = 2 \times 4 = 8$ velocities 
were required. 

In order to recover the analytic solution of the free-streaming (ballistic) limit 
of the relativistic Boltzmann equation, the quadrature order $Q_\xi$ had to be 
increased to $110$ when $1000$ nodes were used. At smaller values of $Q_\xi$, 
the flow progresses in a 
staircaselike fashion, the populations corresponding to each discrete velocity 
propagating independently. Our simulations show that increasing the number of 
grid points also requires the increase of $Q_\xi$ in order to maintain 
smooth macroscopic profiles.

Our study of the dissipative regime starts while investigating the flow 
close to the inviscid regime. 
Due to the kinetic nature of our numerical approach, the systems in our simulations 
always exhibit some dissipation, since the transport coefficients are proportional 
to the relaxation time $\tau_{\rtin{A-W}}$, which has to remain finite in order for 
the simulations to be stable. In the vicinity of the shock front and contact 
discontinuity, the flow profiles can develop discontinuities and the
non-equilibrium quantities (heat flux $q$ and shear stress $\Pi$) become non-negligible, 
since they are proportional to the gradients of the macroscopic fields. The magnitudes 
of $q$ and $\Pi$ can be estimated analytically by considering their integrated value over 
the range where they are non-negligible. A comparison between our numerical results 
and the analytic prediction indicates that our numerical scheme is consistent with 
the Chapman-Enskog prediction for the transport coefficients (heat conductivity and 
shear viscosity), while the analytic results corresponding to the Grad prediction 
for the transport coefficients are in visible disagreement with our numerical results.

The final validation test that we considered was a comparison with the data obtained using 
the BAMPS stochastic method and the vSHASTA hydrodynamics solver 
\cite{bouras10,bouras09prl,bouras09nucl}. In order to correlate our 
simulations with the simulations performed in these references, the relaxation time 
was implemented such that the viscosity to entropy ratio $\etas$, expressed 
in Planck units, matched the one employed therein. This required as an input the 
analytic expression for the shear viscosity. Based on our previous investigation 
of the dissipative effects in the nearlyinviscid regime, our model was implemented
starting from the Chapman-Enskog values for the shear viscosity. 
Good agreement was in general observed for $\etas$ up to $0.5$. There is some 
discrepancy between our simulation results and the BAMPS data for the heat flux,
our results being closer to the vSHASTA prediction.
This can be explained since in the vSHASTA implementation, the heat conductivity 
is fixed at the value predicted through the Chapman-Enskog procedure 
for the Anderson-Witting model for ultrarelativistic particles, which 
coincides with the model employed in our implementation.
The number of velocities required to obtain good agreement with the 
BAMPS data is less than $20$ at $\etas \lesssim 0.1$. At $\etas \simeq 0.5$, 
the flow enters the rarefied regime and a significantly larger number of 
quadrature points is required, as indicated by Figs.~\ref{fig:Qxiconv}
and \ref{fig:Qxiconv_etas05}.

\section{One-dimensional boost-invariant expansion}\label{sec:bjorken}

We now consider the evolution with respect to the proper time $\tau$ of the distribution function 
corresponding to a fluid which is homogeneous with respect to the spatial coordinates of the 
Milne coordinate system introduced in Sec.~\ref{sec:boltz:milne} from an initial equilibrium state,
described by the Maxwell-J\"uttner distribution. 
This assumption represents the implementation of the 
one-dimensional boost-invariant expansion model (Bjorken flow).
This setup is used to test the LB models introduced in Sec.~\ref{sec:LB} 
throughout the whole rarefaction spectrum. 
Our models are validated against analytic solutions in the inviscid and 
ballistic regimes, as well as against numerical solutions of the 
second-order hydrodynamics equations in the viscous regime.
In the transition regime, our numerical results are validated 
by comparison with the semianalytic results reported in 
Ref.~\cite{florkowski13}.

The nondimensionalization convention and initial state 
are discussed in Sec.~\ref{sec:bjorken:setup}.
The macroscopic equations are discussed in Sec.~\ref{sec:bjorken:eqs}.
The subsequent analysis is performed separately for the 
parton gas (flow at vanishing chemical potential) and 
ideal gas ($P =nT$) assumptions, which are discussed 
in Secs.~\ref{sec:bjorken:parton} and \ref{sec:bjorken:idgas}, respectively.
A test of the convergence of our numerical results with respect to the 
quadrature order is performed in Sec.~\ref{sec:bjorken:conv}.
Section~\ref{sec:bjorken:conc} summarizes our results and 
concludes this section.

\subsection{Nondimensionalization convention and initial setup}\label{sec:bjorken:setup}

The nondimensionalization procedure employed in this section uses as reference 
quantities the initial proper time 
$\widetilde{\tau}_{\rm ref} = \widetilde{\tau}_0 = (0.25/\widetilde{c})\ {\rm fm}$
and the initial system temperature $\widetilde{T}_{\rm ref} = \widetilde{T}_0$. 
Two particular values for
$\widetilde{T}_0$ are considered, namely 
$\widetilde{k}_B \widetilde{T}_0 = 0.3\ {\rm GeV}$ and 
$\widetilde{k}_B \widetilde{T}_0 = 0.6\ {\rm GeV}$.
The relaxation time $\tau_{\rtin{A-W}}$ is fixed by imposing a constant 
shear viscosity to entropy ratio $\etas$, expressed in Planck units,
such that:
\begin{equation}
 \tau_{\rtin{A-W}} = \frac{\tau_{\rtin{A-W}; 0}}{T} 
 \left[1 - \frac{\ln(\overline{\lambda} \lambda_{\rtin{ref}})}{4}\right], \quad 
 \tau_{\rtin{A-W}; 0} = \frac{5 \widetilde{\hbar} \etas}
 {\widetilde{\tau}_0 \widetilde{k}_B \widetilde{T}_0},
 \label{eq:bjorken_tauAW}
\end{equation}
where the relative fugacity $\overline{\lambda}$ and reference fugacity 
$\lambda_{\rtin{ref}}$ are defined in Eq.~\eqref{eq:lambda}. More details 
on how to arrive at Eq.~\eqref{eq:bjorken_tauAW} can be found in 
Sec.~\ref{sec:Sod:setup}.
The reference particle number density $\widetilde{n}_{\rtin{ref}}  = \widetilde{n}_0$ 
is taken to be the particle number density at initial time, such 
that $n_0 = 1$ and $\overline{\lambda}(\tau = 1) = 1$. 
Throughout this section, only the case when 
the chemical potential vanishes at initial time is considered, 
i.e., $\lambda_{\rtin{ref}} = 1$, such 
that the initial (reference) particle number density is determined 
by the initial temperature through:
\begin{equation}
 \widetilde{n}_{\rtin{ref}} = \frac{g_s}{\pi^2} 
 \left(\frac{\widetilde{k}_B \widetilde{T}_{\rtin{ref}}}{\widetilde{\hbar} 
 \widetilde{c}} \right)^3,
\end{equation}
where $g_s = 16$ is the gluon number of degrees of freedom.

Taking into account the above non-dimensionalization conventions, the distribution 
function at initial time takes the following form:
\begin{equation}
 f(\tau = 1) = \frac{1}{8\pi} e^{-p}.
 \label{eq:bjorken_f0}
\end{equation}
A generalization of the above initial condition is to account for 
possible anisotropies in the initial particle distribution function,
e.g., by employing the Romatschke-Strickland anisotropic distribution 
\cite{romatschke03,florkowski13}. For simplicity, we only consider the isotropic 
initial state given by Eq.~\eqref{eq:bjorken_f0} and defer the study 
of the anisotropic case for future work.

\subsection{Macroscopic equations}\label{sec:bjorken:eqs}

\begin{figure}
\begin{tabular}{c}
 \includegraphics[angle=0,width=0.9\linewidth]{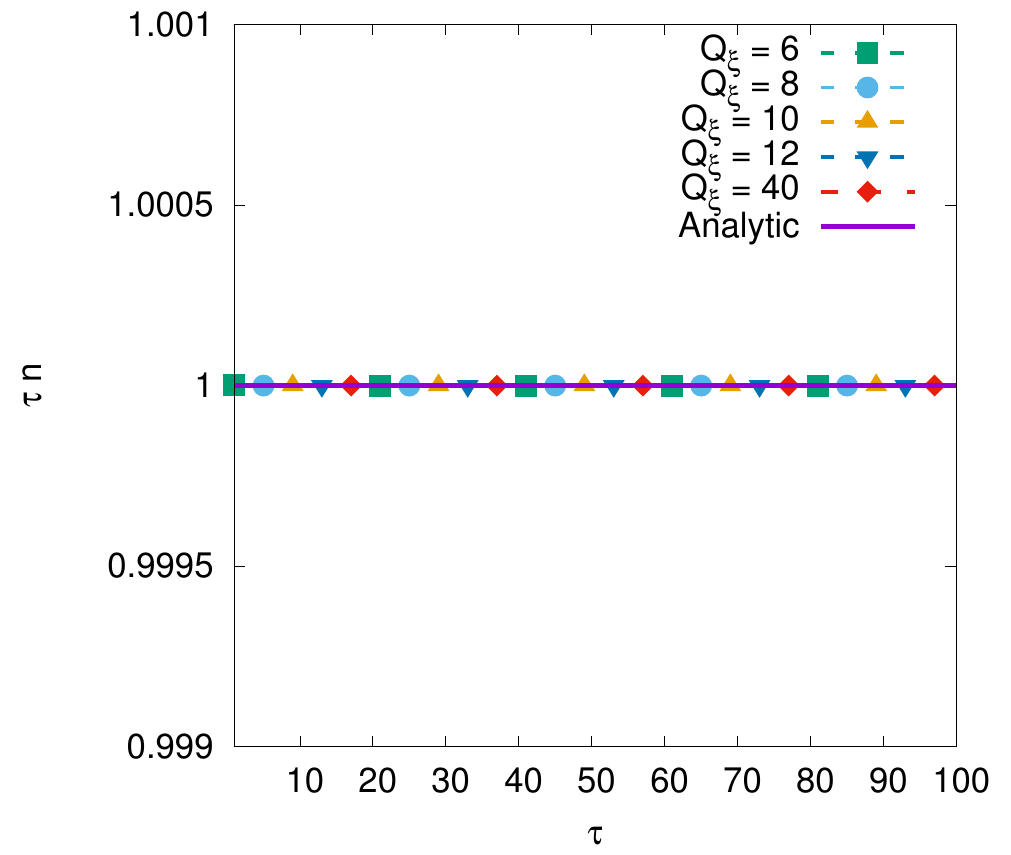} \\
 \includegraphics[angle=0,width=0.9\linewidth]{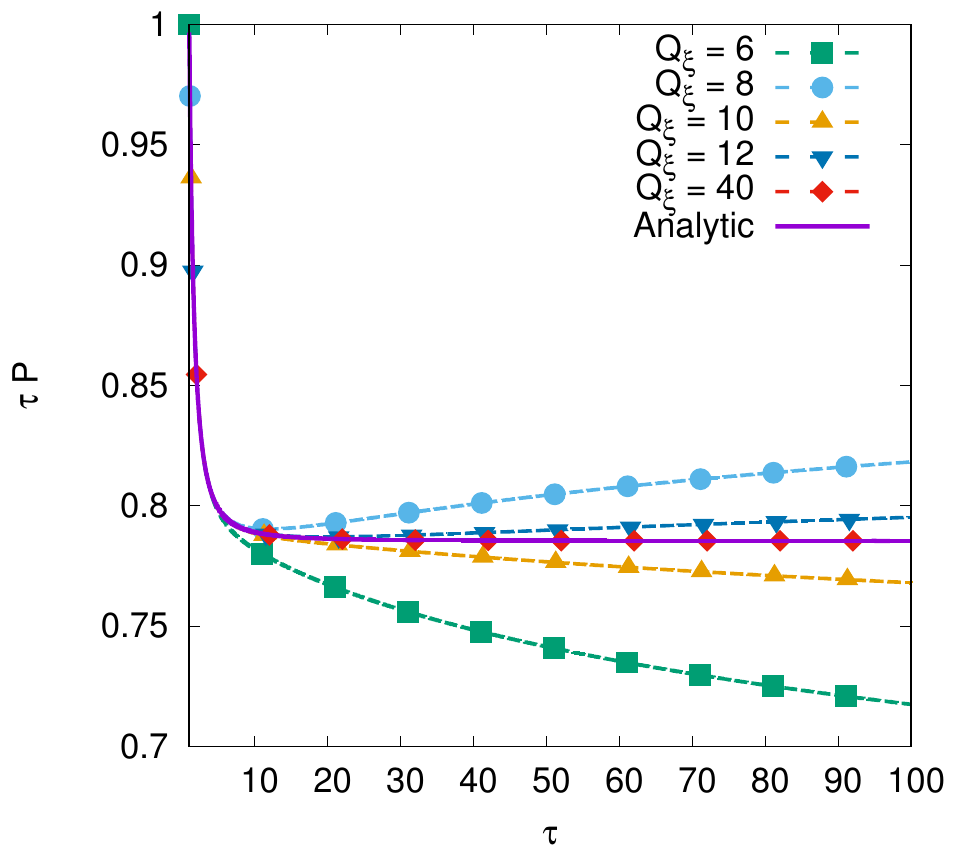} \\
 \includegraphics[angle=0,width=0.9\linewidth]{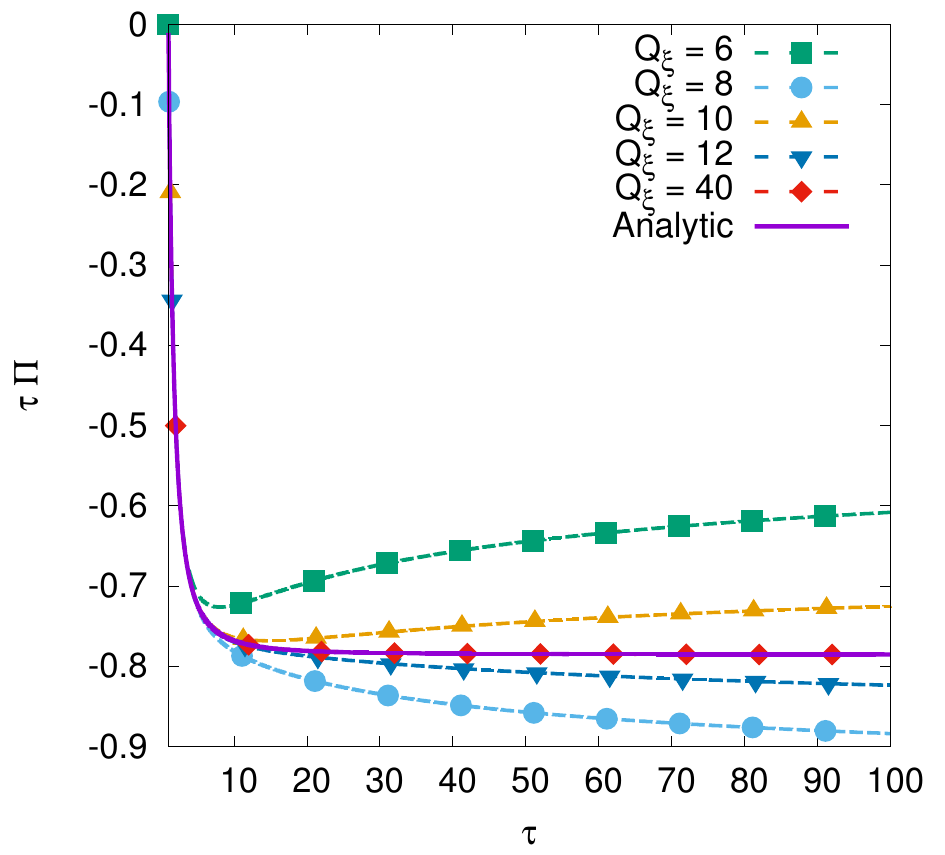}
\end{tabular}
\caption{
Evolution of $\tau n$ (a), $\tau P$ (b), and $\tau \Pi$ (c) 
with respect to $\tau$. The $\tau$ factor is included in order to 
highlight the asymptotic limits \eqref{eq:bjorken_bal_as}.
The numerical results corresponding to various $Q_\xi$ (dashed lines 
and points) are compared to the analytic solutions (solid lines) 
given in Eqs.~\eqref{eq:bjorken_n} and \eqref{eq:bjorken_bal}, which are 
overlapped with the $Q_\xi = 40$ curves.
}
\label{fig:bjorken_bal}
\end{figure}

Using the connection coefficients given in Eq.~\eqref{eq:bjorken_Gamma},
the conservation equations $\nabla_{\halpha} N^\halpha = 0$ and 
$\nabla_{\hsigma} T^{\halpha \hsigma} = 0$ reduce to:
\begin{align}
 \partial_\tau (N^\htau \tau) =&\ 0, \nonumber\\
 \partial_\tau (T^{\htau\htau} \tau) + T^{\heta_M \heta_M} =&\ 0, \nn\\ 
 \partial_\tau (T^{\htau \hatx} \tau) = \partial_\tau (T^{\htau \haty} \tau) =&\ 0, \nn\\
 \partial_\tau (T^{\htau \heta_M} \tau^2) =&\ 0.\label{eq:bjorken_macro}
\end{align}
The equations on the last two lines above are automatically satisfied 
by setting $T^{\htau \hatx} = T^{\htau \haty} = T^{\htau \heta_M} = 0$, 
such that the stress-energy tensor takes the form in Eq.~\eqref{eq:bjorken_SET}
and the Landau velocity is $u^\halpha = (1,0,0,0)^T$ \eqref{eq:bjorken_u}
at all times $\tau > \tau_0$. In this case, the Landau and Eckart frames 
coincide, such that $N^\htau = n$ is just the particle number density.
Furthermore, the first relation in Eq.~\eqref{eq:bjorken_macro} yields
\begin{equation}
 n = \frac{1}{\tau},\label{eq:bjorken_n}
\end{equation}
while the second line in \eqref{eq:bjorken_macro} reduces to:
\begin{equation}
 4 P^{1/4} \partial_\tau(\tau P^{3/4}) + \Pi = 0.
 \label{eq:bjorken_macro_P}
\end{equation}

In the inviscid regime, Eq.~\eqref{eq:bjorken_macro_P} can be solved 
by setting $\Pi = 0$, giving the Bjorken ideal flow solution \cite{bjorken83}:
\begin{equation}
 P = \frac{1}{\tau^{4/3}}, \qquad T = \frac{1}{\tau^{1/3}},
 \label{eq:bjorken_inv}
\end{equation}
where the expression for $T= P / n$ was obtained by noting that 
$n$ is given by Eq.~\eqref{eq:bjorken_n}. It can be 
seen that $\lambda = 1$ for all values of $\tau$, 
such that in the inviscid regime, the chemical potential vanishes 
at all times. 

In the free-streaming regime, the Boltzmann equation \eqref{eq:boltz_Milne} 
subject to the initial condition \eqref{eq:bjorken_f0} imposed at $\tau = 1$
has the following exact solution:
\begin{equation}
 f(\tau; p, \xi) = \frac{1}{8\pi} 
 \exp\left(-p\sqrt{1 + (\tau^2 - 1) \xi^2}\right).\label{eq:milne_bal_f}
\end{equation}
The macroscopic quantities $P = \frac{1}{3} T^{\htau\htau}$ and 
$\mathcal{P}_L = T^{\heta_M \heta_M} = P - \Pi$ can be computed by 
direct integration of Eq.~\eqref{eq:milne_bal_f}:
\begin{align}
 P =& \frac{1}{2} \left[ 
 \frac{\arctan \left(\sqrt{\tau^2 - 1}\right)}{ \sqrt{\tau^2 - 1} }+ \frac{1}{\tau^2}\right],\nonumber\\
 \mathcal{P}_L =& \frac{3}{2} \frac{1}{\tau^2 - 1} \left[
 \frac{\arctan \left(\sqrt{\tau^2 - 1}\right)}{ \sqrt{\tau^2 - 1} } - \frac{1}{\tau^2}\right],
 \label{eq:bjorken_bal}
\end{align}
As $\tau \rightarrow \infty$, the following asymptotic behavior is achieved:
\begin{equation}
 \lim_{\tau \rightarrow \infty} (\tau P) = 
 -\lim_{\tau \rightarrow \infty} (\tau \Pi) = \frac{\pi}{4}.
 \label{eq:bjorken_bal_as}
\end{equation}
The solution~\eqref{eq:bjorken_bal} is compared with our numerical results in 
Fig.~\ref{fig:bjorken_bal}, highlighting the asymptotic behavior \eqref{eq:bjorken_bal} 
by representing the quantities $\tau n$, $\tau P$, and $\tau \Pi$. It can be seen 
that the numerical results improve as the quadrature $Q_\xi$ is increased,
with clear deviations from the analytic solutions of $P$ and $\Pi$ 
at $Q_\xi \lesssim 12$, while at $Q_\xi = 40$, the agreement with the 
analytic solution is excellent. It is noteworthy that the evolution 
of $\tau n$ is exactly recovered even when $Q_\xi = 6$. The expansion order 
of $\feq$ is $N_\Omega = 5$ and the time step was set to $\delta \tau= 10^{-3}$.

In the viscous regime, the Chapman-Enskog procedure shows that the 
shear stress $\Pi$ is given up to first order by:
\begin{equation}
 \Pi = 4\eta \nabla^{\langle\hatx} u^{\hatx\rangle} = -\frac{4\eta}{3} \Gamma^\halpha{}_{\htau\halpha} 
 = -\frac{4\eta}{3\tau},
 \label{eq:bjorken_hydro1_Pi}
\end{equation}
where $\Gamma^\halpha{}_{\htau\halpha} = 1/\tau$ according to Eq.~\eqref{eq:bjorken_Gamma}.
The equation for the pressure \eqref{eq:bjorken_macro_P} becomes:
\begin{equation}
 4 P^{1/4} \partial_\tau(\tau P^{3/4}) = \frac{16P}{15\tau} \tau_{\rtin{A-W}},
 \label{eq:bjorken_macro_P_h1}
\end{equation}
where the Chapman-Enskog expression \eqref{eq:tcoeff_ce} for the shear viscosity $\eta$
was employed.
It can be seen that $\Pi$ takes the value $-4\eta / 3$ at initial time ($\tau = 1$), 
which is incompatible with the initial condition \eqref{eq:bjorken_f0}, according to 
which the fluid is in thermal equilibrium when $\tau = 1$.
Thus, the constitutive equation for $\Pi$ must be upgraded to an evolution 
equation by employing a secondorder hydrodynamics formulation, giving 
rise to the following system of equations \cite{jaiswal2013d,florkowski13}:
\begin{align}
 3\tau \partial_\tau P + 4P =& -\Pi,\nonumber\\
 \tau_{\Pi} \partial_\tau \Pi + \Pi =& -\frac{4\eta}{3\tau} - \beta_\Pi \frac{\tau_\Pi}{\tau} \Pi, 
 \label{eq:bjorken_hydro2_Pi}
\end{align}
where $\tau_{\Pi} = \tau_{\rm A-W}$ and $\beta_\Pi = 38 / 21$ for the Anderson-Witting 
approximation of the collision term \cite{jaiswal2013d,florkowski13}. For completeness,
a derivation of the above relations is presented in Appendix~\ref{app:bjork2} 
using a moment-based method.

\subsection{Parton gas model}\label{sec:bjorken:parton}

\begin{figure*}
\begin{tabular}{cc}
\includegraphics[width=0.48\linewidth]{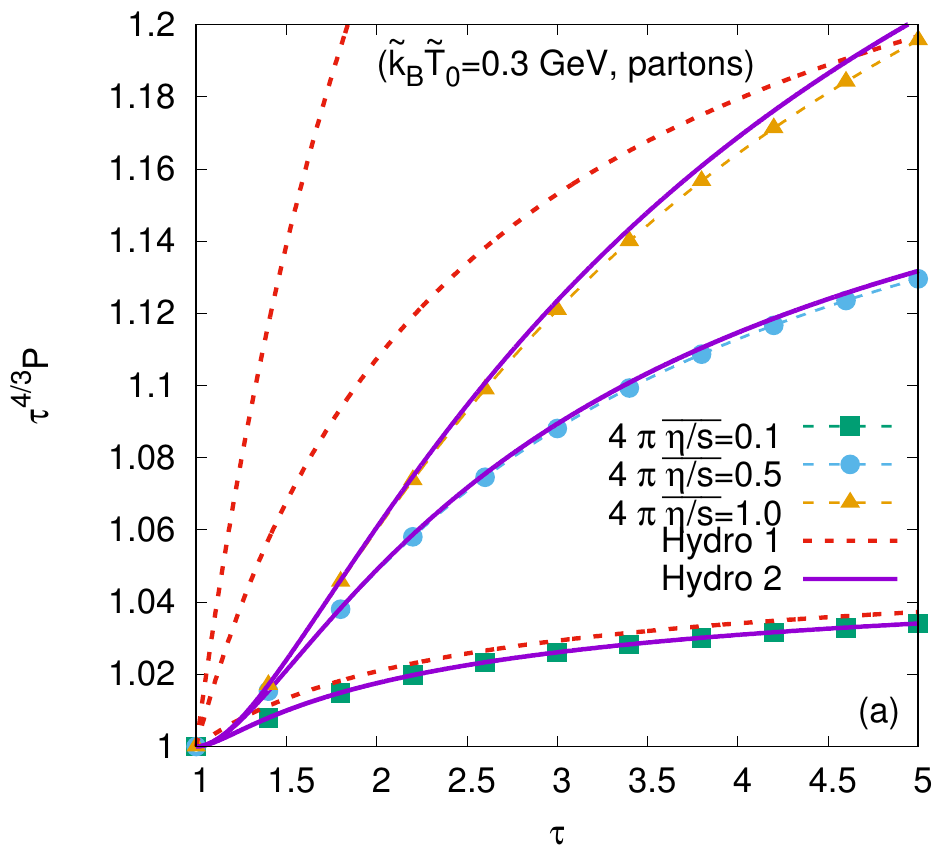} &
\includegraphics[width=0.48\linewidth]{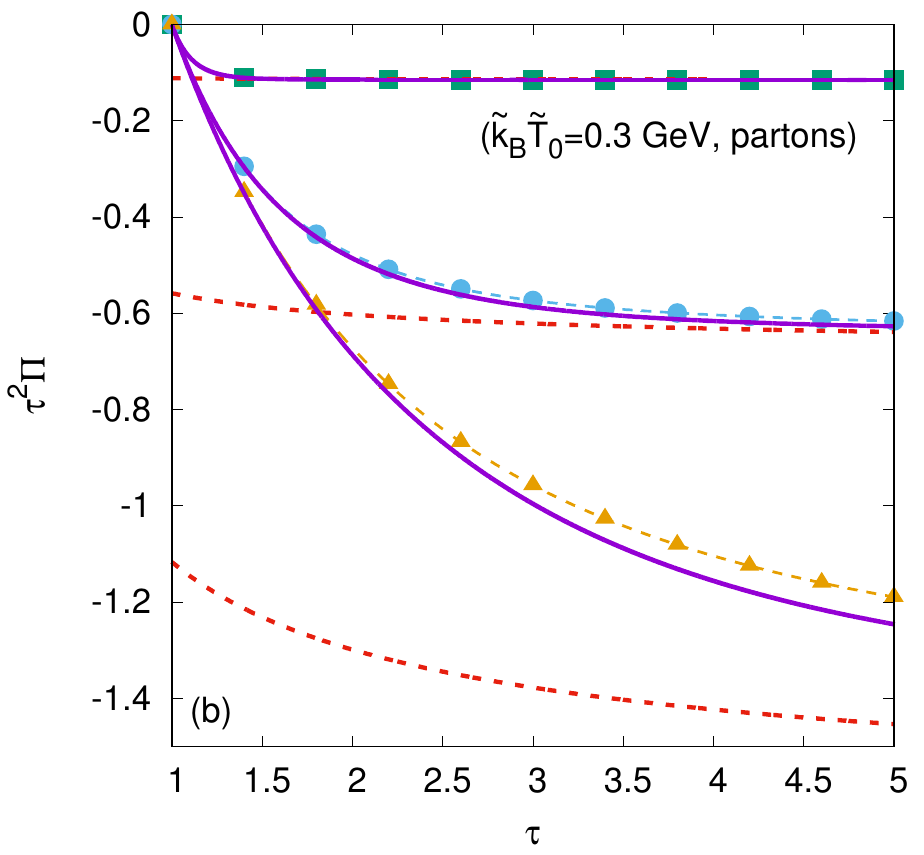} \\
\includegraphics[width=0.48\linewidth]{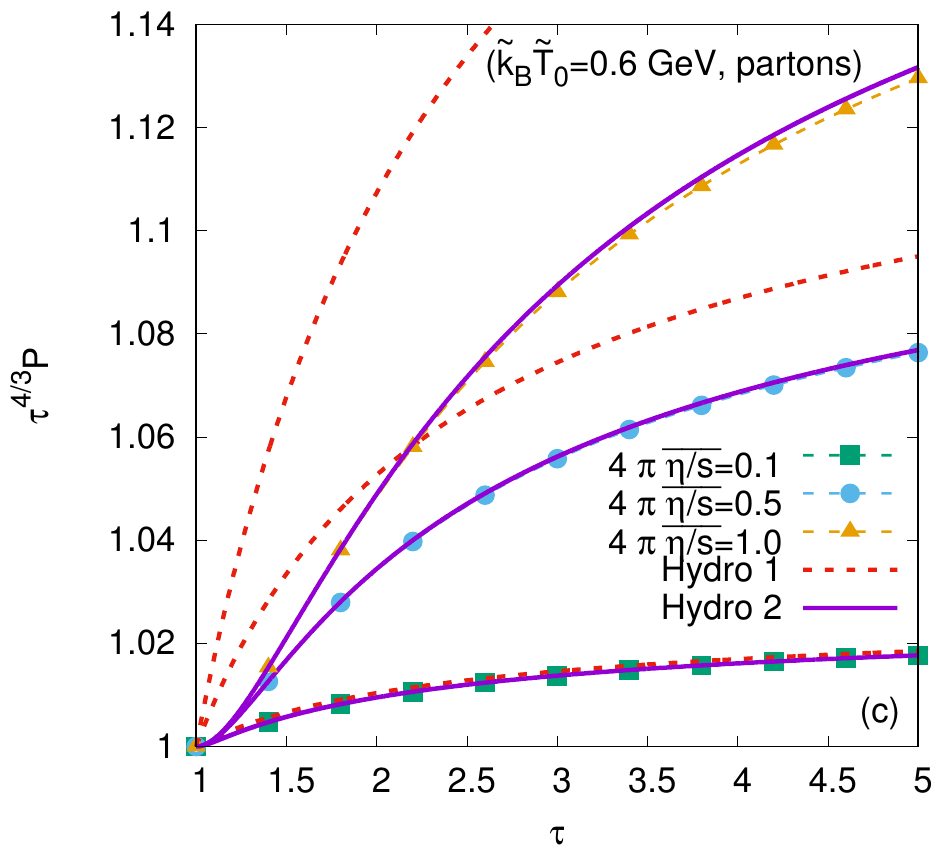} &
\includegraphics[width=0.48\linewidth]{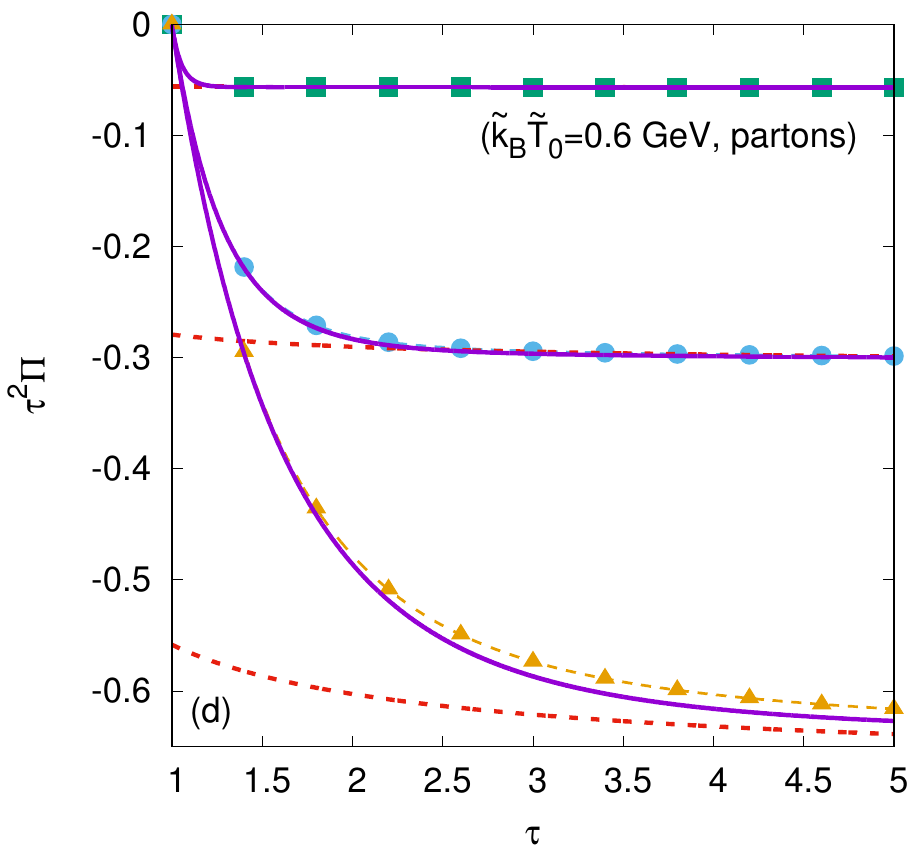} 
\end{tabular}
\caption{
Time evolution of (left) $\tau^{4/3} P$ and (right) $\tau^2 \Pi$ for 
$\widetilde{k}_B \widetilde{T}_0 = 0.3\ {\rm GeV}$ (top) 
and $0.6\ {\rm GeV}$ (bottom) for $4\pi \etas \in \{0.1, 0.5, 1.0\}$ 
for partons ($\mu = 0$). 
The numerical results (dotted lines and points) are compared with the 
first-order (dotted lines) and second-order (solid lines) hydrodynamics 
results obtained from Eqs.~\eqref{eq:bjorken_mu0_hydro1} and 
\eqref{eq:bjorken_mu0_hydro2}, respectively.
The factors $\tau^{4/3}$ and $\tau^2$ multiplying $P$ and $\Pi$ 
are inspired from the form of the first-order hydrodynamics solutions
\eqref{eq:bjorken_mu0_hydro1}.
\label{fig:bjorken_hydro}
}
\end{figure*}

\begin{figure*}
\begin{tabular}{cc}
\includegraphics[width=0.48\linewidth]{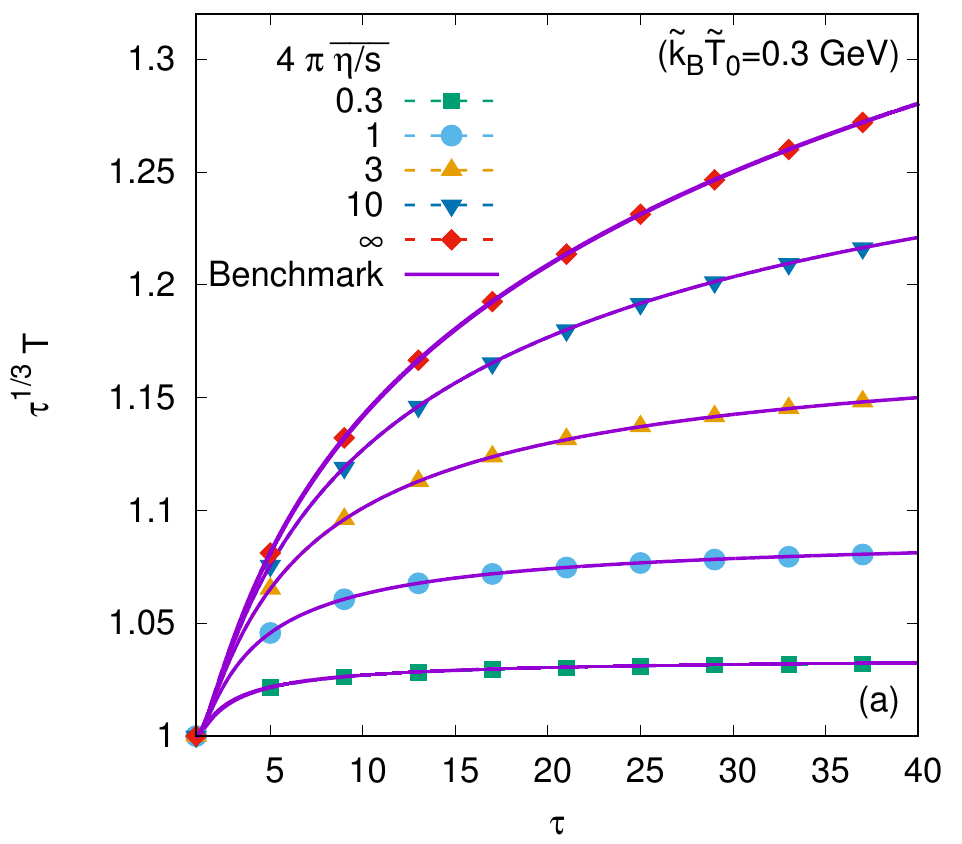} &
\includegraphics[width=0.48\linewidth]{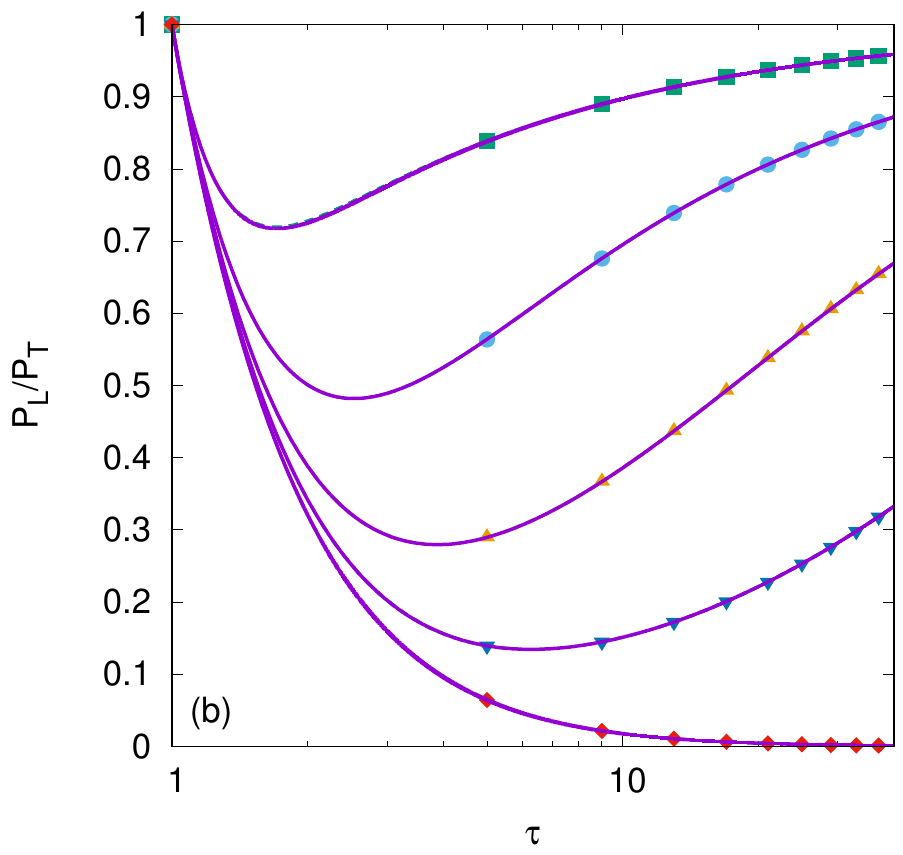}\\
\includegraphics[width=0.48\linewidth]{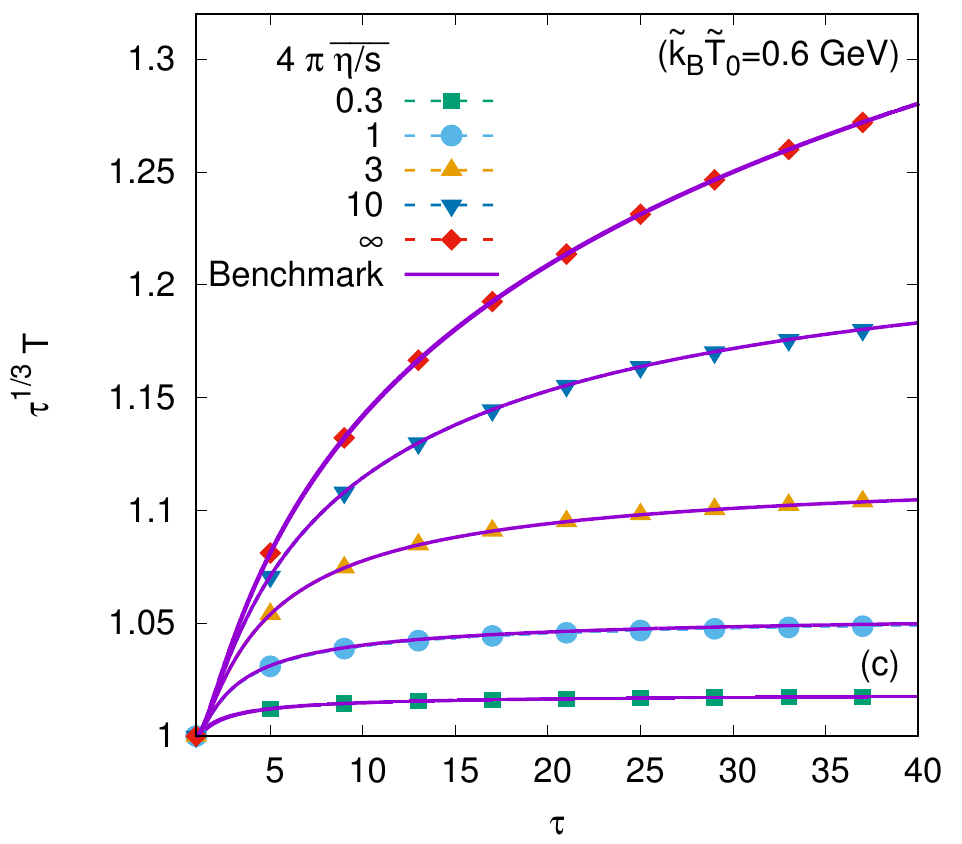} &
\includegraphics[width=0.48\linewidth]{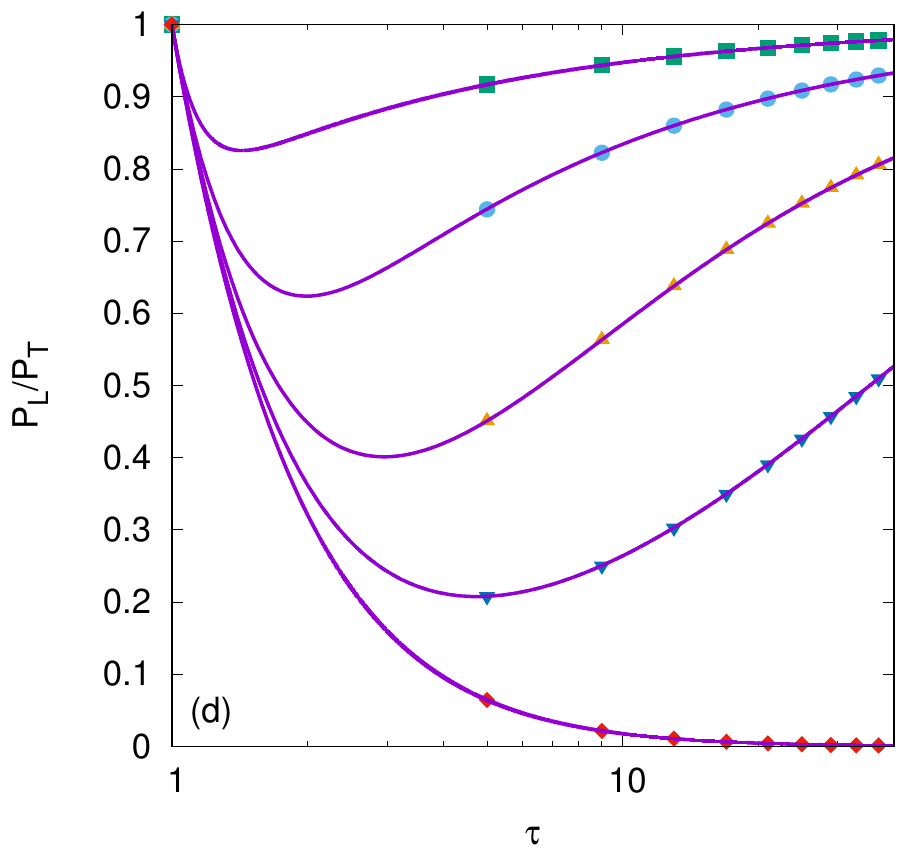}
\end{tabular}
\caption{
Evolution with respect to $\tau$
of (left) $T$ 
and (right) ratio $\mathcal{P}_L / \mathcal{P}_T$ 
of the longitudinal to the transverse pressure, 
for $\widetilde{k}_B \widetilde{T}_0 = 0.3\ {\rm GeV}$ (top) and 
$0.6\ {\rm GeV}$ (bottom), for various values of $\etas$, 
under the assumption of vanishing chemical potential 
discussed in Sec.~\ref{sec:bjorken:parton}.
The numerical results (dotted lines and points) 
are validated using benchmark results (solid lines) as follows:
for $\etas = \infty$, the analytic formulas
\eqref{eq:bjorken_bal} and \eqref{eq:bjorken_mu0} are used;
at $4\pi \etas = 0.3$, the numerical solution of the second order 
hydrodynamics equations \eqref{eq:bjorken_mu0_hydro2} is used;
when $4\pi \etas \in \{1,3,10\}$, the semi-analytic solution reported 
in Ref.~\cite{florkowski13} is represented.
\label{fig:bjorken_flork}
}
\end{figure*}

In the parton gas model, the fluid undergoing boost-invariant 
expansion is dominated by gluons, which are considered to evolve 
at vanishing chemical potential \cite{florkowski13}.
In this model, $\overline{\lambda} = 1$ for all $\tau \ge 1$.
This can be achieved by 
constructing $\feq$ and $\tau_{\rtin{A-W}}$ with 
$n^{\rtin{(eq)}} = T^3$, where the temperature 
is related to the isotropic pressure 
$P = \frac{1}{3} T^{\htau\htau}$ through \cite{florkowski13}:
\begin{gather}
 T^{\rtin{(eq)}} = P^{1/4} \equiv T, \qquad n^{\rtin{(eq)}} = P^{3/4}, \nonumber\\
 \feq = \frac{1}{8\pi} \exp\left(-\frac{p^\htau}{P^{1/4}}\right),
 \label{eq:bjorken_mu0}
\end{gather}
where the first relation also defines the fluid temperature $T$,
such that $\tau_{\rtin{A-W}}$ \eqref{eq:bjorken_tauAW} becomes
\begin{equation}
 \tau_{\rtin{A-W}} = \frac{\tau_{\rtin{A-W}; 0}}{P^{1/4}}, \qquad 
 \tau_{\rtin{A-W}; 0} = \frac{5 \widetilde{\hbar} \etas}
 {\widetilde{\tau}_0 \widetilde{k}_B \widetilde{T}_0}.
 \label{eq:bjorken_tauAW0}
\end{equation}
It is noteworthy that the parton density $n$ is not conserved. Its evolution 
is governed by \eqref{eq:bjorken_hydro2_Pi_moment}:
\begin{equation}
 \partial_\tau (\tau n) = -\frac{P^{1/4}}{\tau_{\rtin{A-W}; 0}} (n - P^{3/4}).
 \label{eq:bjork_n_eq}
\end{equation}

In the context of first-order hydrodynamics, Eq.~\eqref{eq:bjorken_macro_P}
can be simplified by substituting $\Pi$ from Eq.~\eqref{eq:bjorken_hydro1_Pi}:
\begin{equation}
 \frac{\partial_\tau (\tau P^{3/4})}{(\tau P^{3/4})^{2/3}} = 
 \frac{4\tau_{\rtin{A-W}; 0}}{15 \tau^{5/3}}.
\end{equation}
The above equation can be solved analytically, such that $P$ and $\Pi$ are given by
\begin{align}
 P =& \frac{1}{\tau^{4/3}} \left[1 + \frac{2\tau_{\rtin{A-W};0}}{15}\left( 1 - 
 \frac{1}{\tau^{2/3}}\right) \right]^4,\nonumber\\
 \Pi =& -\frac{16 \tau_{\rtin{A-W};0}}{15 \tau^2}\left[1 + \frac{2\tau_{\rtin{A-W};0}}{15}\left( 1 - 
 \frac{1}{\tau^{2/3}}\right) \right]^3.\label{eq:bjorken_mu0_hydro1}
\end{align}

In the case of the second-order hydrodynamics equations, the following system must be solved:
\begin{align}
 \partial_\tau P =& -\frac{4}{3\tau } P - \frac{1}{3\tau} \Pi, \nonumber\\
 \partial_\tau \Pi =& - \frac{P^{1/4}}{\tau_{\rtin{A-W};0}} \Pi - 
 \frac{38}{21 \tau} \Pi - 
 \frac{16}{15 \tau} P.
 \label{eq:bjorken_mu0_hydro2}
\end{align}
Further analytic development seems difficult to achieve, however 
the above system can be easily integrated numerically.

Finally, outside the hydrodynamic regime, a semianalytic formula was 
derived in Ref.~\cite{florkowski13}, which allows the temperature $T$,
parton number density $n$, and 
longitudinal pressure $\mathcal{P}_L$ to be computed iteratively via 
the following equations:
\begin{align}
 T^4(\tau) =& D(\tau, 1) \frac{\mathcal{H}(\tau^{-1})}{\mathcal{H}(1)} \nonumber\\
 &+ \int_1^\tau \frac{T^4(\tau') d\tau'}{2\tau_{\rtin{A-W}}(\tau')}\, 
 D(\tau, \tau') \mathcal{H}\left(\frac{\tau'}{\tau}\right),\nonumber\\
 n(\tau) =& \frac{1}{\tau} \left[D(\tau, 1) + \int_1^\tau \frac{T^3(\tau') \tau' d\tau'}
 {\tau_{\rtin{A-W}}(\tau')} D(\tau, \tau')\right],\nonumber\\ 
 \mathcal{P}_L(\tau) =& 3 \left[D(\tau, 1) 
 \frac{\mathcal{H}_L(\tau^{-1})}{\mathcal{H}(1)}.\right.\nonumber\\
 &\left.+ \int_1^\tau \frac{T^4(\tau') d\tau'}{2\tau_{\rtin{A-W}}(\tau')}\, 
 D(\tau, \tau') \mathcal{H}_L\left(\frac{\tau'}{\tau}\right)\right],
 \label{eq:flork_sol}
\end{align}
where the damping function $D(\tau_2, \tau_1)$ and the $\mathcal{H}$ and 
$\mathcal{H}_L$ functions are given by:
\begin{align}
 D(\tau_2, \tau_1) =& \exp\left[-\frac{1}{\tau_{\rtin{A-W;0}}} 
 \int_{\tau_1}^{\tau_2} d\tau''\, T(\tau'')\right],\nonumber\\
 \mathcal{H}(y) =& y^2 + \frac{\arctan\sqrt{y^{-2} - 1}}{\sqrt{y^{-2} - 1}},\nonumber\\
 \mathcal{H}_L(y) =& y^2 \frac{d}{dy} \left[\frac{\mathcal{H}(y)}{y}\right].
 \label{eq:flork_sol_aux}
\end{align}
The temperature profile $T_{(n)}$ at the $n$th iteration is found by 
inserting $T_{(n-1)}$ on the right-hand side of the first of 
Eqs.~\eqref{eq:flork_sol}. This iterative method was started with 
the ideal flow solution $T_{(0)} = \tau^{-1/3}$. The numerical integrals 
were computed using the trapezoidal rule and the integration domain comprised
$975$ equidistant intervals of length $\delta \tau = 0.04$ distributed between 
$\tau = 1$ and $\tau = \tau_f = 40$. The algorithm was stopped when the $L_2$ norm 
of the difference $T_{(n)} - T_{(n-1)}$ between the temperature profiles obtained 
after $n$ and $n - 1$ iterations decreased below the tolerance threshold 
$10^{-7}$:
\begin{equation}
 L_2^{(n)} = \sqrt{\int_1^{\tau_f} \frac{d\tau}{\tau_f - 1}  [T_{(n)}(\tau) 
 - T_{(n-1)}(\tau)]^2} \le 10^{-7}.
\end{equation}

The validity of the first- and second-order hydrodynamics theories is 
tested in Fig.~\ref{fig:bjorken_hydro}. It can be seen that the second-order 
theory accurately captures the early time relaxation of $\Pi$ (right column) 
from $\Pi = 0$ at $\tau = 1$ to the value predicted by the first-order theory, 
which becomes accurate at $\tau \gtrsim \tau_0 + 5[4\pi \etas]$ for 
sufficiently small values of $4\pi \etas$. The limitations of the 
first-order hydrodynamics theory become evident when considering the 
evolution of the pressure (left column in Fig.~\ref{fig:bjorken_hydro}).
At $4\pi \etas \gtrsim 0.5$, the first-order theory presents significant 
deviations from the second-order theory and the numerical results.

While accurate at $4\pi \etas \lesssim 0.5$, the second-order hydrodynamics 
formulation seems to lose validity when $4\pi \etas \gtrsim 1$, as shown 
in Fig.~\ref{fig:bjorken_hydro}.
For $4\pi \etas \in \{1, 3, 10\}$, Fig.~\ref{fig:bjorken_flork} 
shows a comparison of our results for the temperature $T$ and the 
ratio $\mathcal{P}_L / \mathcal{P}_T$ between the longitudinal 
and the numerical iterative solution of Eq.~\eqref{eq:flork_sol}. 
Excellent agreement is found at both
$\widetilde{k}_B \widetilde{T}_0 = 0.3\ {\rm GeV}$ and $0.6\ {\rm GeV}$. Also in 
Fig.~\ref{fig:bjorken_flork}, a comparison with the analytic solution 
\eqref{eq:bjorken_bal} in the ballistic regime is presented, showing that 
our models can recover the flow features in this regime as well.

The parameters for the simulations discussed in this section 
were $Q_\xi = 40$, $N_\Omega = 5$ and $\delta \tau = 10^{-3}$.

\subsection{Ideal gas model}\label{sec:bjorken:idgas}

\begin{figure*}
\begin{tabular}{cc}
\includegraphics[width=0.48\linewidth]{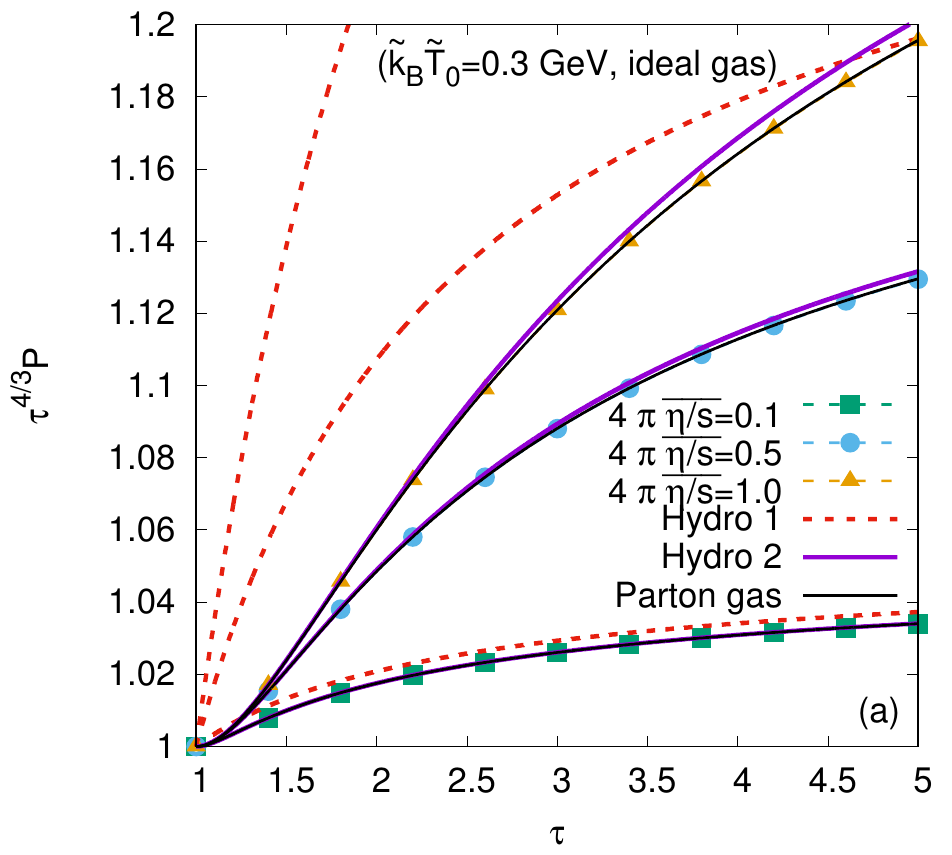} &
\includegraphics[width=0.48\linewidth]{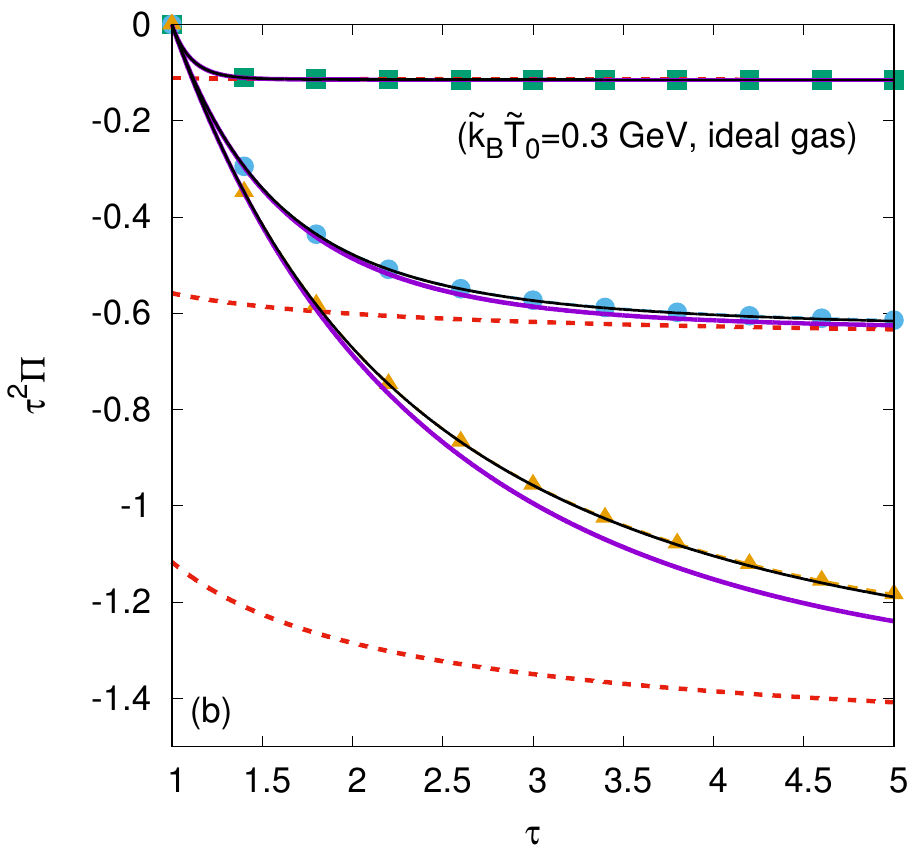} \\
\includegraphics[width=0.48\linewidth]{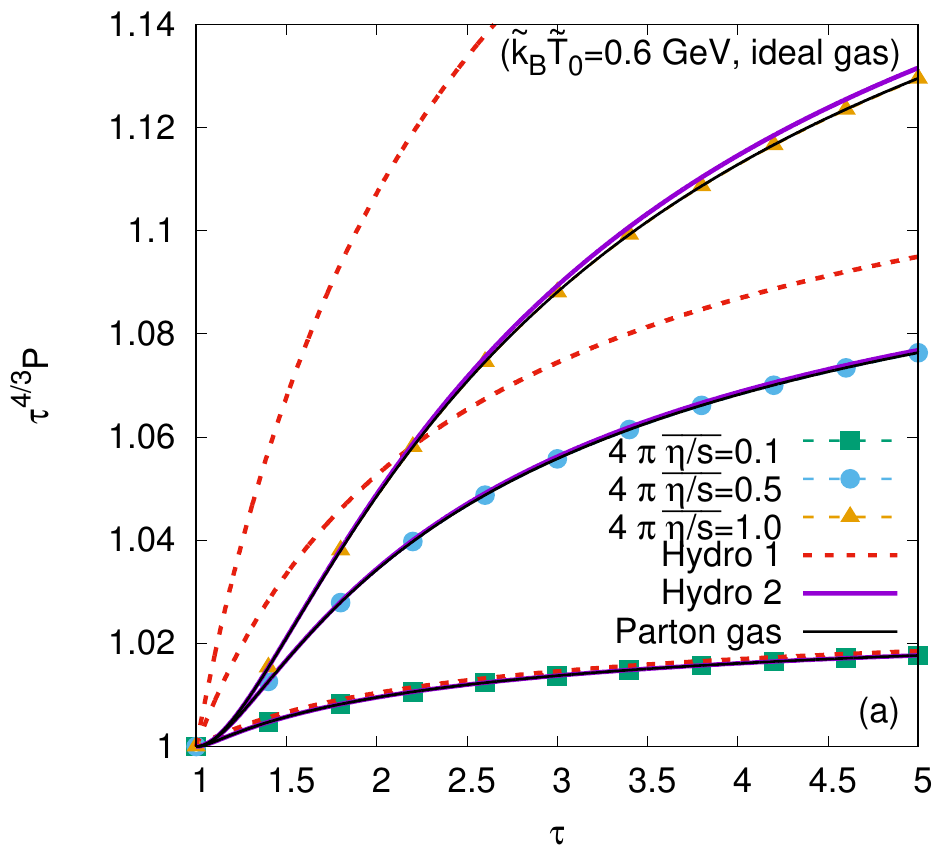} &
\includegraphics[width=0.48\linewidth]{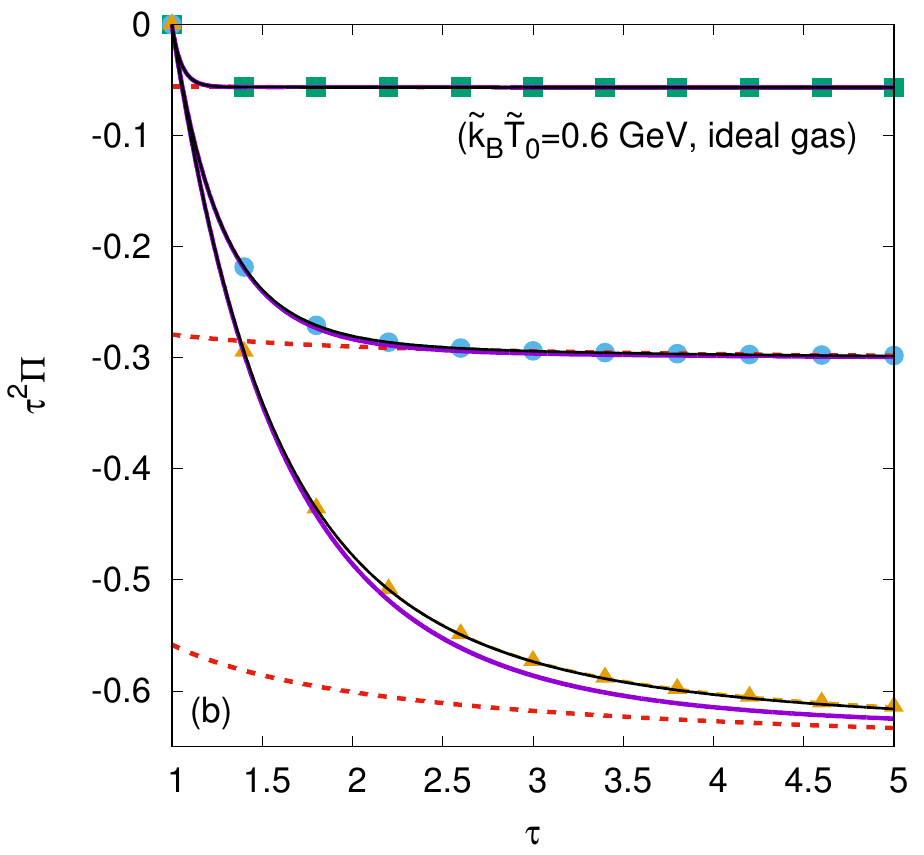}
\end{tabular}
\caption{
Time evolution of (left) $P$ and (right) $\Pi$ for 
$4\pi \etas \in \{0.1,0.5,1\}$ for initial temperatures 
$\widetilde{k}_B \widetilde{T}_0 = 0.3\ {\rm GeV}$ (top) and 
$0.6\ {\rm GeV}$ (bottom). The various curves show the 
numerical results obtained using LB for the cases of the ideal gas 
(dotted lines and points) and of the gas of partons (solid black lines),
as well as the solutions of the first (dotted lines) and second 
(solid purple lines) hydrodynamics equations \eqref{eq:milne_hydro1} 
and \eqref{eq:milne_hydro2} for the ideal gas.
The ideal and parton gas results seem to be overlapped for all 
tested values of $\etas$ and $\widetilde{T}_0$.
\label{fig:milne_hydro}
}
\end{figure*}

\begin{figure*}
\begin{tabular}{cc}
\includegraphics[width=0.48\linewidth]{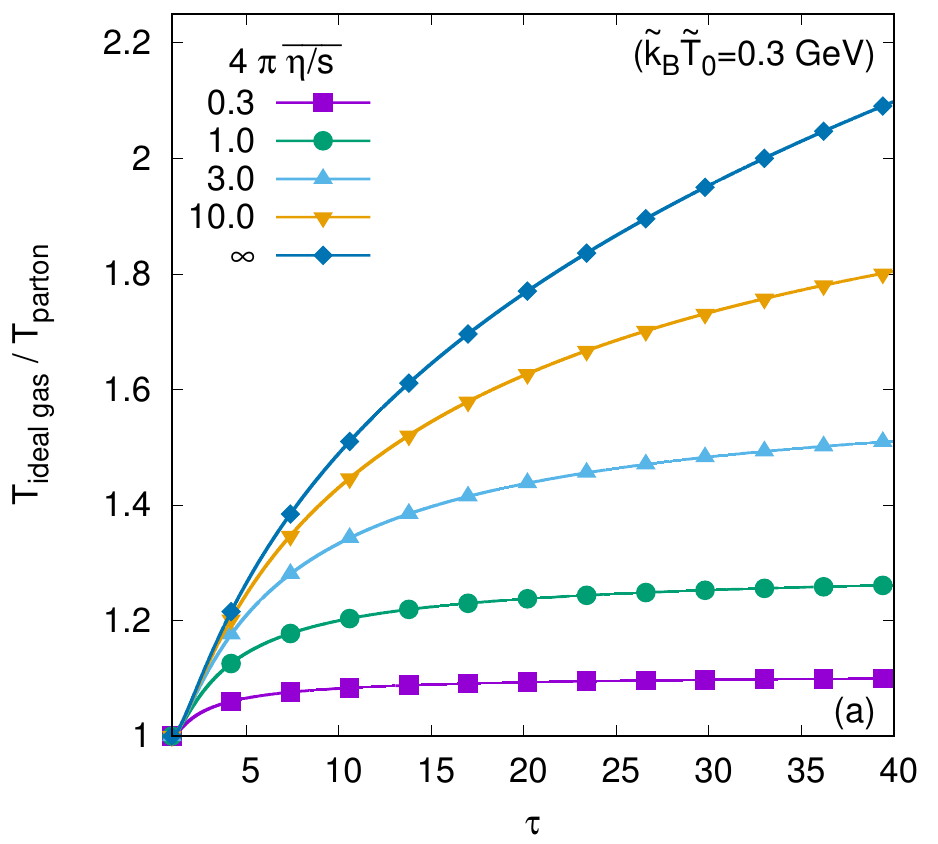} & 
\includegraphics[width=0.48\linewidth]{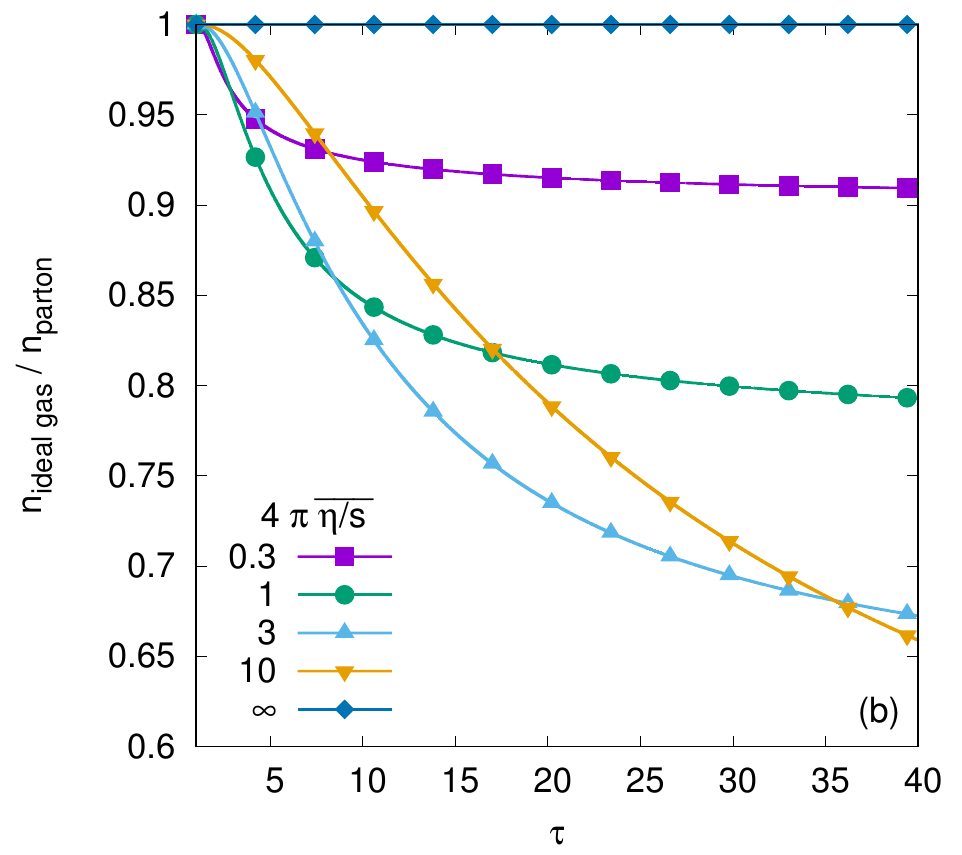} \\
\includegraphics[width=0.48\linewidth]{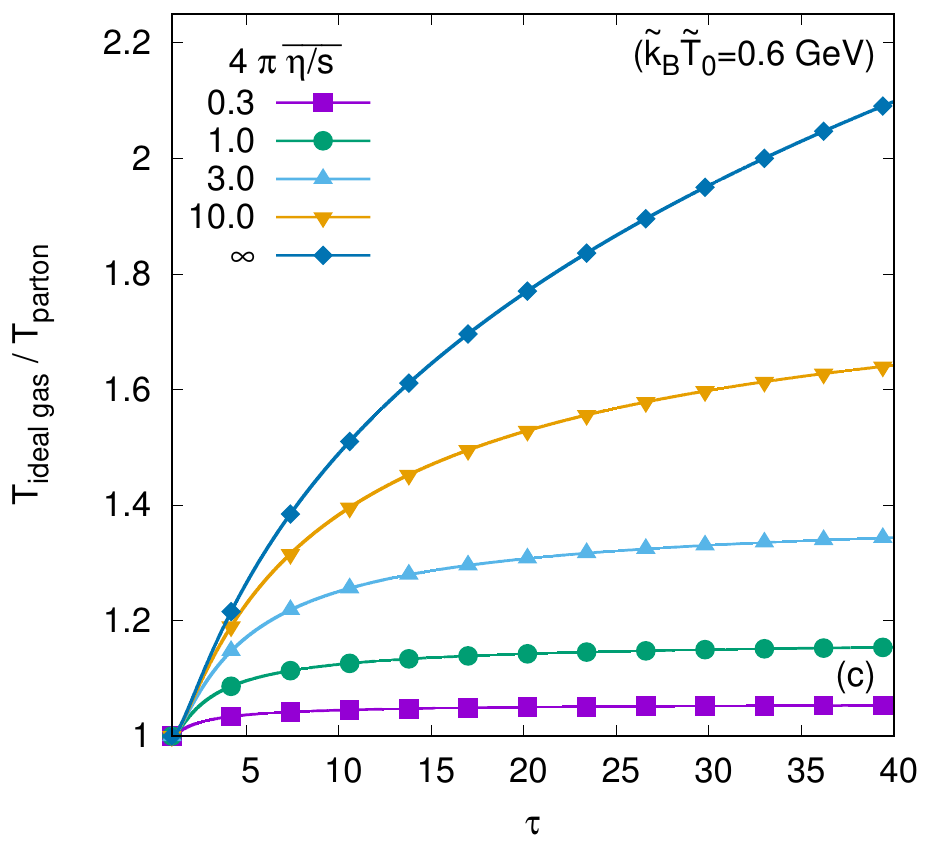} & 
\includegraphics[width=0.48\linewidth]{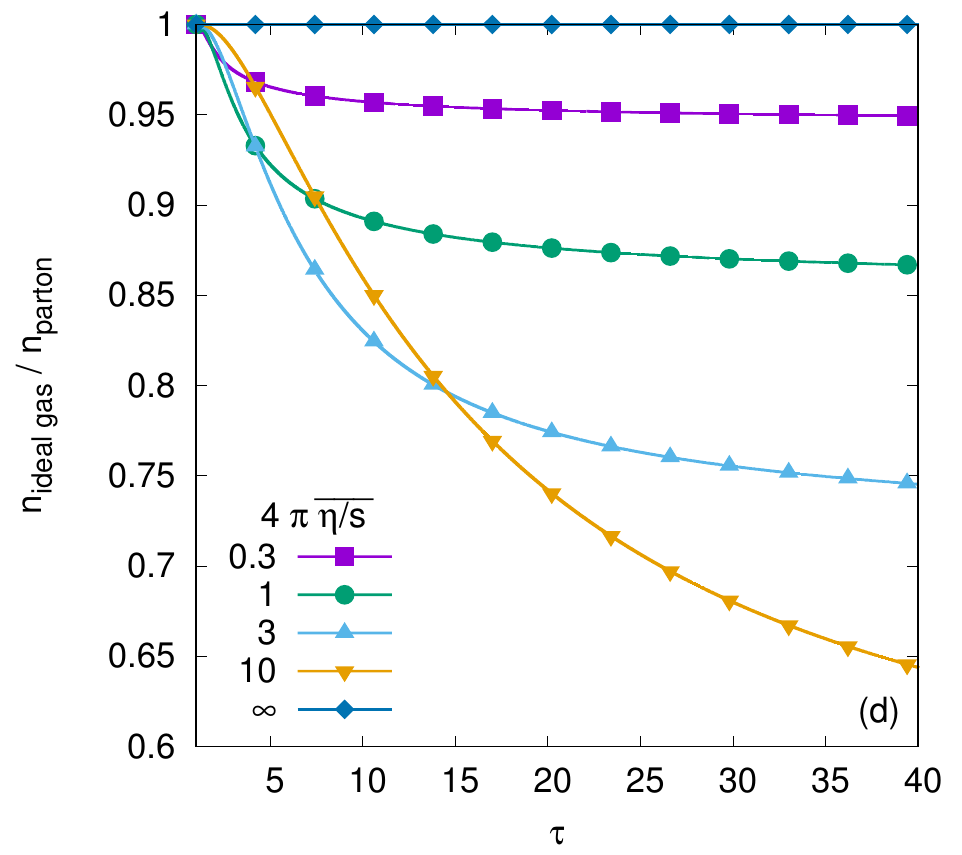} 
\end{tabular}
\caption{
Evolution of the ratios (left) $T_{\text{ideal gas}} / T_{\text{partons}}$ 
and (right) $n_{\text{ideal gas}} / n_{\text{partons}}$ of
the temperatures and densities obtained for the ideal ($P = nT$) and 
parton ($\mu_{\rm eq} = 0$) gases, for 
$\widetilde{k}_B \widetilde{T}_0 = 0.3\ {\rm GeV}$ (top)
and $0.6\ {\rm GeV}$ (bottom) for $4\pi \etas \in \{0.3, 1, 3, 10, \infty\}$.
\label{fig:milnevsbjork}
}
\end{figure*}

Let us now consider that the medium is an ideal gas, such that 
$P = nT$ and $n^{\rtin{(eq)}} = n$ in Eq.~\eqref{eq:bjorken_hydro2_Pi_moment}, 
where $n$ is given by Eq.~\eqref{eq:bjorken_n}, while $T = P / n = \tau P$. 
In this case, $\feq$ is computed via Eq.~\eqref{eq:MJ_tetrad}, 
while $\tau_{\rtin{A-W}}$ \eqref{eq:bjorken_tauAW} becomes
\begin{equation}
 \tau_{\rtin{A-W}} = \frac{\tau_{\rtin{A-W}; 0}}{\tau P} 
 \left[1 + \ln(\tau P^{3/4})\right],
 \quad 
 \tau_{\rtin{A-W}; 0} = \frac{5 \widetilde{\hbar} \etas}
 {\widetilde{\tau}_0 \widetilde{k}_B \widetilde{T}_0}.
 \label{eq:bjorken_tauAW_mu}
\end{equation}

In order to derive the time evolution of $P$ predicted via 
the first-order dissipative hydrodynamics approximation, Eq.~\eqref{eq:bjorken_tauAW_mu} 
can be substituted into Eq.~\eqref{eq:bjorken_macro_P_h1}, giving
the following system of equations:
\begin{align}
 \frac{\partial_\tau (P \tau^{4/3})}{\frac{4}{3} + \ln(P \tau^{4/3})} =&
 \frac{4 \tau_{\rtin{A-W}; 0}}{15\tau^{5/3}}, \nonumber\\
 \Pi =& -\frac{4 \tau_{\rtin{A-W};0}}{5\tau^2} \left[\frac{4}{3} + \ln(P \tau^{4/3})\right].
 \label{eq:milne_hydro1}
\end{align}

In the second-order hydrodynamics approach, the following system of equations must be solved:
\begin{align}
 \partial_\tau P =& -\frac{4}{3\tau } P - \frac{1}{3\tau} \Pi, \nonumber\\
 \partial_\tau \Pi =& - \frac{\tau P}{\tau_{\rtin{A-W};0}[1 + \ln(\tau P^{3/4})]} \Pi - 
 \frac{38}{21 \tau} \Pi - 
 \frac{16}{15 \tau} P.
 \label{eq:milne_hydro2} 
\end{align}
The solution of the above system is obtained via numerical integration.

Figure~\ref{fig:milne_hydro} validates our implementation by comparing 
the numerical results obtained with our LB models with the predictions of the 
first- and second-order hydrodynamics equations \eqref{eq:milne_hydro1}
and \eqref{eq:milne_hydro2}, respectively. As in the case of the
parton gas, the first-order hydrodynamics formulation cannot account for a
vanishing $\Pi$ (right column) at initial time $\tau = 1$ and is thus 
inaccurate for $\tau \lesssim 1 + 5 [4\pi \etas]$. Furthermore, it loses 
applicability when $4\pi \etas \gtrsim 0.5$, especially when considering the 
evolution of the isotropic pressure $P$ (left column). The second-order 
hydrodynamics 
formulation provides a good match for the early time evolution of 
both $\Pi$ and $P$ and seems to lose validity when $4\pi \etas \gtrsim 1$.
Also in Fig.~\ref{fig:milne_hydro}, a comparison between the ideal gas 
(dotted lines and points) and the parton gas (solid black lines) shows that 
the evolution of $P$ and $\Pi$ is (surprisingly) extremely similar in 
the two models.

In Fig.~\ref{fig:milnevsbjork}, the time evolution predicted in the ideal gas ($P = nT$) 
and parton gas ($\mu = 0$) models is further compared by considering the evolution of 
the temperature $T$ (left) and number density $n$ (right). It can be seen that the 
temperature of the ideal gas undergoing the longitudinal boost-invariant 
expansion always exceeds the temperature of the parton gas.
The ratio $n_{\text{ideal gas}} / n_{\text{partons}}$ at fixed values for 
$\tau$ decreases when $4\pi \etas \lesssim 3$ 
from $1$ in the perfect fluid regime [$\etas \rightarrow 0$]
towards a minimum value, after which it 
increases for $4\pi \etas \gtrsim 10$ towards $1$ in the ballistic regime 
[$\etas \rightarrow \infty$].

\subsection{Convergence test} \label{sec:bjorken:conv}

\begin{figure}
\begin{tabular}{c}
 \includegraphics[angle=0,width=0.86\linewidth]{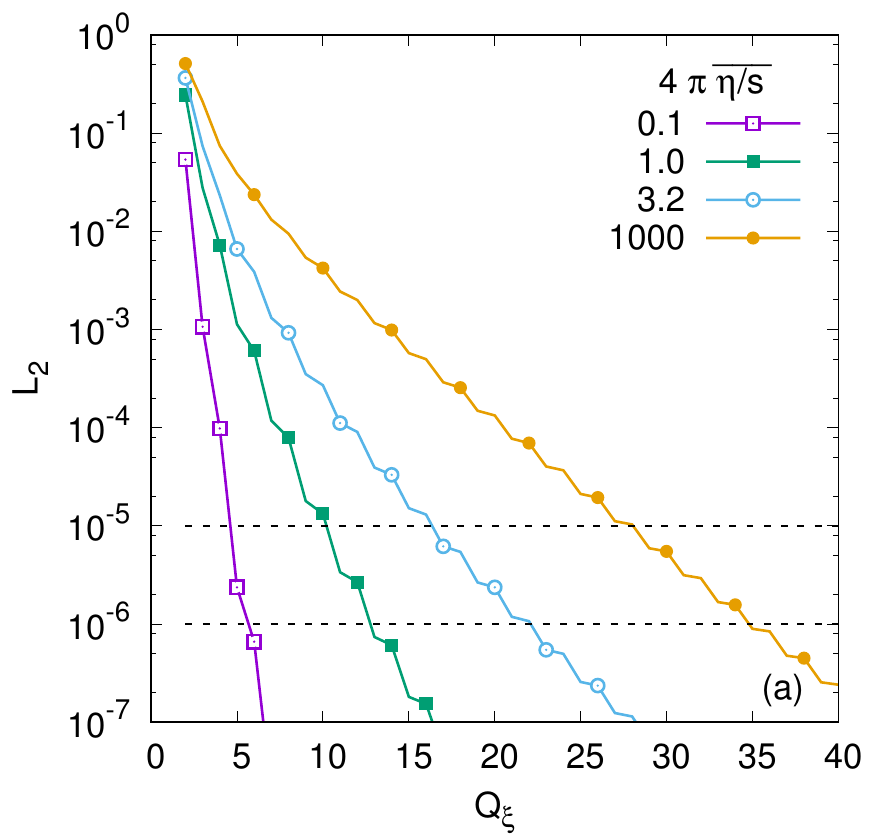} \\
 \includegraphics[angle=0,width=0.86\linewidth]{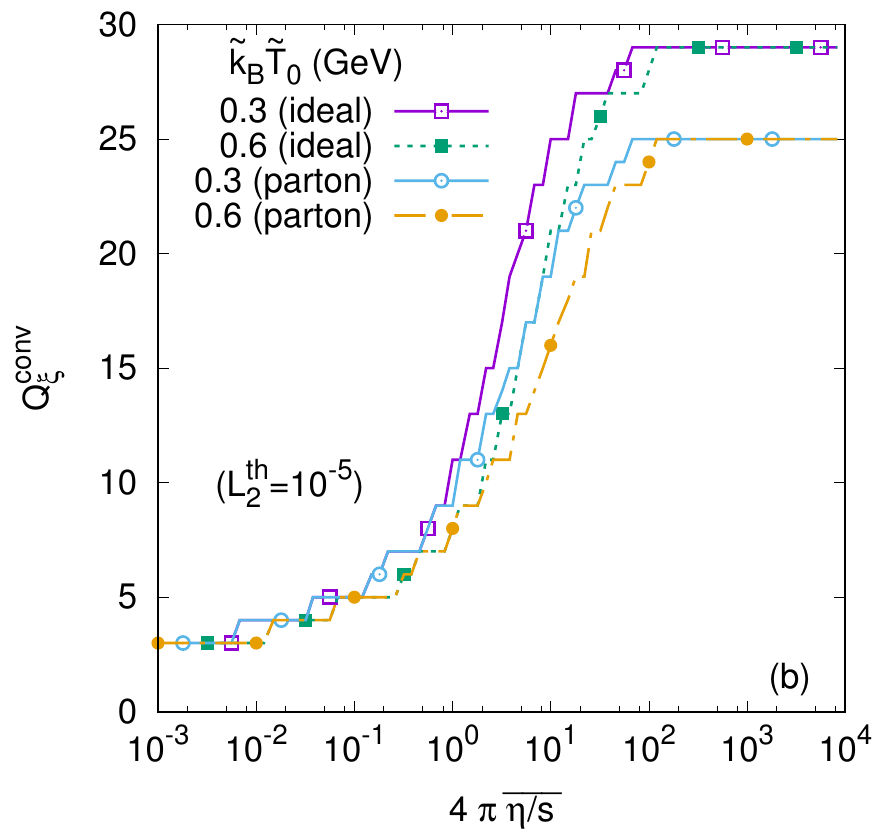} \\
 \includegraphics[angle=0,width=0.86\linewidth]{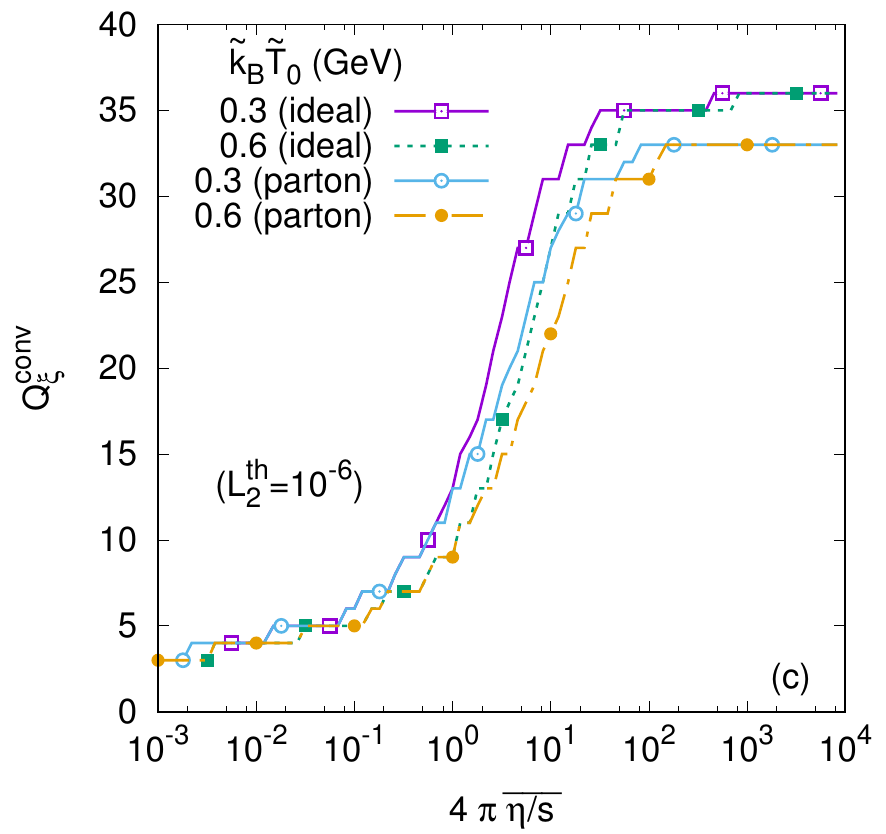}
\end{tabular}
\caption{(a) Maximum $L_2$ norm \eqref{eq:bjork_L2} as a function of $Q_\xi$ 
for $\widetilde{k}_B \widetilde{T}_0 = 0.3\ {\rm GeV}$,
under the assumption of a parton gas ($\mu = 0)$. (b),(c) Dependence 
with respect to $4\pi \etas$ of $Q_\xi^{\rm conv}$ for 
$L_2^{\rtin{th}} = 10^{-5}$ (b) and $L_2^{\rtin{th}} =10^{-6}$ (c), 
obtained for $\widetilde{k}_B \widetilde{T}_0 \in \{0.3\ {\rm GeV}, 0.6\ {\rm GeV}\}$, 
for both the parton and ideal gases. The $L_2$ norm is computed with respect to 
the reference model $\text{R-SLB}(5; 100)$.
\label{fig:bjork_conv}}
\end{figure}

\begin{table}
\begin{ruledtabular}
\begin{tabular}{l|l|r}
 Regime & Range of $4\pi \etas$ & $N_\rtin{vel} $ \\\hline\hline
Inviscid & $ 4\pi \etas \lesssim 10^{-3}$ & $ 6 $ \\
Hydrodynamic & $10^{-3} \lesssim 4\pi \etas \lesssim 0.05$ & $10$ \\
Transition  & $0.05 \lesssim 4\pi\etas \lesssim 1000$ & $10$--$66$ \\
Ballistic & $4\pi\etas \gtrsim 1000$ & $ 66$
\end{tabular}
\end{ruledtabular}
\caption{Number of velocities $N_{\rtin{vel}}$ 
required in the simulation of the Bjorken flow in 
order to reduce below $L_2^{\rtin{th}} = 10^{-6}$ the $L_2$ norm 
for the evolution of $n$, $P$, $T$, and $\Pi$ for a parton gas 
($\mu = 0$) at $\widetilde{k}_B \widetilde{T}_0 = 0.3\ {\rm GeV}$ 
for various flow regimes. The reference model is 
$\text{R-SLB}(5;100)$.
\label{tab:bjork_conv}}
\end{table}

In order to assess the efficiency of our implementation, we consider 
a convergence test similar to that employed in Sec.~\ref{sec:Sod:conv}
in the context of the Sod shock tube. Since the spatial dependence 
of the macroscopic fields is trivial, it is convenient to consider 
the $L_2$ norm of the time evolution of the macroscopic fields 
$n$, $P$, $T$, and $\Pi$ with respect to the results obtained with 
the reference model $\text{R-SLB}(5;100)$ having $N_\Omega = 5$ 
and $Q_\xi = 100$. Denoting by $n_{\rm ref}$, etc.,
the profiles obtained using the reference model, the $L_2$ norm 
for the evolution of the macroscopic field $A \in \{n, T, P, \Pi\}$ 
is computed using 
\begin{equation}
 L_2(A) = \sqrt{\int_1^{\tau_f} \frac{d\tau}{\tau_f - 1}\, [A(\tau) - A_{\rm ref}(\tau)]^2},
\end{equation}
where $\tau_f = 40$.
We considered that convergence was achieved when 
\begin{equation}
 L_2 \le L_2^{\rtin{th}}, \qquad L_2 = {\rm max}_A[L_2(A)],
 \label{eq:bjork_L2}
\end{equation}
where $L_2^{\rtin{th}} \in \{10^{-5}, 10^{-6}\}$.

Figure~\ref{fig:bjork_conv}(a) shows the decrease of $L_2$ with $Q_\xi$ for 
various values of $\etas$. It can be seen that as $\etas$ is increased,
more quadrature points are required in order to achieve the same $L_2$ error norm.
Figures~\ref{fig:bjork_conv}(b) and \ref{fig:bjork_conv}(c) show the minimum quadrature 
order $Q_\xi^{\rm conv}$ required to achieve \eqref{eq:bjork_L2} as a function 
of $4\pi \etas$ between the inviscid limit $10^{-3}$, where $Q_\xi^{\rm conv} = 3$,
and the ballistic limit $10^4$, where $Q_\xi^{\rm conv} = 25$ for 
$L_2^{\rtin{th}} = 10^{-5}$ (b) and $Q_\xi^{\rm conv} = 33$ 
for $L_2^{\rtin{th}} = 10^{-6}$ (c). The convergence
orders are computed separately for $\widetilde{k}_B \widetilde{T}_0 = 0.3\ {\rm GeV}$
and $0.6\ {\rm Gev}$, as well as for the ideal and parton gases.
The results for the convergence test with $L_2^{\rtin{th}} = 10^{-6}$
corresponding to a parton gas at initial temperature 
$\widetilde{k}_B \widetilde{T}_0 = 300\ {\rm GeV}$ 
are summarized in Table~\ref{tab:bjork_conv}.

\subsection{Summary}\label{sec:bjorken:conc}

In this section, the LB models introduced in Sec.~\ref{sec:LB} were
employed to investigate the one-dimensional boost invariant (Bjorken) 
flow in the Milne coordinate system. Our models were validated by 
comparison with analytic solutions in the inviscid and ballistic 
regimes, as well as with the numerical solution of the second-order 
hydrodynamics at finite relaxation times. In the transition regime,
our models were validated by comparison with a semi-analytic solution,
constructed using an iterative process. All comparisons 
show that our models are capable of accurately reproducing the 
benchmark data. Aside from the vanishing chemical potential 
case considered in Ref.~\cite{florkowski13}, corresponding to 
a parton gas comprised of gluons, we presented 
for comparison an analysis of the case when the fluid obeys the 
ideal gas law $P = n T$, highlighting a slower temperature 
decrease compared to the parton gas model.

\section{Summary and conclusion}\label{sec:conc}

In this paper, a relativistic lattice Boltzmann (LB) model was developed for the simulation of flows of ultrarelativistic particles. 
The momentum space is discretized by using a set of quadrature rules, giving rise to off-grid velocity sets. As the procedure for 
selecting the velocities is analytical, the set can be extended up to arbitrarily high orders. This in turn allows  
flows with any degree of rarefaction (i.e., the entire range of the relaxation time) to be described accurately. 
The model is validated in the contexts of the Sod shock tube problem 
and of the Bjorken flow.

In the inviscid limit, both the Sod problem and the Bjorken flow
have analytic solutions (the inviscid limit of the latter case is the 
ideal Bjorken flow discussed in Ref.~\cite{bjorken83}).
In the Anderson-Witting model for the relativistic Boltzmann equation, 
the relaxation time depends linearly on the shear viscosity 
and thus the inviscid result is recovered as the ratio $\etas$ between the 
shear viscosity and entropy density 
(expressed in Planck units) goes to $0$. Our results show a 
clear convergence towards the analytic profiles as $\etas$ is decreased. 

Close to equilibrium, the heat flux and shear pressure can be related to the gradients of 
the temperature, velocity and pressure 
via the constitutive equations of the first order hydrodynamics theory. 
In the case of the Sod shock tube 
problem, the nonequilibrium quantities develop large spikes where the 
macroscopic fields exhibit discontinuities, as confirmed in our plots. 
These spikes can be characterized quantitatively by integrating the heat 
flux and shear pressure around the discontinuities. The numerical results 
for the integrated heat flux and shear pressure 
are in remarkably good agreement with the theoretical estimates 
obtained based on the Chapman-Enskog prediction 
for the transport coefficients (thus invalidating the Grad moment 
expansion values). In the context of the one-dimensional boost-invariant 
expansion, we employed a numerical solution of the second-order 
hydrodynamics theory and found excellent agreement
with our LB results for $4\pi \etas \lesssim 0.5$.

For intermediate relaxation times, our results were validated by comparing 
with the numerical results obtained with the BAMPS 
model \cite{bouras10,bouras09prl,bouras09nucl} in the case of the Sod shock tube problem.
For the Bjorken flow outside the inviscid regime, the semi-analytic results 
reported in Ref.~\cite{florkowski13} were used as benchmark results.
Our models successfully reproduced these results with high accuracy.

Finally, the ballistic regime was considered, where the macroscopic 
profiles can be obtained analytically. 
In this case, the flow constituents stream freely and much 
higher quadrature orders are required in order to 
reproduce the analytic profiles. For the Sod shock tube problem, the smoothness 
of the macroscopic profiles decreases when the grid resolution is increased at 
fixed quadrature order $Q_\xi$, or when $Q_\xi$ is decreased at fixed resolution.
We found that a good agreement with the analytic result is 
obtained with $Z = 1000$ grid points and $Q_\xi = 110$ (220 velocities).
In the case of the one-dimensional boost-invariant expansion, the 
number of quadrature points required increases with the time span of the simulation.
Thus, for a simulation terminated at a time $\tau = 40$ expressed in units of 
the initial proper time, we found that $Q_\xi = 33$ (66 velocities) was 
sufficient to obtain accurate results.

In order to find the combinations of $Q_\xi$ and $N_\Omega$ which yield 
reasonably accurate results for a specific value of $\etas$, a convergence 
test was employed. In the context of the Sod shock tube, 
we tracked the relative error of the profiles of $n$, $P$, and $\gamma$, 
computed with respect to the profiles obtained 
using a reference model with $Q_\xi = 500$ and $N_\Omega = 6$.
The required number of velocities $N_\rtin{vel} = Q_\rtin{L}\times Q_\xi$ 
and expansion order on the angular sector $N_\Omega$,
in the different regimes, are summarized for the Sod shock tube 
problem in Table~\ref{tab:Sod_conv}.
In the context of the Bjorken flow, we considered the evolution of $n$, $P$, $T$,
and $\Pi$ and computed the $L_2$ norm with respect to the result 
obtained using a reference model with $Q_\xi = 100$ for 
$\tau$ between $1$ and $\tau_f = 40$.
The convergence test required that the $L_2$ norm decreased below 
the threshold value of $10^{-6}$. The result of the convergence test
is summarized in Table~\ref{tab:bjork_conv}.

The models introduced in this paper have been successful in reproducing the 
analytic and benchmark results available throughout the whole range of the relaxation time,
making them a reliable and competitive tool for the simulation of relativistic flows of massless 
particles obeying Maxwell-J\"uttner statistics. In the future, we plan to extend our 
models to account for massive particles (with possible density and temperature dependencies 
of the particle effective mass \cite{romatschke12}), as well as quantum statistics (Fermi-Dirac 
and Bose-Einstein statistics \cite{coelho2017}). The current models can be readily applied for 
the study of more complex expansion models for the hydrodynamic phase of the 
quark-gluon plasma produced in heavy ion collision, such as the Gubser 
flow \cite{gubser10} or the radially expanding conformal soliton flow 
\cite{friess07,noronha15}.

{\bf Acknowledgements.} 
The authors are grateful to Dr. M. Mendoza (KTH, Zurich, Switzerland) 
for hospitality during preliminary discussions, 
as well as for suggesting the comparison with the BAMPS data. The authors also thank 
Prof. L.-S. Luo (Old Dominion University, Norfolk, Virginia, USA)
for suggesting the hyperbolicity test. 
The authors are grateful to an anonymous Phys. Rev. C referee for the critical reading 
of the manuscript and for many useful comments.
We are also indebted to the Aptara Editorial Team and to the Physical Review C 
Production Team for their care and patience in preparing this 
manuscript for publication.
This work was supported by a grant of the Romanian National Authority for 
Scientific Research and Innovation, CNCS-UEFISCDI, Project No. PN-II-RU-TE-2014-4-2910.

\appendix

\section{Expansion coefficients for the Maxwell-J\"uttner equilibrium distribution}\label{app:feq}

In this section, the coefficients $a_{\ell, s}^{\rtin{(eq)}}$,
defined in Eq.~\eqref{eq:aeq_def}, are calculated explicitly.
Equation \eqref{eq:aeq_aux} can be brought to a simpler form by changing the integration
variable to $q = 1 - \theta/(u^0 - u \cos\gamma_u)$:
\begin{multline}
 a_{\ell,s}^{\rtin{(eq)}} = \frac{n(\ell + 1)(\ell + 2)}{8\pi u \theta^2} \\
 \times \int_{1 - \theta/(u^0-u)}^{1-\theta/(u^0+u)} dq \, q^\ell (1 - q)
 P_s\left[\frac{1}{u}\left(u^0 - \frac{\theta}{1 - q}\right)\right].
 \label{eq:aeq}
\end{multline}
For simplicity, it is convenient to express $a_{\ell,s}^{\rtin{(eq)}}$ as
\begin{equation}
 a_{\ell,s}^{\rtin{(eq)}} = \frac{n (\ell + 1) (\ell + 2)}{4\pi} u^0 
 \widetilde{a}_{\ell,s}^{\rtin{(eq)}}.
\end{equation}
For the case $\ell = 0$, the following results are obtained:
\begin{widetext}
\begin{align}
 \widetilde{a}_{0,0}^{\rtin{(eq)}} =& 1,\qquad 
 \widetilde{a}_{0,1}^{\rtin{(eq)}} = \frac{u}{u^0},\qquad 
 \widetilde{a}_{0,2}^{\rtin{(eq)}} = \frac{3}{2u^3 u^0} \arcsinh u + 1 - \frac{3}{2u^2},\qquad
 \widetilde{a}_{0,3}^{\rtin{(eq)}} = \frac{15}{2u^4} \arcsinh u -
 \frac{1}{2u^3 u^0}(15 + 5u^2 - 2u^4),\nonumber\\
 \widetilde{a}_{0,4}^{\rtin{(eq)}} =& \frac{15}{4u^5 u^0} (7 + 6u^2) \arcsinh u -
 \frac{1}{4u^4}(105 + 20u^2 - 4u^4),\nonumber\\
 \widetilde{a}_{0,5}^{\rtin{(eq)}} =& \frac{105}{4u^6}(3 + 2u^2) \arcsinh u - 
 \frac{1}{4u^5 u^0}(315 + 315u^2 + 28 u^4 - 4u^6),\nonumber\\
 \widetilde{a}_{0,6}^{\rtin{(eq)}} =& \frac{105}{16 u^7 u^0} (33 + 48 u^2 + 16 u^4) \arcsinh u -  \frac{1}{16u^6}(3465 + 2730 u^2 + 168 u^4 - 16u^6).
\end{align}
When $\ell = 1$, we find the following coefficients:
\begin{align}
 \widetilde{a}_{1,0}^{\rtin{(eq)}} =& \widetilde{a}_{0,0}^{\rtin{(eq)}} - \frac{\theta}{3u^0} (3 + 4 u^2),\qquad 
 \widetilde{a}_{1,1}^{\rtin{(eq)}} = \widetilde{a}_{0,1}^{\rtin{(eq)}} - \frac{4 u \theta}{3},\qquad
 \widetilde{a}_{1,2}^{\rtin{(eq)}} = \widetilde{a}_{0,2}^{\rtin{(eq)}} - \frac{4\theta u^2}{3u^0},\nonumber\\
 \widetilde{a}_{1,3}^{\rtin{(eq)}} =& \widetilde{a}_{0,3}^{\rtin{(eq)}} + \frac{5\theta}{2 u^4 u^0} \arcsinh u - 
 \frac{\theta}{6u^3} (15 - 10u^2 + 8u^4)],\nonumber\\
 \widetilde{a}_{1,4}^{\rtin{(eq)}} =& \widetilde{a}_{0,4}^{\rtin{(eq)}} +\frac{35 \theta}{2u^5} \arcsinh u 
 - \frac{\theta}{6 u^4 u^0}(105 + 35 u^2 - 14 u^4 + 8 u^6),\nonumber\\
 \widetilde{a}_{1,5}^{\rtin{(eq)}} =& \widetilde{a}_{0,5}^{\rtin{(eq)}} + \frac{35\theta}{4u^6 u^0}
 (9 + 8u^2) \arcsinh u -\frac{\theta}{12 u^5} (945 + 210 u^2 - 56 u^4 + 16 u^6),\nonumber\\
 \widetilde{a}_{1,6}^{\rtin{(eq)}} =& \widetilde{a}_{0,6}^{\rtin{(eq)}} + \frac{105\theta}{4u^7} 
 (11 + 8u^2) \arcsinh u -\frac{\theta }{12 u^6 u^0} (3465+3675 u^2 + 378 u^4 - 72 u^6 + 16 u^8).
\end{align}

All coefficients with $s > 1$ are not defined at $u = 0$. To avoid division by zero errors,
in our implementation we used the Maclaurin series expansion of $\widetilde{a}_{\ell,s}^{\rtin{(eq)}}$
up to $u^{11}$ whenever $u < 0.05$:
\begin{gather}
 \widetilde{a}_{0,2}^{\rtin{(eq)}} = \frac{4 u^2}{5}-\frac{24 u^4}{35}+\frac{64 u^6}{105}-\frac{128 u^8}{231}+\frac{512 u^{10}}{1001}+O\left(u^{12}\right),\nonumber\\
 \widetilde{a}_{0,3}^{\rtin{(eq)}} = \frac{4 u^3}{7}-\frac{2 u^5}{3}+\frac{15 u^7}{22}-\frac{35 u^9}{52}+\frac{21 u^{11}}{32}+O\left(u^{12}\right),\qquad
 \widetilde{a}_{0,4}^{\rtin{(eq)}} = \frac{8 u^4}{21}-\frac{128 u^6}{231}+\frac{640 u^8}{1001}-\frac{2048 u^{10}}{3003}+O\left(u^{12}\right),\nonumber\\
 \widetilde{a}_{0,5}^{\rtin{(eq)}} = \frac{8 u^5}{33}-\frac{60 u^7}{143}+\frac{7u^9}{13}-\frac{21 u^{11}}{34}+O\left(u^{12}\right),\qquad 
 \widetilde{a}_{0,6}^{\rtin{(eq)}} = \frac{64 u^6}{429}-\frac{128 u^8}{429}+\frac{1024 u^{10}}{2431}+O\left(u^{12}\right). 
\end{gather}
Similar expansions can be obtained for $\ell = 1$:
\begin{align}
 \widetilde{a}_{1,2}^{\rtin{(eq)}} =& \widetilde{a}_{0,2}^{\rtin{(eq)}} - \frac{4\theta}{3} u^2 + 
 \frac{2\theta}{3} u^4 - \frac{\theta}{2} u^6 + \frac{5\theta}{12} u^8 - \frac{35\theta}{96} u^{10} +O\left(u^{12}\right),\nonumber\\
 \widetilde{a}_{1,3}^{\rtin{(eq)}} =& \widetilde{a}_{0,3}^{\rtin{(eq)}} - \frac{8\theta}{7} u^3 + \frac{64\theta}{63} u^5 - 
 - \frac{640\theta}{693} u^7 + \frac{2560\theta}{3003} u^9 - \frac{1024\theta}{1287} u^{11}+O\left(u^{12}\right),\nonumber\\
 \widetilde{a}_{1,4}^{\rtin{(eq)}} =& \widetilde{a}_{0,4}^{\rtin{(eq)}} - \frac{8\theta}{9} u^4 + \frac{12\theta}{11} u^6
 - \frac{15\theta}{13} u^8 + \frac{7\theta}{6} u^{10} +O\left(u^{12}\right),\nonumber\\
 \widetilde{a}_{1,5}^{\rtin{(eq)}} =& \widetilde{a}_{0,5}^{\rtin{(eq)}} - \frac{64\theta}{99} u^5 + \frac{1280\theta}{1287} u^7
 - \frac{512\theta}{429} u^9 + \frac{28672\theta}{21879} u^{11}+O\left(u^{12}\right),\nonumber\\
 \widetilde{a}_{1,6}^{\rtin{(eq)}} =& \widetilde{a}_{0,6}^{\rtin{(eq)}} -
 \frac{64\theta}{143} u^6 + \frac{32\theta}{39} u^8 - \frac{56 \theta }{51} u^{10}+O\left(u^{12}\right).
\end{align}
\end{widetext}

\section{Hyperbolicity}\label{app:hyp}

In order for a numerical model to provide a reliable numerical solution to 
the relativistic Boltzmann equation, it is desirable that its hyperbolicity 
nature be preserved. In the case of the one dimensional boost-invariant expansion, 
the flow is assumed to be homogeneous with respect to the spatial directions,
thus no advection is performed. In this case, the lattice Boltzmann algorithm 
proposed in Eq.~\eqref{eq:boltz_Milne_disc} is trivially hyperbolic.

In the case of the Sod shock tube, the Boltzmann equation after discretization 
\eqref{eq:boltz_Sod_disc} can be written as:
\begin{equation}
 \partial_t f_{jk} + \xi_j \partial_z f_{jk} = S_{jk},\label{eq:Sod_hyp1}
\end{equation}
where the source term $S_{jk}$ is completely local:
\begin{equation}
 S_{jk} = -\frac{u^0 - u^z \xi_j}{\tau_{\rtin{A-W}}} [f_{jk} - \feq_{jk}].
\end{equation}
Grouping the indices $j$ and $k$ under a single index 
$\kappa$ ($1 \le \kappa \le Q_\xi \times Q_L$), defined such that 
$\kappa = (k - 1) Q_{\xi} + j$, Eq.~\eqref{eq:Sod_hyp1} can 
be written as
\begin{equation}
  \partial_t U_{\kappa} + A_{\kappa,\kappa'} \partial_z U_{\kappa'} = S_{\kappa},
  \label{eq:Sod_hyp2}
\end{equation}
where $U_{\kappa} = f_{jk}$ and 
$A_{\kappa,\kappa'} = \delta_{\kappa,\kappa'} \xi_j$. 
The matrix $A_{\kappa,\kappa'}$ is in diagonal 
form and its eigenvalues $\xi_j$ are the roots of the Legendre polynomial
of order $Q_\xi$, i.e., $P_{Q_\xi}(\xi_j) = 0$. Even though the roots 
of the Legendre polynomials are distinct real numbers satisfying 
$-1 < \xi_j < 1$, each root is repeated $Q_{\rtin{L}} = 2$ times 
due to the radial quadrature. Thus, the system corresponding to 
Eq.~\eqref{eq:Sod_hyp2} is hyperbolic \cite{rezzolla13,toro09}.

\section{Second-order hydrodynamics for the Bjorken flow}\label{app:bjork2}

In order to derive the correct equation obeyed by $\Pi$ at small values of 
the relaxation time $\tau_{\rtin{A-W}}$, a moment expansion 
of Eq.~\eqref{eq:boltz_Milne} 
can be performed, following the procedure described in Ref.~\cite{ambrus18prc}.
In particular, the expansion of $f$ and $\feq$ with respect to the Laguerre 
and Legendre polynomials discussed in Sec.~\ref{sec:LB} is considered:
\begin{align}
 \begin{pmatrix}
  f \\ \feq
 \end{pmatrix}
 =& e^{-p} \sum_{\ell =0}^\infty \frac{L_\ell^{(2)}(p)}{(\ell + 1)(\ell + 2)} 
 \begin{pmatrix}
  \mathcal{F}_{\ell} \\
  \mathcal{F}^{\rtin{(eq)}}_{\ell}
 \end{pmatrix},\nonumber\\
 \begin{pmatrix}
  \mathcal{F}_{\ell} \\
  \mathcal{F}^{\rtin{(eq)}}_{\ell}
 \end{pmatrix} =& 
 \sum_{m= 0}^\infty \frac{2m+1}{2} 
 \begin{pmatrix}
  \mathcal{F}_{\ell,m} \\
  \mathcal{F}^{\rtin{(eq)}}_{\ell,m}
 \end{pmatrix} P_m(\xi),
\end{align}
where $T_0 = 1$ was employed, such that $\wp = p$.
The coefficients $\mathcal{F}_{\ell,m}$ can be written in terms of 
the macroscopic fields $n$, $P$ and $\Pi$ as follows:
\begin{equation}
 \mathcal{F}_{0,0} = n, \quad \mathcal{F}_{1,0} = 3(n - P), \quad 
 3\mathcal{F}_{0,2} - \mathcal{F}_{1,2} = \frac{3}{2}\Pi.
\end{equation}
Furthermore, $\mathcal{F}_{0,1} = \mathcal{F}_{1,1} = 0$ since 
$N^\heta = \mathcal{F}_{0,1}$ and 
$T^{\htau\heta} = 3(\mathcal{F}_{0,1} - \mathcal{F}_{1,1})$ 
are assumed to vanish.
The expansion coefficients of $\feq$ are given by:
\begin{gather}
 \mathcal{F}^{\rtin{(eq)}}_{0,0} = n^{\rtin{(eq)}}, \quad 
 \mathcal{F}^{\rtin{(eq)}}_{1,0} = 3[n^{\rtin{(eq)}} - P^{\rtin{(eq)}}],
\end{gather}
while $\mathcal{F}^{\rtin{(eq)}}_{0,1} = \mathcal{F}^{\rtin{(eq)}}_{1,1} = 
\mathcal{F}_{0,2}^{\rtin{(eq)}} = \mathcal{F}^{\rtin{(eq)}}_{1,2} = 0$.

As remarked in Sec.~\ref{sec:LB:qp}, the evolution of the stress-energy tensor 
is captured exactly even if the expansion with respect to the Laguerre 
polynomials is truncated at $\ell = 1$. 
\begin{equation}
 \begin{pmatrix}
  f \\ \feq
 \end{pmatrix}
 \rightarrow e^{-p} \left[
 \begin{pmatrix}
  \mathcal{F}_0 \\
  \mathcal{F}^{\rtin{(eq)}}_0
 \end{pmatrix} \frac{L_0^{(2)}(p)}{2} + 
 \begin{pmatrix}
  \mathcal{F}_1 \\
  \mathcal{F}^{\rtin{(eq)}}_1
 \end{pmatrix} \frac{L_1^{(2)}(p)}{6} \right],
\end{equation}
where $\mathcal{F}_0$ and $\mathcal{F}_1$ obey Eq.~\eqref{eq:Milne_F}:
\begin{align}
 &\frac{1}{\tau} \partial_\tau (\tau \mathcal{F}_0) - 
 \frac{1}{\tau} \partial_\xi[\xi(1 - \xi^2) \mathcal{F}_0] \nonumber\\
 &\hspace{100pt}= -\frac{1}{\tau_{\rtin{A-W}}} [\mathcal{F}_0 - \mathcal{F}_{0}^{\rtin{(eq)}}], \nonumber\\
 & \frac{1}{\tau} \partial_\tau (\tau \mathcal{F}_1) - \frac{\xi^2}{\tau} 
 [3\mathcal{F}_{0} - \mathcal{F}_1] 
 - \frac{1}{\tau} \partial_\xi[\xi(1 - \xi^2) \mathcal{F}_1] \nonumber\\
 &\hspace{100pt}= -\frac{1}{\tau_{\rtin{A-W}}} [\mathcal{F}_1 - \mathcal{F}_{1}^{\rtin{(eq)}}].
 \label{eq:Milne_F_0_1} 
\end{align}

For simplicity, only the terms up to $m = 2$ are considered
in the expansion with respect to the Legendre polynomials. 
In this case, $\mathcal{F}_\ell$ and 
$\mathcal{F}_{\ell}^{\rtin{(eq)}}$ are approximated as follows:
\begin{equation}
 \begin{pmatrix}
  \mathcal{F}_\ell \\
  \mathcal{F}^{\rtin{(eq)}}_\ell
 \end{pmatrix} \rightarrow \frac{1}{2}
 \begin{pmatrix}
  \mathcal{F}_{\ell,0} \\
  \mathcal{F}^{\rtin{(eq)}}_{\ell,0}
 \end{pmatrix} P_0(\xi) + \frac{5}{2} 
 \begin{pmatrix}
  \mathcal{F}_{\ell,2} \\
  \mathcal{F}^{\rtin{(eq)}}_{\ell,2}
 \end{pmatrix} P_2(\xi).
\end{equation}
Substituting the above expansion into Eq.~\eqref{eq:Milne_F_0_1} gives the following equations:
\begin{align}
 \partial_\tau(\tau \mathcal{F}_{0,0}) + \frac{\tau}{\tau_{\rtin{A-W}}} 
 [\mathcal{F}_{0,0} - \mathcal{F}^{\rtin{(eq)}}_{0,0}] =& 0,\nonumber\\
 \partial_\tau (\tau\mathcal{F}_{1,0}) - \frac{1}{3}(3\mathcal{F}_{0,0} - \mathcal{F}_{1,0})
 -\frac{2}{3}(3\mathcal{F}_{0,2} - \mathcal{F}_{1,2}) & \nonumber\\ 
 + \frac{\tau}{\tau_{\rtin{A-W}}} 
 [\mathcal{F}_{1,0} - \mathcal{F}^{\rtin{(eq)}}_{1,0}] =& 0,\nonumber\\
 \partial_\tau(\tau \mathcal{F}_{0,2}) + \frac{2}{5} \mathcal{F}_{0,0} + 
 \frac{2}{7} \mathcal{F}_{0,2} + \frac{\tau}{\tau_{\rtin{A-W}}} 
 \mathcal{F}_{0,2} =& 0,\nonumber\\
 \partial_\tau (\tau\mathcal{F}_{1,2}) - \frac{2}{5} \mathcal{F}_{0,0} - 
 \frac{11}{7} \mathcal{F}_{0,2} + \frac{8}{15} \mathcal{F}_{1,0}&\nonumber\\
 + \frac{17}{21} \mathcal{F}_{1,2} + \frac{\tau}{\tau_{\rtin{A-W}}} 
 \mathcal{F}_{1,2} =& 0.\label{eq:bjorken_hydro2_FFF}
\end{align}

The above expansion results in the following 
evolution equations:
\begin{align}
 \partial_\tau(\tau n) =& -\frac{\tau}{\tau_{\rtin{A-W}}}[n - n^{\rtin{(eq)}}],\nonumber\\
 4 P^{1/4} \partial_\tau (\tau P^{3/4}) + \Pi =& 
 -\frac{3\tau}{\tau_{\rtin{A-W}}}[P - P^{\rtin{(eq)}}],\nonumber\\ 
 \frac{\tau_{\rtin{A-W}}}{\tau} \partial_\tau (\tau \Pi) + \Pi + 
 \frac{17 \tau_{\rtin{A-W}}}{21\tau} \Pi =& 
 -\frac{16 \tau_{\rtin{A-W}}}{15\tau} P,\label{eq:bjorken_hydro2_Pi_moment}
\end{align}
plus another equation for the fourth degree of freedom in 
Eq.~\eqref{eq:bjorken_hydro2_FFF}.
The conservation equations~\eqref{eq:bjorken_macro} are recovered 
when $\feq$ is defined such that $n^{\rtin{(eq)}} = n$ and 
$P^{\rtin{(eq)}} = P$. While the latter equality is always enforced in 
order to ensure the conservation of the stress-energy tensor, the first 
equality is not valid when the chemical potential is assumed to vanish
\cite{florkowski13}.

\ \\

\bibliography{byblos}

\end{document}